\documentclass[twocolumn, twocolappendix]{aastex701}

\usepackage{amsmath}

\usepackage{graphicx}
\usepackage{txfonts}
\usepackage{hyperref}

\usepackage{float}

\usepackage{subfig}
\usepackage{lipsum}

\begin{document}

\title{Hot Rocks Survey V: Secondary Eclipse Photometry of GJ 3473 b with JWST/MIRI}

\author[orcid=0000-0002-0931-735X]{Måns Holmberg}
\affiliation{Space Telescope Science Institute, 3700 San Martin Drive, Baltimore, MD 21218, USA}
\email[show]{mholmberg@stsci.edu}  

\author[orcid=0000-0001-8274-6639]{Hannah Diamond-Lowe} 
\affiliation{Space Telescope Science Institute, 3700 San Martin Drive, Baltimore, MD 21218, USA}
\email{hdiamondlowe@stsci.edu}  

\author[orcid=0000-0002-6907-4476]{João M. Mendonça}
\affiliation{Department of Space Research and Space Technology, Technical University of Denmark, Elektrovej 328, 2800 Kgs.\,Lyngby, DK}
\affiliation{Department of Physics and Astronomy, University of Southampton, Highfield, Southampton SO17 1BJ, UK}
\affiliation{School of Ocean and Earth Science, University of Southampton, Southampton, SO14 3ZH, UK}
\email{j.mendonca@soton.ac.uk}  

\author[orcid=0000-0003-4269-3311]{Daniel Kitzmann}
\affiliation{Space Research and Planetary Sciences, Physics Institute, University of Bern, Gesellschaftsstrasse 6, 3012 Bern, Switzerland}
\email{daniel.kitzmann@unibe.ch}  

\author[orcid=0000-0001-9513-1449]{Néstor Espinoza}
\affiliation{Space Telescope Science Institute, 3700 San Martin Drive, Baltimore, MD 21218, USA}
\affiliation{Department of Physics and Astronomy, Johns Hopkins University, 3400 N. Charles Street, Baltimore, MD 21218, USA}
\email{nespinoza@stsci.edu}  

\author[orcid=0000-0002-0832-710X]{Natalie H. Allen}
\affiliation{Department of Physics and Astronomy, Johns Hopkins University, 3400 N. Charles Street, Baltimore, MD 21218, USA}
\email{nallen19@jhu.edu}  

\author[orcid=0000-0003-3829-8554]{Prune C. August}
\affiliation{Department of Space Research and Space Technology, Technical University of Denmark, Elektrovej 328, 2800 Kgs.\,Lyngby, DK}
\email{prua@space.dtu.dk}  

\author[orcid=0000-0002-8938-9715]{Mark Fortune}
\affiliation{School of Physics, Trinity College Dublin, University of Dublin, Dublin 2, Ireland}
\email{fortunma@tcd.ie}  

\author[orcid=0000-0003-0854-3002]{Amélie Gressier}
\affiliation{Space Telescope Science Institute, 3700 San Martin Drive, Baltimore, MD 21218, USA}
\email{agressier@stsci.edu}  

\author[orcid=0000-0003-2775-653X]{Jegug Ih}
\affiliation{Space Telescope Science Institute, 3700 San Martin Drive, Baltimore, MD 21218, USA}
\email{jih@stsci.edu}  

\author[orcid=0000-0002-2160-8782]{Erik Meier Valdés}
\affiliation{Department of Physics, University of Oxford, Keble Road, Oxford, OX1 3RH, UK}
\email{erik.meiervaldes@physics.ox.ac.uk}  

\author[orcid=0009-0001-6868-6171]{Merlin Zgraggen}
\affiliation{Center for Space and Habitability, University of Bern, Gesellschaftsstrasse 6, 3012 Bern, Switzerland}
\email{merlin.zgraggen@students.unibe.ch}  

\author[orcid=0000-0003-1605-5666]{Lars A. Buchhave}
\affiliation{Department of Space Research and Space Technology, Technical University of Denmark, Elektrovej 328, 2800 Kgs.\,Lyngby, DK}
\email{buchhave@space.dtu.dk}  

\author[orcid=0000-0002-9355-5165]{Brice-Olivier Demory}
\affiliation{Center for Space and Habitability, University of Bern, Gesellschaftsstrasse 6, 3012 Bern, Switzerland}
\affiliation{Space Research and Planetary Sciences, Physics Institute, University of Bern, Gesellschaftsstrasse 6, 3012 Bern, Switzerland}
\affiliation{ARTORG Center for Biomedical Engineering Research, University of Bern, Murtenstrasse 50, CH-3008, Bern, Switzerland}
\email{brice.demory@unibe.ch}  

\author[orcid=0000-0003-0652-2902]{Chloe Fisher}
\affiliation{Department of Physics, University of Oxford, Keble Road, Oxford, OX1 3RH, UK}
\email{chloe.fisher@physics.ox.ac.uk}  

\author[orcid=0000-0002-9308-2353]{Neale P. Gibson}
\affiliation{School of Physics, Trinity College Dublin, University of Dublin, Dublin 2, Ireland}
\email{n.gibson@tcd.ie}  

\author[orcid=0000-0003-1907-5910]{Kevin Heng}
\affiliation{Ludwig Maximilian University, Faculty of Physics, Scheinerstr. 1, Munich D-81679, Germany}
\affiliation{Munich Center for Geoastronomy, Ludwig Maximilian University, Theresienstrasse 41, D-80333, Munich, Bavaria, Germany}
\affiliation{University College London, Department of Physics \& Astronomy, Gower St, London, WC1E 6BT, United Kingdom}
\email{Kevin.Heng@physik.lmu.de}  

\author[orcid=0000-0001-7216-4846]{Bibiana Prinoth}
\affiliation{Lund Observatory, Division of Astrophysics, Department of Physics, Lund University, Box 118, 221 00 Lund, Sweden}
\email{bibiana.prinoth@fysik.lu.se}  

\author[orcid=0000-0002-6523-9536]{Adam J. Burgasser}
\affiliation{Department of Astronomy \& Astrophysics, UC San Diego, 9500 Gilman Drive, La Jolla, CA 92093, USA}
\email{aburgasser@ucsd.edu}  

\begin{abstract}

JWST is transforming our ability to characterise small exoplanets, from sub-Neptunes to rocky worlds. A key open question is whether highly irradiated rocky planets can retain atmospheres or are stripped bare by stellar irradiation -- a boundary that remains to be mapped observationally. Here we present the first JWST secondary eclipse observations of the rocky exoplanet GJ~3473~b, obtained with MIRI F1500W photometry. Using four visits, we confidently detect the eclipse at an average depth of $186\pm45$~ppm, somewhat lower than expected for a blackbody. We test a wide range of data reduction and analysis assumptions and provide new insights into MIRI detector settling behaviour that will benefit future observations. We model a suite of airless surfaces with varied compositions, textures, and degrees of space weathering, as well as idealised atmospheric scenarios including the possibility of atmospheric collapse. Both atmospheric and bare-rock interpretations remain consistent with the data, but we exclude thick CO$_2$ atmospheres, placing a 95\% credible upper limit of 1.2-6.5 bar on the surface pressure. We also find tentative evidence for visit-to-visit variability in eclipse depth ($33-371$~ppm), though additional data are required to confirm this. Our results highlight the challenges and intrinsic degeneracies in interpreting MIRI F1500W eclipse measurements of rocky exoplanets, indicating that such observations alone may not uniquely distinguish between bare-rock and atmospheric scenarios. Future spectroscopic or phase-curve observations will be required to determine whether or not GJ 3473 b hosts a substantial atmosphere

\end{abstract}

%\keywords{\uat{Galaxies}{573} --- \uat{Cosmology}{343} --- \uat{High Energy astrophysics}{739} --- \uat{Interstellar medium}{847} --- \uat{Stellar astronomy}{1583} --- \uat{Solar physics}{1476}}


\section{Introduction} \label{sec:intro}

The question of whether rocky exoplanets can retain atmospheres in the face of intense stellar irradiation remains a central challenge in exoplanetary science. Although planets like Earth, Venus, and Mars offer examples of secondary atmospheres shaped by outgassing and surface-atmosphere interactions, extrapolating these processes to exoplanets -- especially those orbiting low-mass stars -- comes with many uncertainties. The harsh radiation environments of close-in M-dwarf planets may erode primordial and secondary atmospheres through mechanisms of atmospheric escape, such as photoevaporation, stellar wind stripping, and core-powered mass loss \citep{Lopez2013, luger2015, Owen2017, Ginzburg2018}. Based on the solar system, \cite{Zahnle2017} proposed the idea of a "cosmic shoreline", a boundary that separates bodies with atmospheres from those without depending on the irradiation and escape velocity. While this concept offers a useful organizing principle, recent work has emphasized that the boundary between atmosphere-bearing and airless worlds is not sharp \citep[][Ih et al., in prep]{Ji2025, VanLooveren2025, Pass2025}. Variations in initial volatile inventories, stellar XUV evolution, escape and outgassing mechanisms all contribute to large uncertainties in where the boundary lies for any given planet. These uncertainties make it difficult to predict atmospheric retention from theory alone. Ultimately, observations are needed to anchor and refine our understanding of the different mechanisms at play.

Recent advances with the James Webb Space Telescope \citep[JWST;][]{Gardner2006}, particularly using the Mid-Infrared Instrument \citep[MIRI;][]{Wright2023}, have opened a new window into the atmospheres of small exoplanets, from sub-Neptunes to rocky worlds \citep[e.g.,][]{Greene2023, Zieba2023, Gao2023, Zhang2024, Xue2024, Hu2024, August2025, Tusay2025, Madhusudhan2025}. In the mid-infrared, secondary eclipse photometry and spectroscopy enable direct measurements of a planet's thermal emission, which can be compared with theoretical expectations for bare-rock and atmosphere-bearing scenarios \citep[e.g.,][]{Hu2012, Koll2019b, Mansfield2019}. Because an atmosphere can redistribute heat or have a non-isothermal temperature profile, deviations from the expected thermal flux of an airless, typically dark surface can serve as indirect evidence for atmospheric retention. In addition, absorption by molecules such as CO$_2$ can further decrease the thermal flux at specific wavelengths, e.g. at 15~$\mu$m, if present in the atmosphere. This technique has already been applied to a growing sample of rocky exoplanets orbiting M dwarfs with JWST (see below).

Secondary eclipse photometric observations can serve as a method for identifying candidate atmospheres of hot, tidally-locked, rocky exoplanets \citep{Koll2019b}, especially when compared to spectroscopic or phase-curve observations, which often require a more significant time investment. This relies on the fact that hot bare-rock surfaces are expected to have a low albedo \citep{Koll2019b, Mansfield2019}, coming from materials such as basaltic rock, in addition to the darkening effects of space weathering \citep[e.g.,][]{Hapke2001, Pieters2000, Pieters2016}. Consequently, a deep eclipse depth, as measured by photometry, may suggest that an exoplanet is airless, while a shallow eclipse could indicate the presence of an atmosphere. This makes secondary eclipse observations a valuable tool for screening potential targets for further study. This is the goal of the Hot Rocks Survey (JWST GO program 3730, PI: H. Diamond-Lowe, Co-PI: J. M. Mendonça), designed to search for atmospheres on nine rocky exoplanets orbiting M dwarfs using secondary eclipse photometry with MIRI.

Recent MIRI photometry of TRAPPIST-1~b and c has shown eclipse depths consistent with bare-rock surfaces, disfavouring the presence of thick CO$_2$ atmospheres \citep{Greene2023, Ih2023, Zieba2023, Lincowski2023}. However, \citet{Ducrot2025} proposed that the thermal emission of TRAPPIST-1~b could also be explained by a CO$_2$-dominated atmosphere with a thermal inversion, potentially caused by photochemical hazes. Observations of LHS~1140~c \citep{Fortune2025}, TOI-1468~b \citep{MeierValdes2025}, and LTT 3780~b \citep{Allen2025} are likewise consistent with atmosphere-free, rocky surfaces. In contrast, LHS~1478~b exhibits a notably shallow eclipse depth at 15~$\mu$m \citep{August2025}, potentially suggesting the presence of an atmosphere. However, the two observations of LHS~1478~b were found to be inconsistent due to systematics. This planet is scheduled for follow-up observations with both MIRI LRS and additional F1500W photometry under JWST GO program 7675 (PI: P. August), which aims to determine whether it possesses an atmosphere. In addition, several other hot rocky exoplanets orbiting M-dwarfs -- LTT~1445A~b \citep{Wachiraphan2025}, GJ~1132~b \citep{Xue2024}, GJ~486~b \citep{WeinerMansfield2024}, and GJ~367~b \citep{Zhang2024} -- have been observed using MIRI LRS during secondary eclipse. The thermal-emission spectra for these exoplanets are reported to be consistent with bare-rock scenarios, showing no strong evidence of substantial atmospheres. On the other hand, transmission spectroscopy observations of the small M-dwarf planets L 98-59 b and d have provided tentative evidence for the presence of atmospheres \citep{Gressier2024, Banerjee2024, BelloArufe2025}. However, several other transmission spectroscopy observations of small planets orbiting M-dwarfs have reported non-detections or inconclusive results regarding atmospheric spectral features \citep[e.g.,][]{LustigYaeger2023, moran_high_2023, Lim2023, May2023, Scarsdale2024, Alam2025, Alderson2024, Alderson2025}.

In this work, we present four secondary eclipse observations of GJ~3473~b, a rocky exoplanet orbiting a slowly-rotating M dwarf ($T_\mathrm{eff} = 3347\pm54$~K, $J = 9.62$) with an orbital period of 1.2~days, as part of the Hot Rocks Survey. We use JWST MIRI F1500W photometry to assess whether the planet exhibits thermal emission consistent with a bare, atmosphere-free surface, or whether the observed eclipse depth suggests the presence of an atmosphere. GJ~3473~b has a radius of $1.264\pm 0.050 \,R_\oplus$ and a mass of $1.86\pm0.30 \,M_\oplus$ \citep{Kemmer2020}, with a zero-albedo equilibrium dayside temperature of approximately  $T_\mathrm{day} = 1003\,\mathrm{K}$ (assuming no redistribution). A recent analysis by \citet{Ji2025} suggests that GJ~3473~b may retain a CO$_2$ atmosphere, provided it formed with a sufficiently large initial volatile inventory; however, it is less likely to have an atmosphere compared to most planets in their work. Our observations represent the first JWST characterization of GJ~3473~b and provide direct constraints on its dayside emission, contributing to the broader effort to determine if these worlds retain atmospheres under extreme irradiation.

In Section~\ref{sec:obs}, we describe the JWST observations, data reduction, and light curve fitting. Section~\ref{sec:modeling} presents our forward modelling of the planet's thermal emission, exploring a wide range of surface and atmospheric scenarios, including space weathering for airless surfaces. In Section~\ref{sec:results}, we systematically compare these models with the measured eclipse depth of GJ~3473~b to assess the plausibility of each scenario. In Section~\ref{sec:discussion} we discuss the possibility of atmospheric collapse and the potential evidence for eclipse depth variability. Finally, in Section~\ref{sec:conclusion}, we summarise our work and discuss the broader implications of our findings for atmospheric retention on GJ~3473~b.

\section{Observations and data analysis} 
\label{sec:obs}

We observed four secondary eclipses of the exoplanet GJ~3473~b using MIRI F1500W photometry as part of the JWST GO program 3730. The observations took place on 12 March 2024, 13 March 2024, 30 March 2024, and 20 October 2024, with each observation lasting 3.4 hours (corresponding to a total charge time of 19.1 hours). We utilised the SUB256 subarray in FASTR1 readout mode with 39 groups per integration, resulting in a total of 1017 integrations per visit. Across the observations, the peak flux was kept below 50\% detector saturation. Notably, there was no high-gain antenna movement during any of the visits. In addition to the MIRI observations, we also refined the orbital parameters of GJ~3473~b using TESS sectors 34, 61, and 88, as described in Appendix~\ref{app:TESS}. Below, we describe the data reduction and light curve fitting of the MIRI data, for which we used a version of the \texttt{JExoRES} pipeline \citep{holmberg2023}. Leveraging our four visits, we also characterise the detector settling of MIRI to inform aspects of the analysis. 

\subsection{Data reduction} \label{sec:reduction}

We begin by performing Stage~1 and Stage~2 processing using the JWST pipeline \citep{Bushouse2020}. This includes the data quality initialization, emicorr, saturation, first-frame, last-frame, linearity, RSCD, dark current, and ramp fitting steps. We omit the jump detection step in Stage~1 and instead identify outliers during Stage~3. Following this, we run the source-type, assign wcs and flat-field corrections in Stage~2. Stage~3 processing includes outlier correction, background subtraction, and aperture photometry. For the outlier and bad pixel correction step, we first mask all pixels flagged by the pipeline. We then identify temporal outliers by comparing each pixel’s light curve to a median-filtered version, flagging any points that deviate by more than 5$\sigma$. To account for cosmic ray hits that can affect neighbouring pixels, we additionally mask adjacent pixels around each flagged outlier. Affected values are interpolated linearly in time. Bad pixels are similarly corrected by interpolating along detector columns.

We perform the background subtraction in two steps. First, we perform a common background subtraction derived from the time-average image. To do this, we iteratively fit a cubic polynomial to each detector column while removing 3$\sigma$ outliers. To start the process, we initially mask pixels within a 15-pixel radius of the point-spread-function (PSF) centre. In the end, we obtain a background model and a mask of background pixels. Second, after subtracting this model from each image, we perform a time-dependent background subtraction using the average of all background pixels. After background subtraction, the standard deviation of the background pixels in the average image is less than 1~DN/s, with the largest deviation being 3.1~DN/s. Thus, in a worst-case scenario, where all pixels in the aperture have an excess flux of 3.1~DN/s, the eclipse depth would be diluted by less than 0.7~\%\footnote{The dilution factor can be calculated via $1 / (1 + \mathrm{F}_\mathrm{bgr} / \mathrm{F}_\star)$, where $\mathrm{F}_\mathrm{bgr}$ correspond to the total excess background flux within the aperture, and $\mathrm{F}_\star$ is the stellar flux.}. As an additional method, we employed traditional aperture photometry by subtracting the average flux within an annulus that spans 20 to 40 pixels \citep[e.g.,][]{August2025} from the centroid position (on a per-integration basis). We find that the eclipse depth is consistent between the two methods, differing only by a few ppm. Thus, we conclude that inaccuracies in background subtraction do not significantly affect the inferred eclipse depth.

Next, we fit the PSF of each image using a Gaussian to obtain an estimate of the centroid and width as a function of time. Using the average centroid position, we then create the extraction mask with a nominal radius of 6~pixels, as discussed below and in Appendix~\ref{app:aperture}. After extracting the flux, we remove integrations with more than 1\% of the flux masked (within the aperture) to limit the influence of interpolated data. For robustness, we additionally search for outliers in the extracted light curves and mask integrations with 4$\sigma$ outliers, using the same method as above. This results in an average of 33 discarded integrations per visit. Furthermore, we note that there is a large cosmic ray hit at the edge of the aperture in the second observation, occurring after the end of the eclipse. This caused a persistence signal in which the flux of a single pixel increased by 1-2\% for the remaining of the observation \citep[similar events have been noted by][]{Xue2024, Fortune2025}. However, given that the affected pixel accounts for less than 0.1\% of the flux within the 6-pixel aperture, it does not significantly affect the inferred eclipse depth.

\subsection{Detector settling} \label{sec:detector_settling}

\begin{figure}
\centering
	\includegraphics[width=0.48\textwidth]{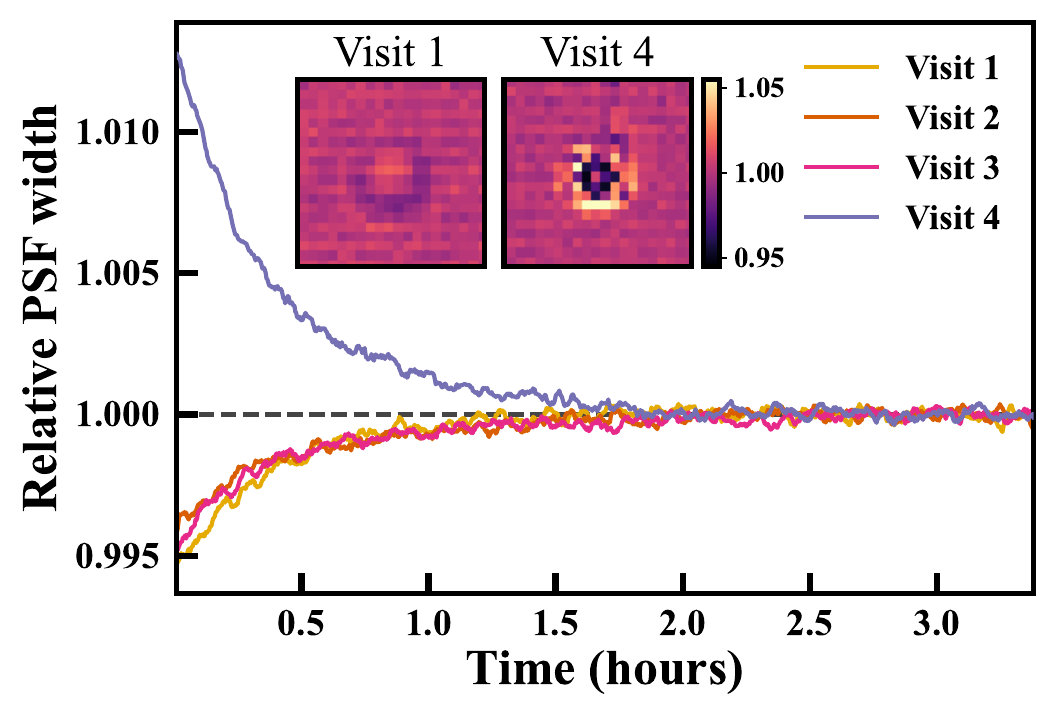}
    \vspace{-5mm}\caption{ Settling behaviour of the point-spread function (PSF) across the four visits. Visits 1–3 exhibit an increase in PSF width over time, while visit 4 shows the opposite trend. The two images display the flux ratio between the average of the first 20 integrations and the average of the final quarter of integrations for visits 1 and 4, respectively, prior to background subtraction. In visits 1–3, the central pixels are initially brighter and the surrounding pixels fainter, resulting in a narrow PSF at the start of the observations. In visit~4, the opposite pattern is observed, with a broader initial PSF that narrows over time.  }
    \label{fig:systematics} 
\end{figure}

MIRI time-series observations reveal a range of detector settling behaviours \citep[e.g.,][]{Greene2023, Bouwman2023, Bell2024, Zhang2024}, which appear as exponential-like ramps in the light curves with either positive or negative amplitudes. In this analysis, we characterize the settling behaviour observed in our four observations to enhance our understanding of this phenomenon and to inform the data reduction and light curve fitting processes.

By inspecting individual pixel light curves, we find that the detector settling behaviour varies across the PSF, as mentioned by \citet{Fortune2025}. Specifically, we find that the settling appears to depend on a pixel’s location within the PSF. We identify three distinct behaviours: (1) pixels at the edge of the PSF, which receive low illumination, show little to no settling; (2) pixels near the PSF core, which are highly illuminated, exhibit strong exponential-like ramps; (3) pixels in an intermediate annulus display exponential trends with amplitudes opposite in sign to those of the central pixels. This spatial variation results in time-dependent changes in the overall PSF width, either broadening or narrowing, depending on the visit. In observations 1–3, the PSF width increases over time, whereas in observation 4 it decreases, stabilizing after approximately two hours, as illustrated in Figure~\ref{fig:systematics}. Although the amplitude of this effect varies, the timescale appears similar across visits. We measure an average e-folding time of $22.1\pm2.1$ minutes, in good agreement with the inferred light curve settling timescale of $24.0\pm2.6$ minutes (from the canonical case; see Section~\ref{sec:light_curves}). Moreover, we find that the direction of the PSF width evolution correlates with the sign of the light curve ramp\footnote{We measured the flux at the beginning of each observation to be $1950\pm220$~ppm, $1450\pm240$~ppm, $1900\pm230$~ppm, and $-2940\pm240$~ppm relative to the baseline flux for visits 1-4, respectively. These values correspond to $p_1$ in the first case of equation~\eqref{eq:sys}. }: increasing width corresponds to a negative flux ramp, and vice versa. This suggests that temporal variations in the PSF may be related to the observed systematics in the light curves. Consequently, the PSF width time series may be valuable for light curve modelling, either as a direct input or to inform priors. Finally, the variation in settling behaviour between visits may be linked to persistence effects and differing illumination histories \citep{Dicken2024, Fortune2025}. Notably, prior to the fourth observation, the target was observed with the F560W filter for 45 minutes, delivering significantly more flux than with the F1500W filter. In contrast, the other visits were preceded by either the P750L prism or the F1500W filter. This discrepancy may explain why the fourth visit exhibits distinct settling characteristics compared to the others.

Due to the spatially dependent settling effects described above, the choice of aperture size can directly influence the shape of the extracted light curves \citep{Allen2025} and consequently, potentially affect the inferred eclipse depth. We find that smaller aperture radii produce stronger exponential ramps, while larger apertures tend to suppress these trends. This occurs because the opposing exponential behaviours seen in different parts of the PSF (e.g., positive near the centre and negative in surrounding annuli) partially cancel out when a broader region is integrated. However, this cancellation is incomplete; residual systematics remain even at large aperture sizes. To assess the impact of aperture size on the eclipse depth, we explore a range of aperture radii in Appendix~\ref{app:aperture}. We find that the uncertainty in the inferred eclipse depth is larger for smaller apertures, but stabilizes for apertures beyond a certain size. This highlights the importance of aperture selection in the presence of spatially variable systematics, particularly for high-precision eclipse measurements.

\subsection{Light curve fitting} \label{sec:light_curves}

\begin{figure}
\centering
	\includegraphics[width=0.49\textwidth]{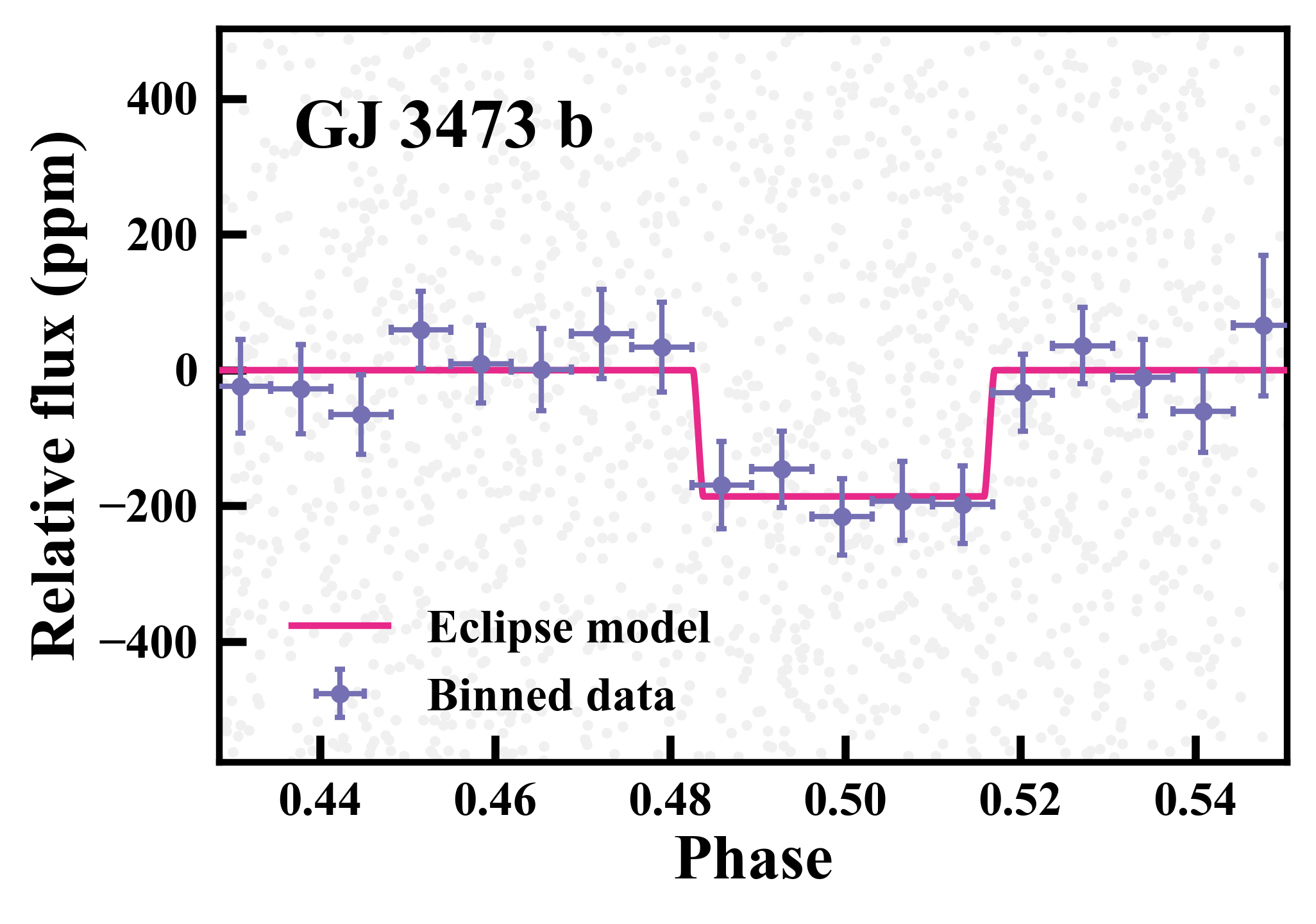}
    \caption{ Phase-folded and systematics-corrected MIRI data and secondary eclipse model of GJ~3473~b. The unbinned and binned data are shown in light grey and purple, respectively. To detrend the data, we subtracted the predicted GP models and divided by the systematic models. The pink curve corresponds to an eclipse model with parameters from Table~\ref{tab:MIRI_params}, corresponding to our canonical case. }
    \label{fig:miri_phase_folded} 
\end{figure}

To perform the light curve fitting, we utilise \texttt{batman} to model the eclipses \citep{kreidberg_batman_2015} and \texttt{MultiNest} \citep{Feroz2009} to sample from the posterior distributions using nested sampling \citep{Skilling2004}. We jointly fit the light curves from the four visits using a variety of different assumptions, as outlined in Table~\ref{tab:eclipse_depth}. For all cases, we sampled the posterior distributions using 1000 live points. We show the systematics-corrected and phase-folded data along with our canonical model fit in Figure~\ref{fig:miri_phase_folded}.

\begin{table*}
\renewcommand{\arraystretch}{1.25} % Default value: 1
\setlength{\tabcolsep}{7pt}
\centering
\begin{tabular}{llllcc}
\hline \hline
Case & Trend & GP kernel & Orbit & $\ln B$ & Eclipse depth \\ \hline
Canonical case & Exp & Exp & Circular & $0$ & $186_{-45}^{+45}$ ppm\\  \hline %\hdashline
Effects of trend assumptions & Exp \& linear function & Exp & Circular & $-3.5$ & $185_{-48}^{+49}$ ppm \\ 
 &  Exp, separate time-scales & Exp & Circular & $-4.3$ & $192_{-51}^{+50}$ ppm \\ 
& PSF width & Exp & Circular & $-6.7$ & $189_{-54}^{+53}$ ppm \\ 
& Exp \& PSF position & Exp & Circular & $-17.9$ & $174_{-38}^{+39}$ ppm \\ \hline %\hdashline
Effects of GP assumptions & Exp & Matern-3/2 & Circular & $-1.0$ & $185_{-45}^{+45}$ ppm \\
& Exp & Exp, separate & Circular & $-0.9$ & $160_{-45}^{+46}$ ppm\\
& Exp & None  & Circular & $-1.8$ & $185_{-34}^{+34}$ ppm \\  \hline %\hdashline
Eccentric orbit & Exp & Exp & Eccentric & $-2.4$ & $184_{-52}^{+52}$ ppm \\ \hline %\hdashline
Variable eclipse depth & Exp & Exp & Circular & $2.1$ & see Table~\ref{tab:variability} \\ 
Variable eclipse depth, no GP & Exp & None & Circular & $1.5$ & see Table~\ref{tab:variability} \\ 
No eclipse depth & Exp & Exp & Circular & $-5.1$ & Fixed at 0 \\ \hline
\end{tabular}
\vspace{1mm}
\caption{The MIRI F1500W eclipse depth of GJ~3473~b given different model assumptions. The model evidence for the canonical case is $\ln Z_0 = -19443.0 $. The natural logarithm of the Bayes factor for each model is defined as $\ln B = \ln Z - \ln Z_0$. We consider $|\ln B\,| > 2.5$ and $|\ln B\,| > 5.0$ as moderate and strong evidence \citep{Trotta2008}, respectively.
}
\vspace{-5mm}
\label{tab:eclipse_depth}
\end{table*}

\begin{figure*}
        \subfloat{%
            \includegraphics[width=.5\linewidth]{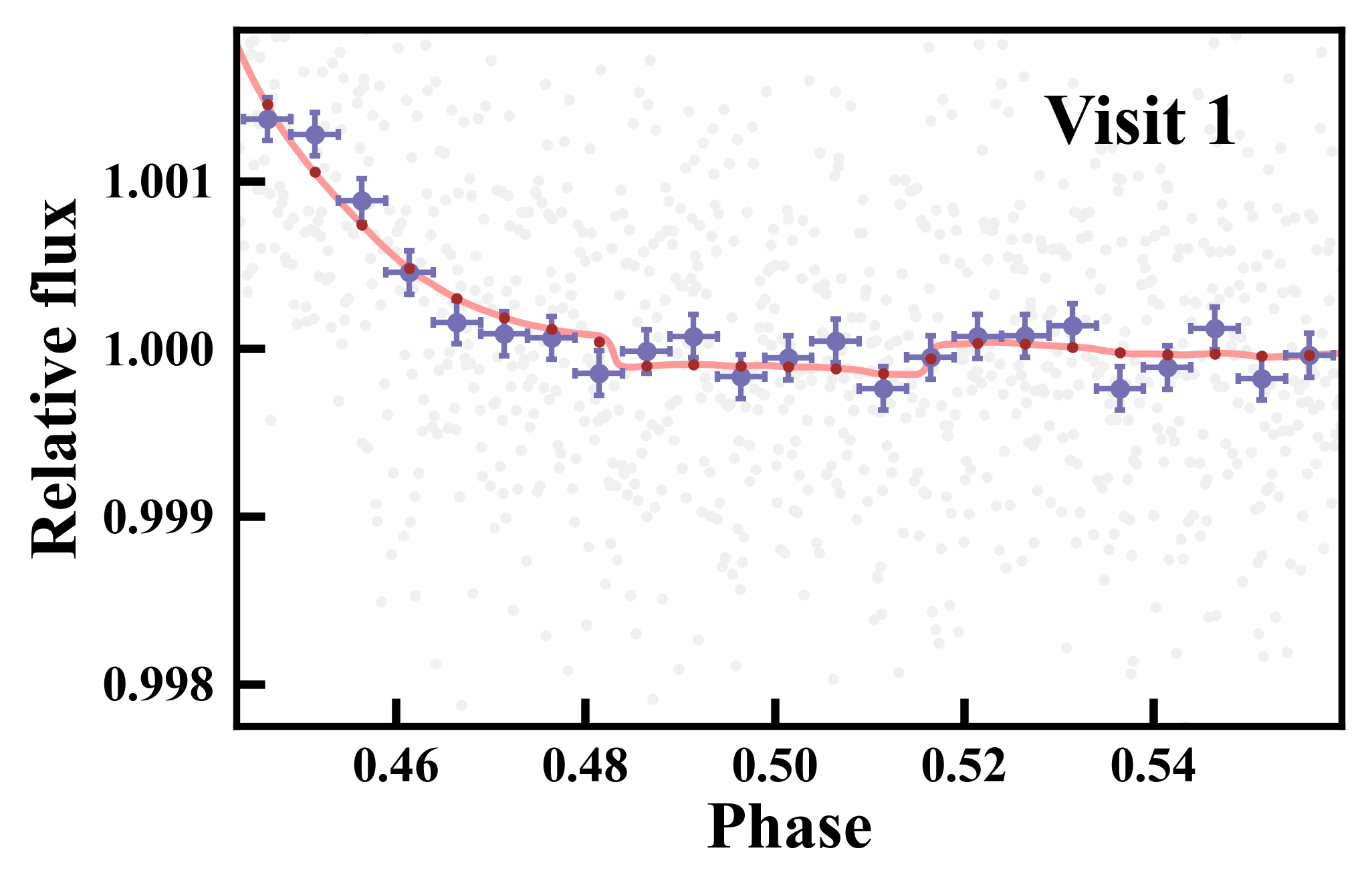}%
            \label{subfig:a}%
        }\hfill
        \subfloat{%
            \includegraphics[width=.5\linewidth]{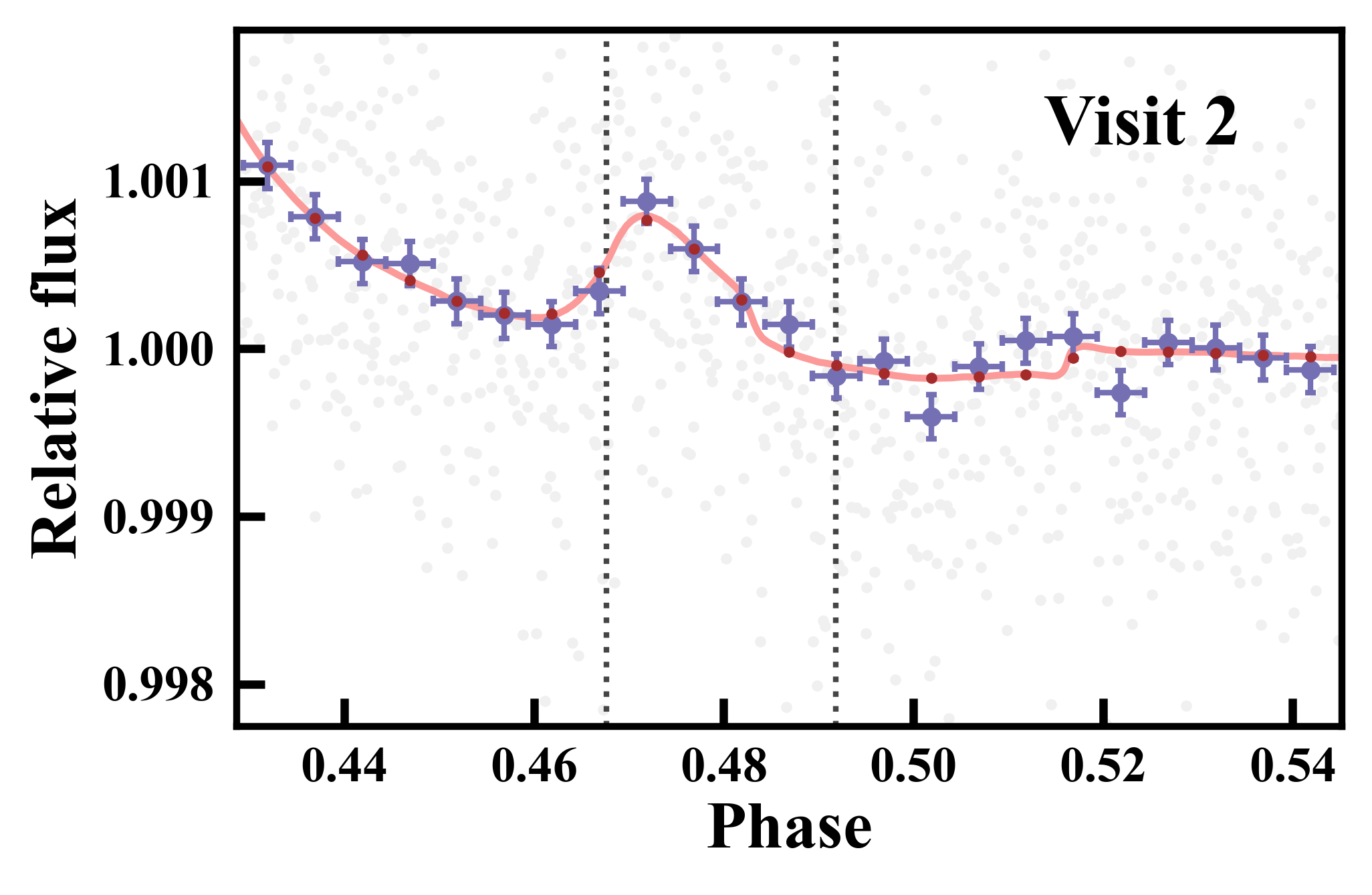}%
            \label{subfig:b}%
        }\\[-1.5em]
        \subfloat{%
            \includegraphics[width=.5\linewidth]{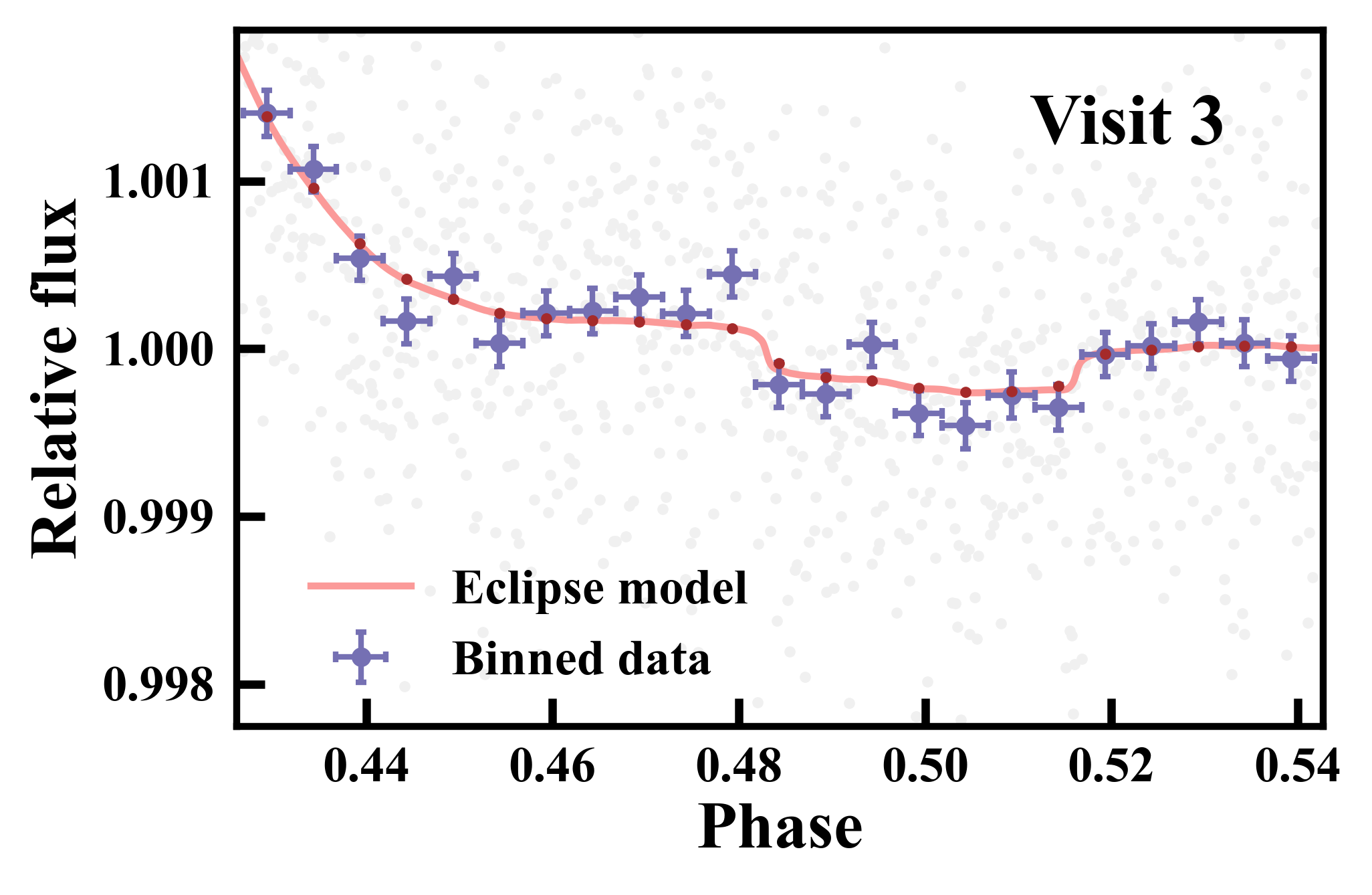}%
            \label{subfig:c}%
        }\hfill
        \subfloat{%
            \includegraphics[width=.5\linewidth]{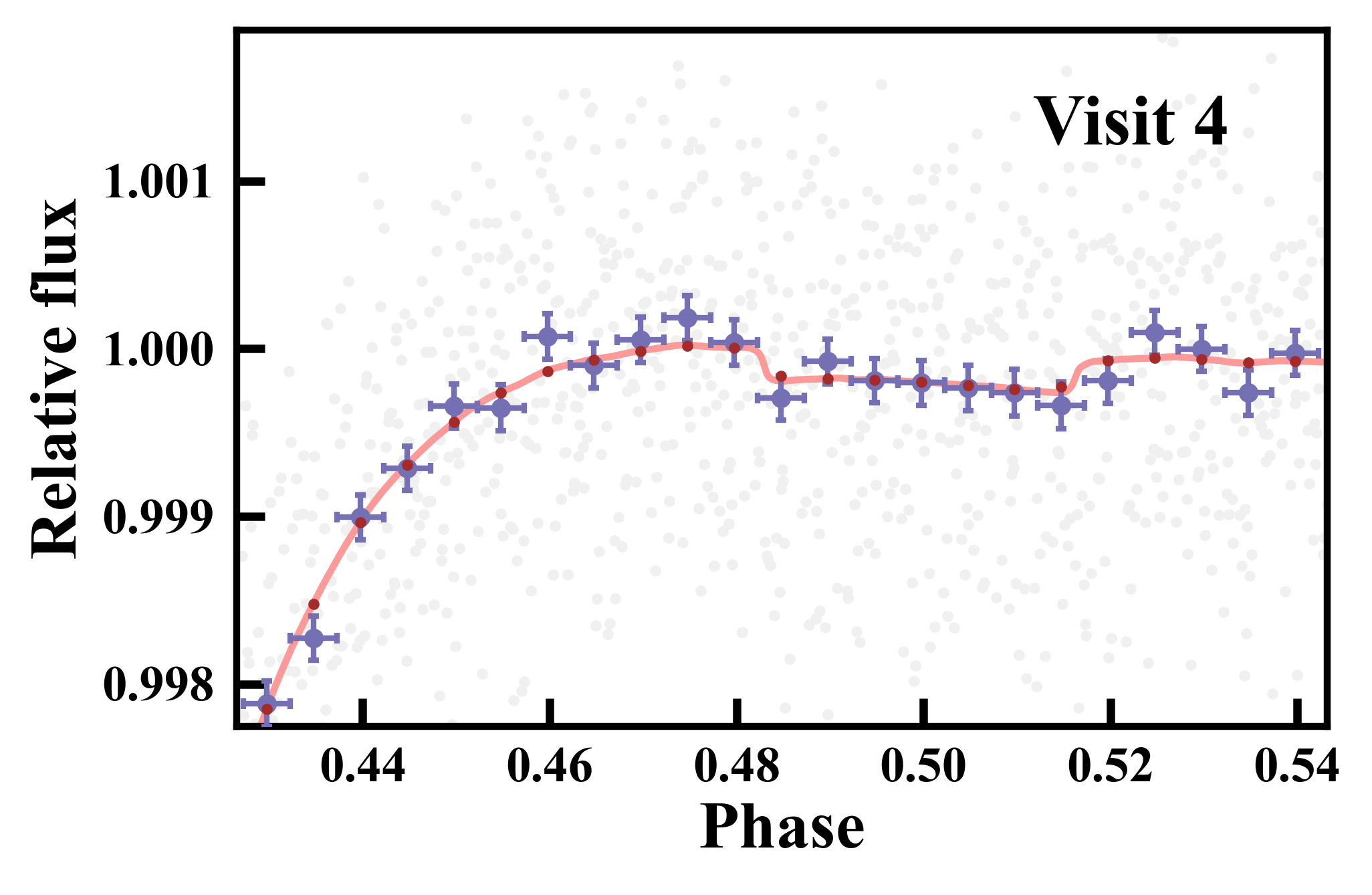}%
            \label{subfig:d}%
        }
        \vspace{-3.5mm}
\caption{MIRI F1500W light curves of GJ 3473~b at secondary eclipse. The raw and binned data are shown in light grey and purple, respectively. The pink curves correspond to the median model predictions from the joint fit, while the brown data points depict the binned model. The model shown here corresponds to the \cite{TovarMendoza2022} flare model case, as described in Appendix~\ref{app:flare}, used to model the flare-like feature during the second visit. The dotted vertical lines show the nominally masked region in the second visit. Compared to Figure~\ref{fig:miri_phase_folded}, this figure displays the data without correcting for systematics.}
        \label{fig:MIRI_all}

\end{figure*}

In general, we model the data for each visit as
\begin{equation} \label{eq:obs_flux}
    F_\mathrm{obs}(t) = F_\star \, \epsilon_\mathrm{sys}(t)\left(f_\mathrm{eclipse}(t) + f_\mathrm{flare}(t) \right)\,,
\end{equation}
where $F_\star$ is the stellar flux; $\epsilon_\mathrm{sys}$ corresponds to our systematic model; $f_\mathrm{eclipse}$ is the eclipse model; and $f_\mathrm{flare}$ is a flare model \citep{TovarMendoza2022}, described in Appendix~\ref{app:flare}. To model the detector settling, we employ a range of possible functions:

\begin{equation} \label{eq:sys}
\epsilon_\mathrm{sys}(t) = \begin{cases}
          1 + p_1 \, e^{-(t - t_0) / p_2}  \\[0.5em]
          (1 + p_1 \, e^{-(t - t_0) / p_2}) (1 + p_3 (t - t_0))  \\[0.5em]
          1 + p_1 w(t) \\[0.5em]
          (1 + p_1 \, e^{-(t - t_0) / p_2}) (1 + p_3 \, x(t) + p_4 \, y(t))
     \end{cases}
\end{equation} 
where $p_i$ are free parameters, $t_0$ is the time at the start of the visit, and $w$, $x$, and $y$ are the time series of the position coordinates and width of the PSF\footnote{These time series are filtered with a median filter to reduce noise, using a window size of 15 integrations.}, respectively. For our canonical case, we use an exponential function (first case in equation~\eqref{eq:sys}) with a common time-scale across all visits, while allowing the amplitudes to differ, following the discussion in Section~\ref{sec:detector_settling}. Such a model is preferred over a model where the time-scales are independent with a log Bayes factor of $|\ln B\,| = 4.3$.  Using any of the other functions in equation~\eqref{eq:sys} yields lower evidence, as shown in Table~\ref{tab:eclipse_depth}. However, the inferred eclipse depth remains consistent among these different choices.

We mask the first few integrations (2 minutes) of each observation to remove any short-lived settling effects and, for the second visit, we also nominally mask integrations numbered 325 to 524. These additional integrations are masked to mitigate the effects of a temporary $\sim$700~ppm increase in the flux just before the start of the eclipse during the second visit, as shown in Figure~\ref{fig:MIRI_all} -- potentially caused by a stellar flare. As discussed in Appendix~\ref{app:flare}, we also modelled this excess flux with a flare model instead of masking it. However, we find consistent results between the two approaches. Additionally, we investigated the impact of masking different numbers of integrations at the beginning of the observations, as pursued in other works \citep[e.g.,][]{August2025}. We found that removing 2, 15, 30, or 60 minutes (corresponding to 10, 75, 150, and 300 integrations) does not significantly affect the inferred eclipse depth, as these result in average eclipse depths of $186_{-45}^{+45}$~ppm ($\chi^2_\mathrm{\nu}$\,=\,$1.003$), $192_{-41}^{+42}$~ppm ($\chi^2_\mathrm{\nu}$\,=\,$1.005$), $200_{-41}^{+43}$~ppm ($\chi^2_\mathrm{\nu}$\,=\,$1.007$), and $195_{-44}^{+45}$~ppm ($\chi^2_\mathrm{\nu}$\,=\,$1.009$), respectively. These values are statistically consistent within $<0.34\sigma$, comparable to the variation introduced by alternative trend parameterizations (Table~\ref{tab:eclipse_depth}). As shown in Figure~\ref{fig:MIRI_all}, the systematics model provides a good fit to the data from the start of each observation. Because removing additional data does not improve the fit quality but reduces the available baseline for constraining the exponential ramps and noise properties, we adopt the minimal 2-minute cut.

We account for potential time-correlated noise by using a Gaussian process (GP) model, implemented via \texttt{celerite} \citep{ForemanMackey2017}. For our canonical case, we use an exponential kernel with a common amplitude and time scale across all visits. We also included a jitter term (added in quadrature to the diagonal elements of the covariance matrix) to account for additional white noise. Given that the light curve scatter varies between the different visits, ranging from 847-879~ppm, we kept the jitter term separate for each visit (i.e., adding four parameters). Doing this, we find that the white noise is about 20\% higher compared to the expected uncertainties obtained by propagating the photon and read noise. In context, the measured white noise corresponds to signal-to-noise ratios (SNR) of 1140-1180, which is close, but somewhat higher, than the expected SNR of 1050 from the ETC\footnote{We normalised the ETC model flux density to match the measured flux density of 7.974 mJy in the F1500W bandpass, as described in Section~\ref{sec:absolute_flux}.}. In addition to our canonical case, we also vary the GP assumptions as described in Table~\ref{tab:eclipse_depth}. We find a slight preference for an exponential kernel with common parameters compared to using separate kernel parameters for each visit, as well as compared to using a Matern-3/2 kernel. We note that the lower eclipse depth in the case with separate kernel parameters is a result of non-uniform weighting of the visits in combination with the somewhat variable eclipse depth, as discussed in Section~\ref{sec:variability}. Ultimately, we determine that the overall preference for time-correlated noise is $|\ln B\,| = 1.8$. While this evidence is weak, we choose to use a GP due to its more conservative estimate of the eclipse depth uncertainty.

We confirm that we detect the secondary eclipse with a log Bayes factor of $|\ln B\,| = 5.1$ when compared to a model where the eclipse depth is fixed at zero. For this comparison, we assume a circular orbit which, as noted below, is preferred by the data. This level of evidence is considered strong, according to \cite{Trotta2008}. Additionally, we examine whether the eclipse depth remains consistent across the four observations by allowing it to vary. Among all the tests we considered, this was the only scenario that provided stronger evidence than our canonical model. However, the difference in evidence is relatively weak, with a log Bayes factor of only $|\ln B\,| = 2.1$. We discuss this further in Section~\ref{sec:variability}. 

We also examine whether the data is consistent with a circular orbit for GJ 3473 b or if an eccentric orbit is preferred. For this, we fit the eccentricity $e$ and the argument of periastron $\omega$ using uniform priors ranging between $0-0.5$ and $0-2\pi$ (with periodic boundary conditions), respectively. As shown in Table~\ref{tab:eclipse_depth}, we find that a model with the circular orbit is weakly preferred; however, the eclipse depth remains insensitive to this choice. To accurately model the timing of the eclipse, we consider the effect of light travel time. However, with a stellar radius of $R_\star= 0.364 \, R_\odot$ \citep{Kemmer2020}, this translates to a delay of around 16 seconds, which is only marginally longer than the 12 seconds between integrations. We find that the eclipse timing is consistent with a circular orbit, since $e \cos(\omega) = -0.0010_{-0.0026}^{+0.0028}$ is consistent with zero \citep{Winn2010}. Likewise, the eclipse duration is also consistent with that of a circular orbit, given $e \sin(\omega) = 0.005_{-0.018}^{+0.085}$. Overall, we determine an upper limit for the eccentricity, finding $e < 0.36$ at 95\% confidence. This upper limit is broad because when $|e \cos(\omega)| \gtrsim 0.1$, the eclipse starts occurring outside of the observing windows, at which point we effectively recover the no-eclipse case. By excluding these solutions, the 95\% upper limit becomes $e \lesssim 0.17$, which is primarily set by the upper limit of $e \sin(\omega)$. This result is consistent with previous radial velocity data \citep{Kemmer2020}, which also did not show any preference for an eccentric orbit.

Overall, the canonical case has a total of 20 free parameters used to fit all four visits simultaneously. Among these parameters, 12 vary between the different visits, including the stellar fluxes, settling amplitudes, and the jitter terms. The remaining parameters, which are common across the visits, include the settling time scale, GP amplitude, GP time scale, mid-transit time, orbital period, normalized semi-major axis, inclination, and eclipse depth. The prior distributions and posterior estimates for these parameters are presented in Appendix~\ref{app:parameters}.

\subsection{Absolute flux calibration } \label{sec:absolute_flux}

In addition to measuring the eclipse depth of GJ~3473~b, we also measure the absolute flux density of the star in the MIRI F1500W bandpass. This allows us to compare the stellar flux with that of a model in the present bandpass to ensure consistency. To perform the absolute flux calibration, we follow the method outlined by \cite{Gordon2025}. This involves using a specific aperture radius of 5.69 pixels and a background annulus range of 8.63 to 11.45 pixels, along with an aperture correction factor. The resulting measurement is then converted to physical units (DN/s to Jy) using a time-dependent calibration factor. For each visit, we carry out this flux calibration for all integrations and take the median value. The resulting flux density measurements are $7.974 \pm 0.038$~mJy, $7.976 \pm 0.038$~mJy, $7.975 \pm 0.038$~mJy, and $7.965 \pm 0.038$~mJy, for each visit, respectively. Note that the uncertainty in these measurements arises from a systematic calibration uncertainty of 0.48\%, rather than photon noise, which is negligible in comparison. However, the repeatability among these four measurements appears to be much better, at around 0.07\%, as computed from the standard deviation of the measurements. Furthermore, compared to \cite{Gordon2025}, our Stage~1 processing included the emicorr step, while we did not use the jump step, as described in Section~\ref{sec:reduction}. Nonetheless, we find that these steps do not significantly impact the absolute flux measurements.

Next, we compare our measurements with stellar models. We adopt a BT-Settl stellar model  \citep{Allard2014}, which we linearly interpolated given the stellar parameters from \cite{Kemmer2020}. Along with the distance estimate from GAIA DR3 of $27.315 \pm 0.018$~pc \citep{Gaia2016, Gaia2023} and a stellar radius of $0.364\pm0.012\,R_\odot$ \citep{Kemmer2020}, we obtain a model flux density of $8.3\pm0.5$~mJy. For this estimate, we account for uncertainties in the effective temperature, distance, and stellar radius, with the majority of the uncertainty arising from the stellar radius. In the end, we find that the measured flux is in good agreement with the model prediction. 

We also compute the predicted flux using a SPHINX stellar model \citep{Iyer2023} to evaluate the systematic uncertainty caused by variations in the stellar model. For the SPHINX model, we obtain a model flux density of $8.1\pm0.5$~mJy, which is approximately 3\% different from the BT-Settl stellar model. This small discrepancy in the stellar spectrum has only a slight impact on the derived brightness temperature and eclipse depth model predictions, affecting them by a few per cent. This change is much smaller than the 24\% uncertainty associated with the eclipse depth itself. Therefore, we conclude that the choice of stellar model has minimal influence on the results of this study.

\section{Modelling} \label{sec:modeling}

To assess the likelihood of an atmosphere on GJ~3473~b, we examine various atmospheric and surface model scenarios, assuming that the planet is tidally locked in a 1:1 spin–orbit resonance. Our goal is to explore the potential ranges of eclipse depth within the observed bandpass for these scenarios, allowing us to compare the results with the data. For each scenario, we calculate the eclipse depth as measured in the MIRI F1500W bandpass as follows

\begin{equation} \label{eq:fpfs}
    \left. \frac{F_\mathrm{p}}{F_\star} \right|_{\mathrm{F1500W}} = \left( \frac{R_\mathrm{p}}{R_\star} \right)^2 \frac{ \int \lambda\, F_\mathrm{p}(\lambda) \, w_{\mathrm{F1500W}}(\lambda) \, d\lambda }{ \int \lambda\, F_\star(\lambda) \, w_{\mathrm{F1500W}}(\lambda) \, d\lambda }\,,
\end{equation}
where $w_{\mathrm{F1500W}}$ is the total throughput of JWST/MIRI using the F1500W filter obtained from Pandeia \citep[version 4.0, February 2025;][]{Pontoppidan2016}, and $F_\mathrm{p}$ and $F_\mathrm{\star}$ are the planetary and stellar flux, respectively.

\subsection{Surface scenarios} \label{sec:surface}

We model the dayside thermal emission of an airless planet using the approach of \cite{Paragas2025}. The surface temperature is set by the balance between the absorbed stellar flux and the emitted thermal radiation: 
\begin{equation}
    f \int \varepsilon_d  F_\mathrm{inc} \, d\lambda = \int \varepsilon_h B_\lambda(T_\mathrm{day}) \,d\lambda\,,
\end{equation}
where $F_\mathrm{inc}$ is the incident stellar irradiance, $B_\lambda$ is the Planck function, and $f$ is a geometric (redistribution) factor that accounts for the longitudinal temperature gradient across the dayside. Here, $\varepsilon_d = 1 - r_h$ is the directional emissivity given by Kirchhoff's law, with $r_h$ being the directional-hemispherical reflectance \citep{Hapke2001, Paragas2025}. The hemispheric emissivity $\varepsilon_h$ is the hemispheric average of the directional emissivity. 
Physically, $\varepsilon_d$ characterizes how efficiently the surface absorbs radiation arriving from the single direction of the star, while $\varepsilon_h$ describes how efficiently the surface emits thermal radiation averaged over all outgoing directions. Both $\varepsilon_h$ and $f$ can be derived from the directional–hemispherical reflectance $r_h$ \citep{Hapke2001, Paragas2025}. Once $T_\mathrm{day}$ is determined, the emergent dayside flux of the planet is 
\begin{equation}
    F_\mathrm{p} = \pi \varepsilon_d B_\lambda(T_\mathrm{day}) + r_h F_\star \left( \frac{ R_\star}{a}\right)^2\,,
\end{equation}
where the first term represents thermal emission and the second term represents reflected starlight. We note that in this formalism $r_h$ is a surface property, whereas reflected light is often expressed in terms of the geometric albedo $A_g$, an observational quantity defined as the disk-integrated reflectivity at full phase. For a Lambertian surface, these are related by $A_g = 2/3 \, r_h$. Furthermore, as described in Section~\ref{sec:absolute_flux}, we use a BT-Settl stellar model \citep{Allard2014}, which provides a good match to the measured flux in the MIRI F1500W bandpass. We use the normalised semi-major axis from the TESS data, which gives $a/R_\star = 9.32$, as described in Appendix~\ref{app:TESS}.

We consider a wide range of surface compositions and textures mainly drawn from the recent library of \cite{Paragas2025}. These include Fe-oxidized materials (hematite), ultramafic rocks (dunite xenolith, olivine clinopyroxenite), mafic rocks (olivine pyroxenite, basaltic andesite, Kilauea basalt, olivine gabbronorite), felsic rocks (dalmatian granite, orlando gold granite), and a feldspathic lunar anorthosite sample \citep[Feldspathic;][]{Cheek2009, Hu2012}. Textures range from solid (low reflectance) to crushed or powdered (higher reflectance). While these materials provide a range of possible surface compositions motivated by solar system bodies, this list is not exhaustive. Other materials not considered here may exist on the surface of exoplanets like GJ~3473~b. Furthermore, we note that although we model the surface as being composed of a single material, a realistic surface could consist of a variety of different materials.

In addition to grain-scale texture, macroscopic roughness (e.g. craters) can produce thermal beaming, where thermal radiation is emitted anisotropically, often preferentially toward the star \citep{Spencer1990, Lagerros1998}. We do not account for this effect here, in line with previous works \citep[e.g.,][]{Xue2024}. 

\subsection{Space weathering} \label{sec:space_weathering}

\begin{figure}
	\includegraphics[width=0.49\textwidth]{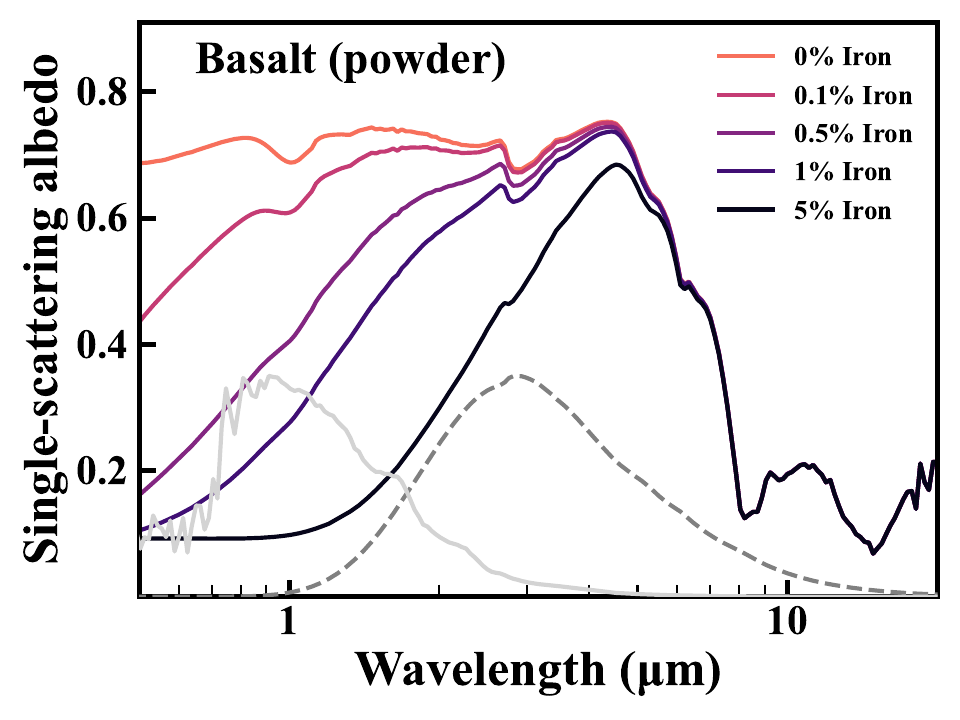}
    \caption{ Single-scattering albedo of Basalt powder at varying degrees of space weathering via the formation of nanophase metallic iron particles. No added space weathering is represented by 0\% iron. The spectra of the stellar irradiation and planetary flux, with 0.5\% iron, are represented by solid light grey and dashed grey lines, respectively; however, these are not to scale for clarity. }
    \label{fig:Basalt_space_weather} 
\end{figure}

We additionally include effects of space weathering \citep[e.g.,][]{Pieters2016}, whereby a surface not protected by an atmosphere is altered by a variety of processes, such as micrometeorite impacts and solar wind irradiation. In the solar system, these processes often lowers the albedo of airless bodies -- darkening the surface and imparting a red slope to the visible and near-infrared reflectance, depending on the mechanism. To capture this effect, we follow \cite{Hapke2001} and \cite{Lyu2024}, modelling space weathering as the addition of small fractions of nanophase metallic iron (npFe$^0$) or graphite into the host surface material. In the solar system, iron explains the low albedo of the Moon \citep{Cassidy1975, Pieters2000}, while graphite has been invoked to explain the darkening of Mercury's surface \citep{Syal2015}. Weathering alters the reflectance of a surface, depending on the volume fraction of iron or graphite, the refractive index of the host material, and the mean photon path length $D$, which we assume is $10$~$\mu$m for all materials -- somewhat smaller than the average grain size of the powdered samples \cite[$\sim$40$~\mu$m;][]{Paragas2025}). The strength of weathering is determined by the volume fraction of these inclusions \citep[up to 5\%;][]{Lyu2024}. We note that the range of eclipse depths at 15$\mu$m resulting from different levels of space weathering is not strongly dependent on the choice of $D$. For the refractive indices, we adopt values for iron and graphite from the Refractive Index database \citep{Polyanskiy2024}. For the different surface types, we obtain the refractive indices from the Aerosol Refractive Index Archive\footnote{\url{https://eodg.atm.ox.ac.uk/ARIA/}}. For ultramafic materials, we used the refractive index of olivine, while for mafic materials, we used the refractive index of basalt. For lunar anorthosite, we instead used a constant (real) refractive index of $1.7$ \citep{Hapke2001}, since we could not find its wavelength-dependent refractive index. In Figure~\ref{fig:Basalt_space_weather}, we show the effects of iron space weathering on the single scattering albedo of basalt powder. In this case, as well as for other materials with a high optical albedo, the reddening caused by iron results in a reduction of the albedo at wavelengths where the incident stellar flux is significant. Meanwhile, it maintains high reflectance (i.e., low emissivity) at longer wavelengths where the planet emits, causing the surface temperature to potentially exceed that of a simple blackbody. 

\subsection{Atmospheric scenarios} \label{sec:atmosphere}

For the atmosphere scenarios, we utilise HELIOS \citep{Malik2017, Malik2019b, Malik2019a} to calculate the temperature structure and emission spectrum of the planet. In this work, we consider well-mixed atmospheres composed entirely of either pure CO$_2$ or H$_2$O, representing end-member cases for potential atmospheric compositions. The model employs k-distribution tables for the opacities, calculated using the HELIOS-K opacity calculator \citep{Grimm2015, Grimm2021}, while integrating radiative fluxes over 386 spectral bands and 20 Gaussian points. We used cross-sections for CO$_2$ \citep{Rothman2010} and H$_2$O \citep{Barber2006}. Additionally, we incorporate Rayleigh scattering \citep{Cox2000, Sneep2005, Thalman2014} and collision-induced absorption cross-sections for CO$_2$ and H$_2$O \citep{Baranov2004, Gruszka1997, Mlawer2023}. In terms of the surface pressure, we consider a wide range of between 0.1 mbar and 100 bar. At the surface, we adopt a constant albedo of 0.1 \citep[e.g.,][]{August2025}. We nominally select an albedo of 0.1, approximately corresponding to basalt, rather than zero, since an atmosphere would protect the surface against space weathering. For the heat redistribution factor, we use the approximate scaling relation  by \cite{Koll2022}. The stellar spectrum is derived from the BT-Settl model grid \citep{Allard2014}, which is linearly interpolated based on the stellar parameters. We also take into account the effect of a non-zero Bond albedo. Rather than using a cloud or haze prescription, we simulate a Bond albedo of A$_B = 0.3$ (similar to that of Earth) by increasing the semi-major axis. Finally, we note that other atmospheres not considered here are certainly possible, in terms of different gas species, possible contributions from clouds or hazes, or dynamics, which could impact the temperature structure and the resulting emission spectrum. However, given the data quality, we treat these idealized cases as distinct end-member scenarios, similar to our approach with the bare rock scenarios. 

\section{Results} \label{sec:results}

In this work, we provide the first measurement of the secondary eclipse depth of GJ~3473~b to assess the likelihood of the planet possessing an atmosphere. Using our four visits with JWST, we find an average eclipse depth of $F_\mathrm{p} / F_\star = 186\pm45$~ppm in the MIRI F1500W bandpass, corresponding to a brightness temperature of $820\pm120$~K\footnote{In the absence of an atmosphere, the brightness temperature serves as a lower limit of the dayside surface temperature, unless the emissivity in the observed bandpass is equal to one, in which case they are the same.}, which we obtained by solving for $T_{\mathrm{day}}$ in equation \eqref{eq:fpfs}. The eclipse is detected with a log Bayes factor of $\ln B = 5.1$, corresponding to strong evidence \citep{Trotta2008}. We find that the ratio of the brightness temperature at 15 $\mu$m to that of a blackbody is $R = 0.82 \pm 0.12$. In this section, we evaluate if the measured eclipse depth of GJ~3473~b is consistent with the planet having a bare rock surface, an atmosphere, or if the data is inconclusive. 

\subsection{Implications for an atmosphere} \label{sec:results_atmosphere}

\begin{figure}
	\includegraphics[width=0.495\textwidth]{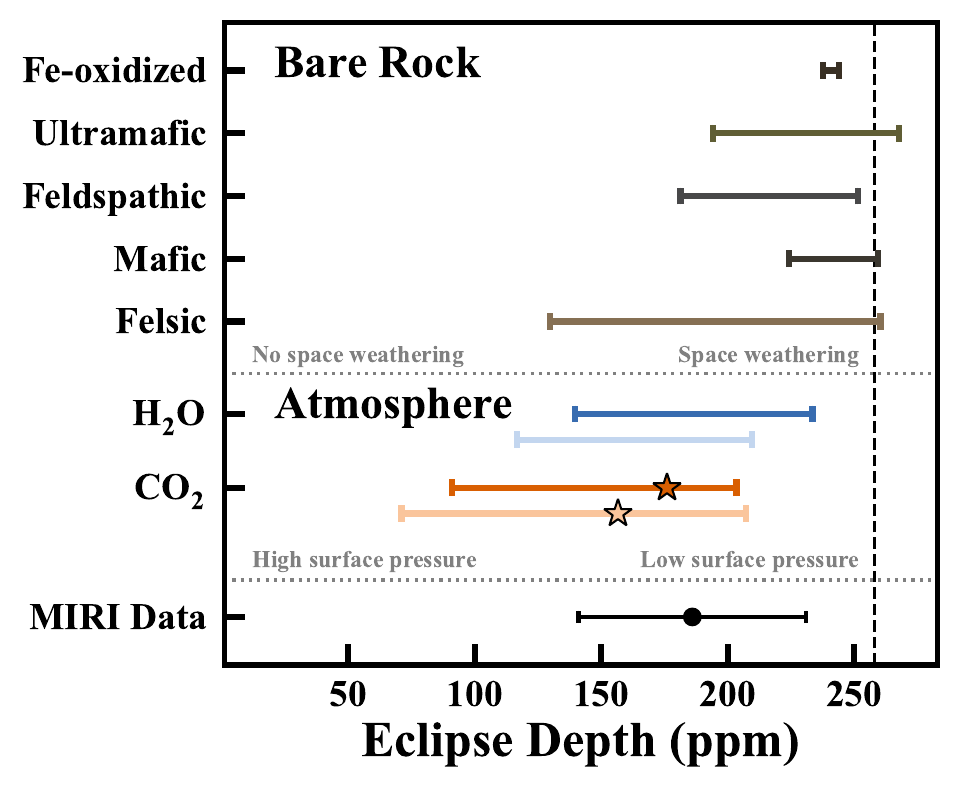}
    \caption{ The span of eclipse depths of GJ~3473~b in the MIRI F1500W bandpass for different bare-rock and atmosphere scenarios, assuming that the planet is tidally locked in a 1:1 spin–orbit resonance. In the bare-rock cases, the limits arise from different materials, textures, and varying degrees of space weathering, as described in Section~\ref{sec:surface}. In contrast, in the atmospheric scenarios, the limits are determined by varying surface pressures, which range from 0.1 mbar to 100 bar, as outlined in Section~\ref{sec:atmosphere}. The two shades of color represent Bond albedos of 0 (darker) and 0.3 (lighter), respectively. The stars indicate the critical surface pressure (0.6-1 mbar) below which a CO$_2$ atmosphere is susceptible to collapse, as discussed in Section~\ref{sec:collapse}. The dashed vertical line represents the eclipse depth of a blackbody. The measured average eclipse depth of GJ 3473 b is shown in black at the bottom of the figure.}
    \label{fig:scenarios} 
\end{figure}

In order to evaluate the likelihood of an atmosphere of GJ~3473~b, we use forward modelling to explore a range of possible atmosphere and surface compositions. The goal is to assess the range of the eclipse depths allowed by different scenarios, both for an atmosphere and a bare rock surface, in order to compare these with our measurement of $F_\mathrm{p} / F_\star = 186\pm45$~ppm in the MIRI F1500W bandpass. For this, we assume that GJ~3473~b is tidally locked. 

Starting with the bare rock scenarios, we compute the predicted eclipse depths for each of the different surface categories: Fe-oxidized, ultramafic, feldspathic, mafic, and felsic, as described in Section~\ref{sec:surface}. These categories contain a range of materials with different textures \citep{Paragas2025}. For each of these, we consider different levels of space weathering by iron and graphite (between 0-5\% volume fraction, as described in Section~\ref{sec:space_weathering}). We then take the minimum and maximum eclipse depths of these cases to represent the possible ranges for each of the surface categories. These limits, along with our measurement, are shown in Figure~\ref{fig:scenarios}. Note that the eclipse depths occasionally exceed that of blackbody for some highly reflective materials when adding space weathering, as explained in Section~\ref{sec:space_weathering}. We find that the measured eclipse depth of GJ~3473~b is consistent with a range of different surface compositions. In fact, all surface compositions that we assess are consistent with the data within 2$\sigma$. However, surfaces that allow for a lower brightness temperature, such as ultramafic, feldspathic, and felsic materials, are somewhat preferred. Among the considered surface compositions, we find that granite, being highly reflective, allows for the lowest brightness temperature in the MIRI F1500W bandpass -- with an eclipse depth of 130~ppm. We note that other highly reflective materials may also be possible \citep{Hammond2025}. Furthermore, while we consider surfaces of a single composition, this may not be realistic. For instance, granite, which is found only on Earth among solar system bodies, is not distributed uniformly on the planet. Overall, we cannot rule out the possibility of GJ~3473~b being a bare rock given our photometric data alone.

We also explore a range of different atmospheric scenarios. We consider two types of atmospheres: 100\% CO$_2$ or H$_2$O with surface pressures between 0.1 mbar and 100 bar, as described in Section~\ref{sec:atmosphere}. For the redistribution factor, we consider the scaling relation by \cite{Koll2022}, resulting in low redistribution at low surface pressures and higher redistribution at higher surface pressures. Out of these scenarios, a CO$_2$ atmosphere has been suggested to potentially be possible on GJ~3473~b, given a sufficiently high initial volatile content \citep{Ji2025}; however, this is less likely compared to most planets in their study. The same study found that N$_2$, H$_2$O, and CH$_4$ atmospheres are less likely to persist compared to CO$_2$. We additionally consider cases that approximate a lower Bond albedo of 0.3, as detailed in Section~\ref{sec:atmosphere}. The ranges on the possible eclipse depths, in the MIRI F1500W bandpass, for these different atmospheric compositions are shown in Figure~\ref{fig:scenarios}. The upper limits are set by atmospheres with low surface pressure, while the lower limits are reached for the highest pressures. Even though we cannot differentiate between a bare rock surface or an atmosphere, the data allows us to rule out a thick 100\% CO$_2$ atmosphere, placing a 95\% credible upper limit of 1.2-6.5 bar on the surface pressure, depending on the Bond albedo (0-0.3). All other atmospheric scenarios that we consider are consistent with the data. Moreover, while pure atmospheres consisting of a single gas are not entirely realistic, they serve as end-member cases given our limited data. In the end, additional observations are required to determine if GJ~3473~b has an atmosphere or not.

Finally, GJ~3473~b might be in a higher-order spin-orbit resonance, similar to Mercury's 3:2 resonance, or it may not be tidally locked. In such a case, stellar irradiation would be distributed more evenly across the planet’s surface, depending on its rotation rate and the thermal inertia of its surface or atmosphere \citep{Lyu2024}. This could reduce the dayside temperature and lower the secondary eclipse depth. As a result, a low measured eclipse depth could be mistakenly interpreted as evidence for an atmosphere, when it may instead reflect the effects of non-synchronous rotation. Although short-period planets are generally assumed to be tidally locked due to the typically rapid timescales for tidal synchronization, this assumption depends on several factors, including orbital eccentricity, internal structure, atmospheric presence, and potential gravitational perturbations from other planets \citep{Barnes2017}. Notably, a higher-order spin-orbit resonance requires a non-zero eccentricity -- typically above $e \gtrsim0.01$ \citep{Turbet2018} -- which remains fully within our 95\% upper limit of $e \lesssim 0.17$ (see Section~\ref{sec:light_curves}). Future phase-curve observations of GJ~3473~b could help break this degeneracy by constraining the longitudinal temperature distribution, as demonstrated for the rocky exoplanet LHS~3844~b, which was shown to be tidally locked in a 1:1 resonance \citep{Lyu2024}.

\subsection{A Bayesian approach} \label{sec:Bayesian}

In Figure~\ref{fig:scenarios}, we present the eclipse depth ranges of different bare-rock and atmosphere scenarios in the MIRI F1500W bandpass, suggesting that many of these cases are degenerate with only one photometric data point. However, the ranges indicated solely reflect the maximum and minimum eclipse depths for each scenario. To better quantify the preference for an atmosphere compared to a bare-rock surface, given our data, we employ a Bayesian model comparison framework. From this analysis, we can derive the posterior probability of an atmosphere by summing the probabilities of our atmospheric models and comparing this total to that of the bare-rock models, under the assumption that the planet is tidally locked in a 1:1 spin–orbit resonance.

\begin{figure}
	\includegraphics[width=0.49\textwidth]{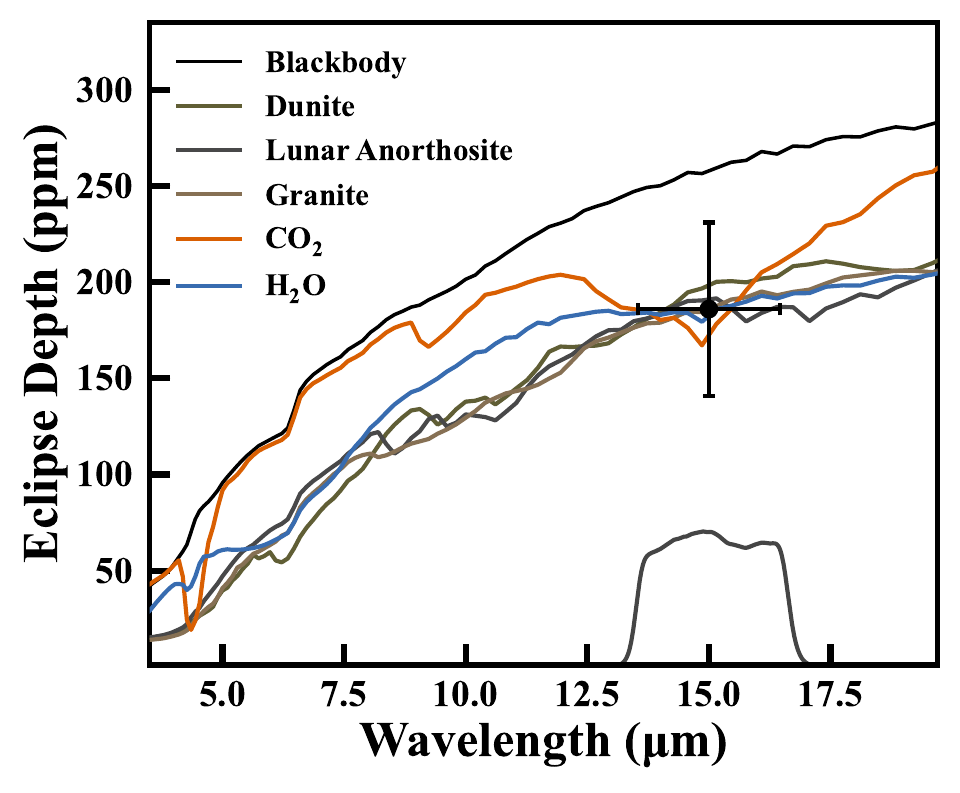}
    \caption{  The eclipse depth of GJ~3473~b as a function of wavelength for a range of best-fit bare rock and atmosphere scenarios that are consistent with the data, as described in Section~\ref{sec:Bayesian}. The data for the dunite (ultramafic) and granite (felsic) surfaces are obtained from \cite{Paragas2025}, while the lunar anorthosite (feldspathic) data are from \cite{Hu2012}. The free parameter for the bare rock surfaces was the level of space weathering (in this case, iron), whereas the surface pressure was the free parameter for the atmospheric scenarios (with best-fit values of 0.4 mbar and 34 mbar for CO$_2$ and H$_2$O, respectively). For comparison, the eclipse depth of a blackbody is shown in black. }
    \label{fig:eclipse_depth_spectra} 
\end{figure}

For each case, we compute the Bayesian evidence by integrating the likelihood over the prior, normalized by the prior volume. For the bare-rock surface cases, we vary the level of space weathering, with log-uniform priors between 0.1~ppm and 5\%. For the atmospheric cases, we vary the surface pressure with a log-uniform distribution between 0.1 mbar and 100 bar (linearly interpolating between the model grid). We show some of the best-fit spectra in Figure~\ref{fig:eclipse_depth_spectra}. We then calculate the probability $p_{i}'$ of each model $i$, using the (log) Bayesian evidence $\ln Z_i$, via
\begin{equation}
    p_{i}' = \pi_i \, \exp(\ln Z_i - \ln Z_\mathrm{max})\,,
\end{equation}
where $\ln Z_\mathrm{max}$ corresponds to the maximum evidence\footnote{We subtract $\ln Z_\mathrm{max}$ for numerical stability.} and where $\pi_i$ is the prior probability of the model itself. Since it is difficult to quantify the prior probability of each model, we construct a set of agnostic priors that ensures: (1) an equal probability of having either an atmosphere or a bare-rock surface, (2) an equal probability for each type of atmosphere or surface (e.g., a mafic and an ultramafic surface are given the same prior probability), and (3) an equal probability for each model within each type. For example, the last condition means that the atmospheric cases with bond albedos of 0 and 0.3 are equally likely. We express this as follows
\begin{equation}
    \pi_i = \frac{1}{2} \, \frac{1}{N_\mathrm{i, types}} \, \frac{1}{N_\mathrm{i, models}}\,,
\end{equation}
where $N_\mathrm{i, types}$ are the number of atmosphere/surface types (2 and 5, respectively), and where $N_\mathrm{i, models}$ are the number of models in each type. For example, for mafic surfaces, we have 20 models which correspond to 10 different samples from \cite{Paragas2025}, with both iron and graphite space weathering. Next, we compute the overall probability of an atmosphere and a bare-rock surface as
\begin{equation}
    P'_\mathrm{atm} = \sum_{i\,\in\, \mathrm{atm}} p_i'\,, \hspace{5mm} P'_\mathrm{rock} = \sum_{i\,\in\, \mathrm{rock}} p_i'\,, 
\end{equation}
which we then normalise,
\begin{equation}
    P_\mathrm{atm} = \frac{P'_\mathrm{atm}}{P'_\mathrm{atm} + P'_\mathrm{rock}}\,, \hspace{5mm} P_\mathrm{rock} = 1 - P_\mathrm{atm}\,.
\end{equation}
For convenience, we convert the probability of an atmosphere to the number of standard deviations $\sigma$ away from the mean of a normal distribution, via $\sigma = \Phi^{-1}(P_\mathrm{atm})$, where $\Phi^{-1}$ is the inverse cumulative distribution function. This defines a one-sided significance, such that $\sigma = 0$ corresponds to equal posterior probabilities, $P_\mathrm{atm} = P_\mathrm{rock} = 1/2$, while positive values indicate a preference for an atmosphere, and negative values indicate a preference for a bare rock surface. We refer to $\pm \sigma$ as the detection significance for either an atmosphere or a bare rock surface. Note that a one-sided significance of $2.5\sigma$, as defined above, corresponds to an odds ratio of~$\sim$$150:1$, which we adopt as the threshold for strong evidence \citep{Trotta2008}.

\begin{figure}
\begin{minipage}{\columnwidth}
\includegraphics[width=0.95\textwidth]{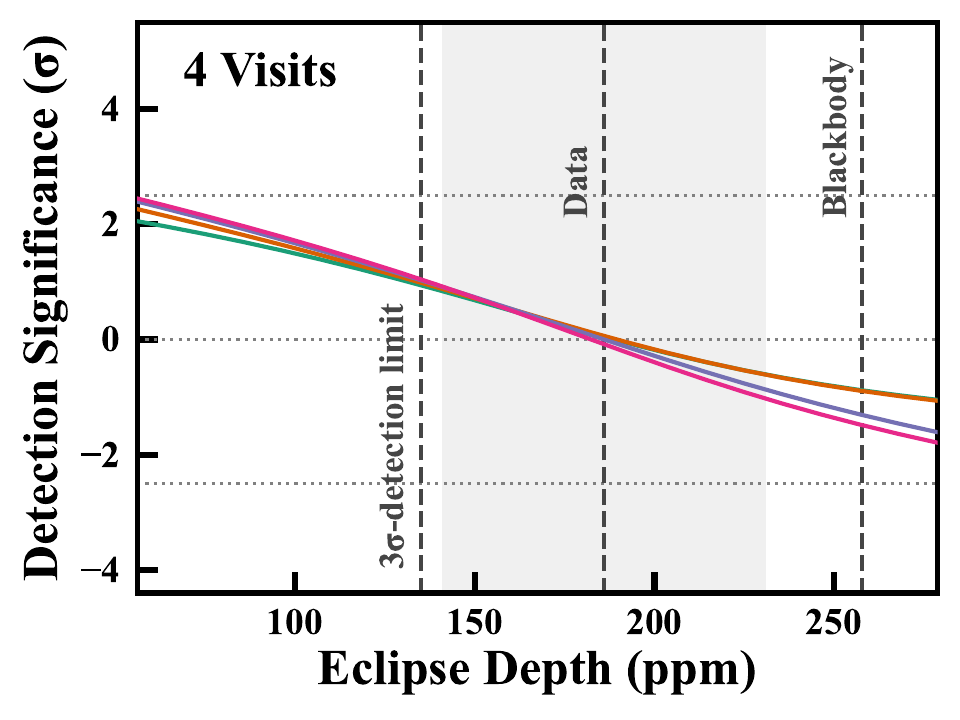}
\end{minipage}
\begin{minipage}{\columnwidth}
\includegraphics[width=0.95\textwidth]{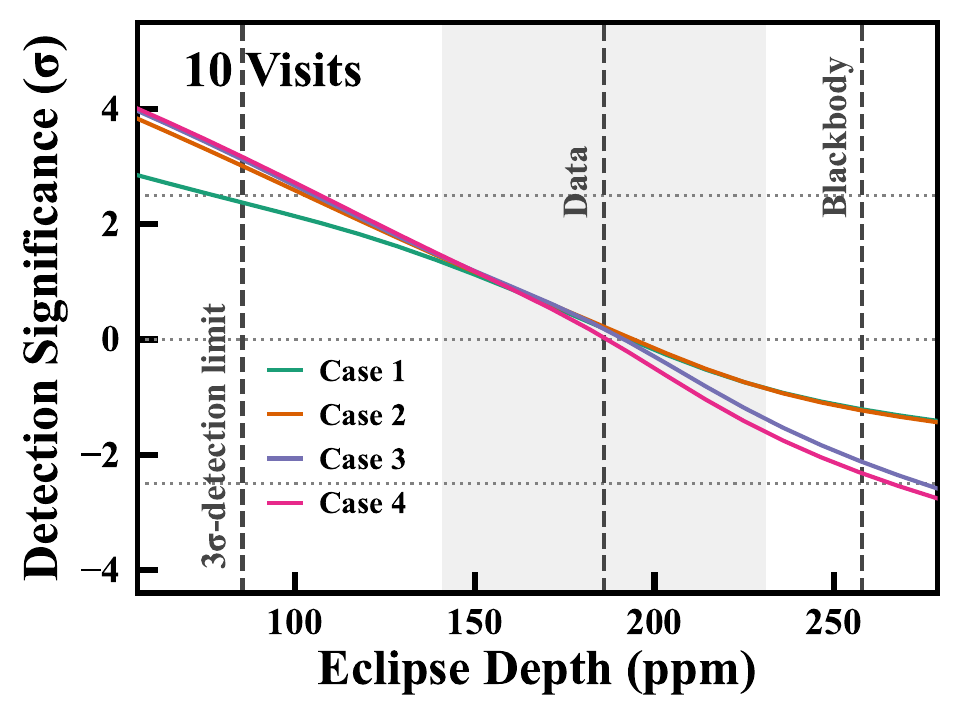}
\end{minipage}
\caption{Detection significance of an atmosphere on GJ~3473~b as a function of the eclipse depth in the MIRI F1500W bandpass, with 4 (top) and 10 visits (bottom). Positive values correspond to the detection significance of an atmosphere, while negative values correspond to the detection significance of a bare rock. At a significance of zero, these two scenarios are equally likely. The four cases are described in Section~\ref{sec:Bayesian}, representing different prior considerations. The vertical dashed lines denote the required eclipse depth for a 3$\sigma$ detection of the eclipse itself, the actual measured eclipse depth, and the eclipse depth of a blackbody, respectively. The light-grey region represents the uncertainty of the present measurement. The horizontal dotted lines show the detection significances at 0 and $\pm$2.5$\sigma$.}
\label{fig:significance}
\end{figure}

Next, we explore various model and prior considerations. Our first case consists of including all our model scenarios, using the five surface types and two atmospheric scenarios discussed above and shown in Figure~\ref{fig:scenarios}. For the second case, we remove felsic surfaces, i.e., highly-reflective granitoids, as these have been suggested to be unlikely to form at the high temperatures of GJ~3473~b \citep{Mansfield2019}. As a third case, we only consider a CO$_2$ atmosphere while also excluding the felsic surfaces (consistent with case 2). We include this case because CO$_2$ atmospheres are expected to be more resilient to the high irradiation compared to other atmospheric compositions, such as H$_2$O or N$_2$ \citep[e.g.,][]{Ji2025}. In our fourth case, we begin with the conditions of case 3 but restrict the lower prior range of the surface pressure to 0.6 and 1 mbar, for bond albedos of 0 and 0.3, respectively, rather than the nominal 0.1 mbar. These pressures correspond to the critical surface pressures for the collapse of a CO$_2$ atmosphere, as discussed later in Section~\ref{sec:collapse}. Thus, these cases progress from considering all scenarios equally to imposing several constraints on the prior space based on theoretical expectations. The purpose of examining these cases is to assess how these different choices influence the potential inference of an atmosphere on GJ~3473~b. Figure~\ref{fig:significance} shows the detection significance of an atmosphere for a range of possible eclipse depths, with either 4 or 10 visits.

Based on the measured eclipse depth of $186\pm45$~ppm using four visits, we find that it is not possible to distinguish whether GJ 3473 b has an atmosphere or a bare-rock surface. The posterior probability of the planet having an atmosphere is estimated to be between 47\% and 52\% (across the four cases), which aligns with the assumed prior probability of 50\%. This finding is consistent with the scenarios presented in Figure~\ref{fig:scenarios}, showing that various atmospheric and bare-rock scenarios are compatible with the data. Furthermore, by considering a range of potential eclipse depths, we find that four observations alone are insufficient to provide a posterior probability that exceeds $2.5\sigma$ for either an atmosphere or a bare-rock surface on GJ~3473~b, regardless of the measurement outcome. While a low eclipse depth could yield a $2\sigma$ preference for the presence of an atmosphere, in such a scenario, the eclipse itself would remain undetected. This could lead to potential ambiguity regarding the existence of an atmosphere and the timing of the eclipse itself, e.g. due to non-zero orbital eccentricity. 

If we instead had a total of 10 MIRI F1500W visits, it would be possible to achieve a stronger preference for an atmosphere. For this experiment, we scaled the measured eclipse depth uncertainty by $\sqrt{4}/\sqrt{10}$, resulting in an uncertainty of 28~ppm with 10 visits. Among the four cases considered, we find that case 1 produces the lowest detection significance for an atmosphere, due to the degeneracy between an atmosphere and a high-albedo surface. As a result, the preference for an atmosphere increases at low eclipse depths when we assume that highly reflective felsic surfaces are not possible (cases 2-4). Furthermore, with ten observations, we find that it is impossible to infer a bare rock surface with high confidence if we consider both H$_2$O and CO$_2$ atmospheres down to a surface pressure of 0.1 mbar (cases 1-2). However, if we can rule out H$_2$O atmospheres in advance and limit the surface pressure to be above 0.6-1 mbar (case 4), we could establish a $2-2.5\sigma$ preference for a bare rock surface, provided that the measured eclipse depth is sufficiently high.

These results emphasise the opportunity and challenges of detecting an atmosphere using MIRI F1500W photometry alone and the importance of carefully considering prior knowledge. By allowing for highly reflective surfaces, atmospheric compositions beyond just CO$_2$, and for surface pressures down to 0.1 mbar, we demonstrate that distinguishing between an atmosphere and a bare-rock surface on GJ~3473~b is difficult, even with ten MIRI F1500W observations. Future research can incorporate more complex models, including atmospheres with a mixture of gases, a broader range of temperature structures, potential clouds or hazes, and additional surface types or a combination of different surfaces. These factors may change the overall constraining power of the data. Additionally, we note that it might be easier to infer the presence of an atmosphere on rocky exoplanets other than GJ~3473~b. Although we show that confidently detecting an atmosphere with only MIRI F1500W photometry can be challenging, our framework provides a useful metric, namely $P_\mathrm{atm}$, to identify promising targets for follow-up observations. Additional observations, e.g. using MIRI LRS, could provide stronger constraints on the presence or absence of an atmosphere \citep[e.g.,][]{Piette2022}.

\section{Discussions} \label{sec:discussion}

In this section, we evaluate whether the eclipse depths from our four visits are consistent and explore whether the potential for atmospheric collapse can be used to constrain the prior on the surface pressure of a CO$_2$ atmosphere for GJ~3473~b or other rocky exoplanets.

\subsection{The possibility of atmospheric collapse} \label{sec:results_collapse}

\label{sec:collapse}
Given that GJ~3473~b is likely tidally locked \citep{Leconte2015}, we consider the possibility of atmospheric collapse in the case of a pure CO$_2$ atmosphere. Collapse can occur if the nightside temperature falls below the condensation temperature of CO$_2$, turning the atmosphere into surface ice on the planet’s permanent night side \citep{Kasting1993}. While GJ~3473~b is strongly irradiated, collapse may still occur if heat redistribution becomes sufficiently inefficient at low surface pressures, allowing the nightside to become a cold trap, assuming no other sources of heating and negligible abundances of other gases. This could provide a useful lower limit of the surface pressure for CO$_2$ atmospheres.

To estimate the critical surface pressure below which the atmosphere is susceptible to collapse, we approximate the nightside temperature following \citet{Wordsworth2015}:
\begin{equation} 
     T_\mathrm{night} = \left( \frac{(1-A_\mathrm{B}) \,\kappa \,p_\mathrm{s}}{4 g}\right)^{1/4} \hspace{-1mm} \sqrt{\frac{R_\star}{a}} \,T_\mathrm{eff} \,,
\end{equation}
where $\kappa = 1.6\times10^{-4}$~m$^2$/kg is the approximate infrared opacity for CO$_2$, and $g=11.15$~m/s$^2$ is the surface gravity of GJ~3473~b. Solving for the surface pressure using the CO$_2$ condensation curve from \cite{Wordsworth2015}, we find a critical threshold of 0.6-1 mbar for Bond albedos between 0 and 0.3. These pressures correspond to eclipse depths between $157$-$176$~ppm in the MIRI F1500W bandpass, as illustrated in Figure~\ref{fig:scenarios}. Since the eclipse depth corresponding to the critical surface pressure is consistent with the data, we cannot rule out the presence of a thin CO$_2$ atmosphere. However, we note that more recent work suggests that the critical surface pressure can exceed the purely radiative estimate used here \citep{Auclair-Desrotour2020}. For other close-in rocky planets with lower irradiation, the critical surface pressure would be higher than for GJ~3473~b, potentially providing a valuable constraint when interpreting observations. 

Among the planets in our program, we find that LHS~1140~c exhibits the highest critical surface pressure, ranging from 13 to 19 mbar for Bond albedos between 0 and 0.3. Notably, this aligns with the $3\sigma$ upper limit of 10 mbar for a CO$_2$ atmosphere reported by \cite{Fortune2025}. Consequently, since lower pressures are not expected to be possible, we may be able to rule out a CO$_2$ atmosphere on LHS~1140~c, if the planet is tidally locked. However, this conclusion depends on the specific temperature structure of the atmosphere, which is affected by dynamics, internal heating, clouds/hazes and the amount of non-condensable gas. 

\subsection{Eclipse depth variability} \label{sec:variability}

\begin{figure}
	\includegraphics[width=0.49\textwidth]{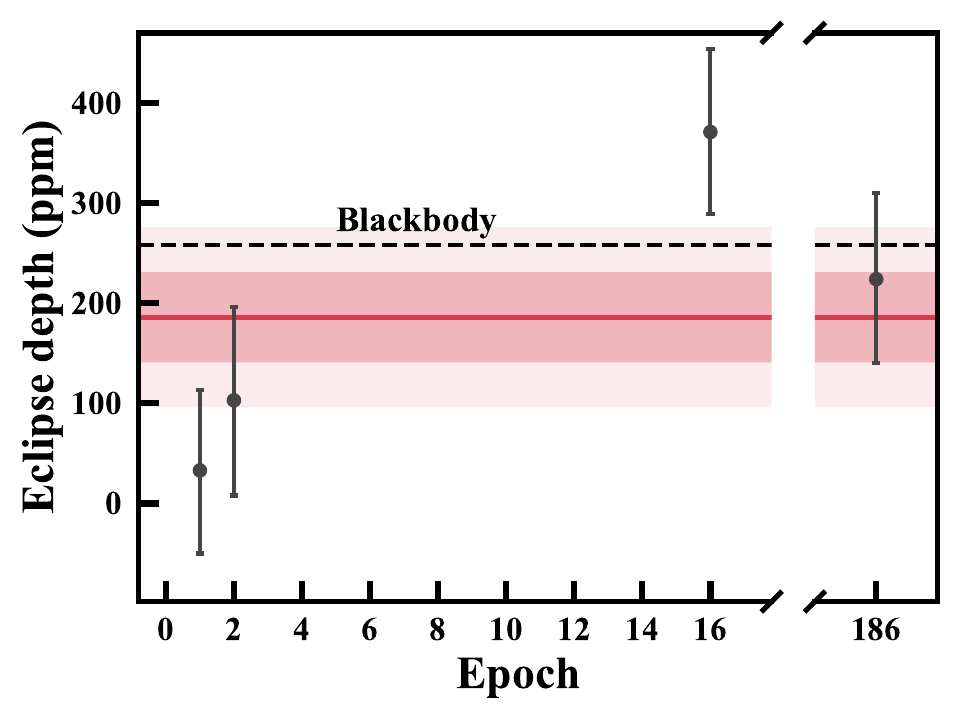}
    \caption{ The eclipse depth of GJ~3473~b as a function of epoch, starting on 12 March 2024. The solid red line corresponds to the average eclipse depth obtained from the canonical case. The two lighter shades of red illustrate the 1$\sigma$ and 2$\sigma$ regions. The dashed black line show the eclipse depth of a blackbody.}
    \label{fig:variability} 
\end{figure}

We find some evidence for a variable eclipse depth of GJ~3473~b at 15~$\mu$m, although with a log Bayes factor of $\ln B = 2.1$ (compared to a constant eclipse depth), this evidence is weak at best \citep{Trotta2008}. The evidence of this model, with varying eclipse depths, was computed using a uniform prior range of 0-500~ppm for the eclipse depths -- the same as in the canonical case. To measure the eclipse depths of the different visits, we instead adopted a uniform prior range of $\pm$1000~ppm, to not be influenced by the edges of the priors. The measured eclipse depths for each visit are listed in Table~\ref{tab:variability} and shown in Figure~\ref{fig:variability}. The weighted average of these depths is $186\pm43$~ppm, consistent with the result from the constant eclipse depth fit (canonical case). Using this average, we obtained $\chi^2 = 9.53$ with three degrees of freedom. This corresponds to a p-value of 2.3\%, indicating some evidence against a constant eclipse depth. We also note that although the eclipse depths from visits 1 and 3 differ from the canonical average by approximately $2\sigma$; excluding these does not significantly affect the overall average. When we consider only visits 2 and 4, we obtain an average of $170\pm63$~ppm, which is consistent with the canonical result.

This apparent variability, if not a statistical outlier, could be planetary in nature or be due to unaccounted systematic errors, such as stellar activity (e.g., minor flares) or unknown instrumental effects. For instance, \cite{August2025} found that two observations of LHS~1478~b -- also using MIRI F1500W photometry -- were inconsistent due to unknown systematics during the second visit. In our case, however, the low evidence for time-correlated noise ($\ln B = 1.8$, with a nominal aperture of 6 pixels), the absence of correlation with the PSF positions ($\ln B = -17.9$), and the stability of the stellar flux between visits ($\sim$0.07\%) suggest that the quality of our observations is high. It is noteworthy that the visits with the most significant difference in eclipse depth -- visits 1 and 3 -- exhibit nearly identical detector settling behaviour. Both observations show an excess flux of around 1900 ppm at the start of the visits (see Section~\ref{sec:detector_settling}). Therefore, we conclude that the differences in the detector settling ramps are unlikely to affect the measured eclipse depths. Moreover, we note that the evidence for time-correlated noise drops from $\ln B = 1.8$ to $\ln B = 0.6$ when allowing for a variable eclipse depth, indicating some degeneracy between the presence of time-correlated noise and variable eclipse depth. We also find that the evidence for variability increases to $\ln B = 3.3$ when we do not use a GP. However, out of all the cases, the model with the highest evidence includes both time-correlated noise and variability in eclipse depth, as described in Section~\ref{sec:light_curves}. In the end, the overall evidence for a variable eclipse depth across our four visits remains weak and may therefore not be real. 

\begin{table}
\renewcommand{\arraystretch}{1.63} % Default value: 1
\setlength{\tabcolsep}{4pt}
\centering
\begin{tabular}{lcc}
\hline \hline
Date & Eclipse depth (GP) & Eclipse depth (no GP) \\ \hline
12 March 2024 & $33_{-83}^{+80}$ ppm & $47_{-62}^{+62}$ ppm\\ 
13 March 2024 &  $103_{-95}^{+93}$ ppm & $114_{-71}^{+72}$ ppm \\ 
30 March 2024 &  $371_{-81}^{+83}$ ppm & $357_{-67}^{+67}$ ppm \\ 
20 October 2024 &  $224_{-84}^{+86}$ ppm & $218_{-70}^{+67}$ ppm \\ \hline
\end{tabular}
\vspace{2mm}
\caption{The eclipse depth of GJ~3473~b for each of the four visits, with and without Gaussian Processes (GP). The visits are provided in the same order as in Figure~\ref{fig:variability}. For these cases, we jointly fit the light curves while allowing for different eclipse depths with a uniform prior between $\pm$1000 ppm. Although the eclipse depths are somewhat different, the evidence for variability is weak, as described in Section~\ref{sec:variability}.
}
\label{tab:variability}
\end{table}

Eclipse depth variability has previously been reported for 55~Cnc~e using Spitzer photometry \citep{Demory2015, Tamburo2018} and recently with JWST \citep{Patel2024}. For 55~Cnc~e, the cause of the variability has not been robustly established; however, several possibilities have been proposed, such as volcanic plumes or circumstellar/circumplanetary material \citep{Demory2015, Heng2023}. If the apparent variability of GJ~3473~b is planetary in nature, this could provide evidence for an atmosphere on the planet, as it is unlikely that a bare rock surface could vary significantly over the time spanned by our observations. On the other hand, circumstellar/circumplanetary material, such as Io's plasma torus, may act to change the apparent eclipse depth over time. Furthermore, the fact that the eclipse depths of the first two observations (only separated by one orbital period) remain consistent within 1$\sigma$, while the other observations (separated by $\gtrsim$20~days) are somewhat different, may be evidence for a gradual change over time. We also note that the measured eclipse depth appears to vary by a factor of a few (e.g., $\sim$3.6 between visits 2 and 3), similar to the variability reported for 55~Cnc~e \citep{Demory2015}. Ultimately, more observations are needed to reliably demonstrate or rule out a variable eclipse depth of GJ~3473~b.

\section{Conclusions} \label{sec:conclusion}

We present the first JWST secondary eclipse observations of GJ~3473~b, a highly irradiated rocky exoplanet orbiting a mid-M dwarf, as part of the Hot Rocks Survey (JWST GO program 3730). Using four MIRI F1500W photometry visits, we detect the eclipse at high confidence ($\ln B=5.1$), with an average eclipse depth of $186\pm45$~ppm. To ensure the robustness of our analysis, we explore an extensive range of data reduction and light curve fitting assumptions, including variations in background subtraction, extraction methods, and the modelling of systematics, as outlined in Section~\ref{sec:obs}. In addition, in the second visit we identify a $\sim$700~ppm flare-like feature (Appendix~\ref{app:flare}), for which we estimate an energy of ${3.24_{-0.86}^{+1.08}} \times 10^{28}$~erg in the F1500W bandpass. We find that the data are consistent with the planet having a circular orbit, and place an upper limit on the eccentricity of $e \lesssim 0.17$ at 95\% confidence. Our observations provide the first constraints on the thermal emission of GJ~3473~b, allowing us to empirically assess the likelihood of an atmosphere.

We find that the measured eclipse depth is consistent with both bare-rock or atmospheric scenarios. Forward modelling of airless surfaces, accounting for a wide range of compositions, textures, and degrees of space weathering, indicates that a variety of materials -- including mafic, ultramafic, feldspathic, and felsic surfaces -- can reproduce the observed eclipse depth within the measurement uncertainties (at less than 2$\sigma$). Notably, highly reflective materials like granite yield the lowest brightness temperatures among the surfaces that we consider, and is compatible with the data. However, we caution that such compositions may be geologically unrealistic for planets like GJ 3473 b \citep[e.g.,][]{Mansfield2019}. However, other surface compositions not considered here may be present, potentially widening the range spanned by the bare-rock scenario. Finally, if the planet is not tidally locked in a 1:1 spin–orbit resonance, the dayside temperature could be significantly cooler, making it more difficult to differentiate between the various scenarios. 

Additionally, we also explore a suite of idealized atmospheric scenarios, including CO$_2$ and H$_2$O atmospheres over a broad range of surface pressures. While the data do not require GJ~3473~b to have an atmosphere, we are able to rule out a thick CO$_2$ atmosphere, placing a 95\% credible upper limit of 1.2-6.5 bar on the surface pressure. Conversely, thin CO$_2$ atmospheres and atmospheres with other compositions remain consistent with the observed eclipse depth. Without further constraints of the prior parameter space, it may be challenging to confidently detect an atmosphere using MIRI F1500W photometry alone, due to the degeneracies and many unknowns involved. Nevertheless, a low eclipse depth may still identify promising candidates for follow-up observations. Ultimately, detecting atmospheric features through spectroscopy can help definitively determine whether an atmosphere is present.

To further constrain the atmospheric parameter space, we consider the possibility of atmospheric collapse for a CO$_2$-dominated atmosphere, assuming that GJ~3473~b is tidally locked in a 1:1 spin–orbit resonance. We estimate that a surface pressure of approximately 1~mbar is required to prevent atmospheric collapse on GJ~3473~b. However, this limit does not allow us to rule out a thin CO$_2$ atmosphere on GJ 3473 b. On the other hand, for exoplanets receiving lower stellar irradiation, the critical pressure would be higher and could therefore serve as a useful prior. In cases where the critical pressure is higher than the limit placed by observations, this could allow us to eliminating CO$_2$-rich scenarios altogether. This approach may be especially valuable for planets where CO$_2$ is the only atmospheric constituent expected to survive long-term atmospheric escape \citep{Ji2025}, offering a path to rule out the presence of thin atmospheres.

We also find some evidence, albeit weak, of eclipse depth variability across the four visits, with a log Bayes factor of $\ln B=2.1$. Although not conclusive with the present data, this result raises the possibility of temporal variability in the dayside emission, potentially linked to dynamic atmospheric processes, volcanism, or circumstellar/circumplanetary material \citep[as proposed for 55~Cnc~e;][]{Demory2015}. However, it is also possible that instrumental systematics or stellar activity (e.g., in the form of unresolved flares) could be contributing to this apparent variability. If the variability is not a statistical outlier, it could bias the interpretations of similar data unless several epochs are observed, regardless of the variability's origin. Additional observations will be essential to establish the nature and origin of this variability, if it is real.

The presence or absence of an atmosphere on GJ 3473 b remains an open question, as the measured eclipse depth is consistent with both a bare rock surface and a secondary atmosphere. The resulting degeneracy between surface and atmospheric emission illustrates the limitations of interpreting MIRI F1500W eclipse photometry in isolation, even with a significant thermal emission detection. Our analysis demonstrates the importance of explicitly accounting for these degeneracies when interpreting current and future MIRI F1500W observations of rocky exoplanets. For GJ~3473~b, additional observations, such as secondary eclipse spectroscopy or a thermal phase curve, will be required to break these degeneracies and place more robust constraints on the planet's atmospheric or surface composition. While our work highlights the difficulties of detecting atmospheres on highly irradiated rocky exoplanets using single-band photometry alone, it establishes a practical framework for interpreting such data and for guiding future follow-up observations.

\begin{acknowledgments}
This work is based on observations with the NASA/ESA/CSA James Webb Space Telescope obtained at the Space Telescope Science Institute, which is operated by the Association of Universities for Research in Astronomy, Incorporated, under NASA contract NAS5-03127. Support for program number 3730 was provided through a grant from the STScI under NASA contract NAS5-03127. JMM acknowledges support from the Horizon Europe Guarantee Fund, grant EP/Z00330X/1. NHA acknowledges support by the National Science Foundation Graduate Research Fellowship under Grant No. DGE1746891. PCA acknowledges support from the Carlsberg Foundation, grant CF22-1254. B.-O. D. acknowledges support from the Swiss State Secretariat for Education, Research and Innovation (SERI) under contract number MB22.00046. CEF acknowledges support from the European Research Council (ERC) under the European Union's Horizon 2020 research and innovation program under grant agreement no. 805445. NPG gratefully acknowledges support from Science Foundation Ireland and the Royal Society through a University Research Fellowship (URF\textbackslash R\textbackslash 201032). BP acknowledges support from the Walter Gyllenberg Foundation. MH thanks Guadalupe Tovar Mendoza for useful discussions regarding stellar flares. We acknowledge the input from the following individuals to the GO 3730 proposal: Kathryn Jones, Jens Hoeijmakers, Anna Lueber, Alexander Rathcke, Can Akin, Matthew Hooton, Andrea Guzmán Mesa, Nicholas Borsato, Jens Hoeijmakers, Meng Tian, Mette Baungaard. 
\end{acknowledgments}

The JWST data presented in this article were obtained from the Mikulski Archive for Space Telescopes (MAST) at the Space Telescope Science Institute. The specific observations analyzed can be accessed via \dataset[doi: 10.17909/n43h-6185]{https://doi.org/10.17909/n43h-6185}.

% \begin{contribution}
% %%This section gives authors the space to recognize author contributions. The text inside this environment is NOT counted towards the total word quanta. At a minimum, manuscripts are expected to include this text:

% All authors contributed equally to the Terra Mater collaboration.

% %% But authors are expected to provide more specific details, e.g. 
% %%
% %%SC was responsible for writing and submitting the manuscript.
% %%WWM came up with the initial research concept and edited the manuscript.
% %%OTS obtained the funding and edited the manuscript.
% %%EBF provided the formal analysis and validation. He also edited the manuscript.
% %%GEH Supervised the undergraduates, wrote the software and administers the project github and Zenodo repositories.
% %%
% %% Authors can use the Contributor Role Taxonomy (CRediT) at
% %% https://credit.niso.org
% %% for ideas on how write a good statement tailored to their needs.

% \end{contribution}

\facilities{JWST (MIRI)}

% \software{astropy \citep{2013A&A...558A..33A,2018AJ....156..123A,2022ApJ...935..167A},  
%           Cloudy \citep{2013RMxAA..49..137F}, 
%           Source Extractor \citep{1996A&AS..117..393B}
%           }


\appendix

\section{TESS light curve fitting} \label{app:TESS}

\begin{figure}
	\includegraphics[width=0.49\textwidth]{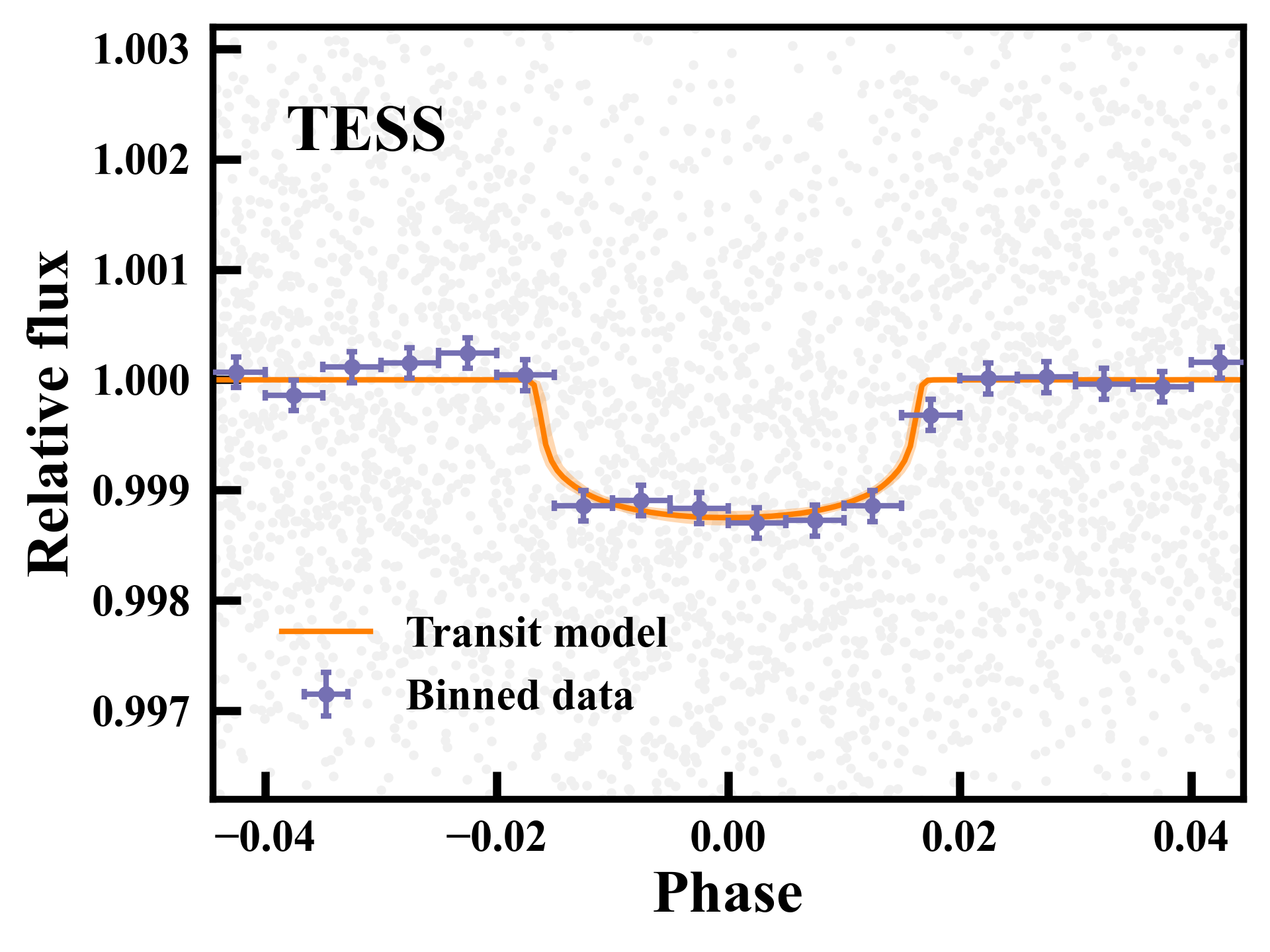}
    \caption{ Phase-folded TESS data and transit model of GJ~3473~b. Prior to phase folding, we subtracted the median GP model. The purple data points are binned for visual clarity. The shaded orange region correspond to the 1$\sigma$ contour.}
    \label{fig:TESS} 
\end{figure}

We improve the orbital parameters of GJ~3473~b by utilising data from TESS sectors 34, 61, and 88. First, we disregard the first two hours from each sector and remove outliers deviating more than 4$\sigma$ from a median-filtered version of the light curves. Next, we employ the \texttt{juliet} framework \citep{Espinoza_juliet2019} to jointly fit the transit light curves. For this, we use priors from \cite{Kemmer2020} for the period $P$, mid-transit time, $R_\mathrm{p}/R_\star$, normalised semimajor axis $a/R_\star$, and impact parameter $b$. This is why we did not include TESS sector 7 in the fit, as that data was used to derive the priors. We assumed a circular orbit of the planet, in line with \cite{Kemmer2020}. This assumption is also consistent with the present MIRI observations, as discussed in Section~\ref{sec:obs}. Additionally, we employed Gaussian processes, using a Matern-3/2 kernel, to account for stellar variability. The refined parameters are presented in Table \ref{tab:TESS_params}, and Figure \ref{fig:TESS} illustrates the phase-folded TESS data and transit model.

\begin{table}
\renewcommand{\arraystretch}{1.25} % Default value: 1
\setlength{\tabcolsep}{5pt}
\centering
\begin{tabular}{lcccc}
\hline \hline
Parameter & Prior & Value \\ \hline
P (days) & N(1.1980035, $1.9\times10^{-6}$) & $1.19800484_{-5.2\times10^{-7}}^{+5.3\times10^{-7}}$ \\
T$_0$ (BJD) & N(58491.70408, 0.00043) &  $58491.70415_{-0.00042}^{+0.00039}$   \\ 
$a / R_\star$ & N(9.39, 0.21) & $9.32_{-0.19}^{+0.18}$   \\ 
$b$  & N(0.336, 0.074) & $0.300_{-0.060}^{+0.055}$   \\ 
$i$ (deg) & - & $88.16_{-0.36}^{+0.38}$   \\ 
$R_\mathrm{p} / R_\star$ & N(0.03184, 0.00069) & $0.03220_{-0.00055}^{+0.00055}$  \\ 
$q_1$ & U(0, 1) & $0.59_{-0.28}^{+0.27}$   \\ 
$q_2$ & U(0, 1)  & $0.23_{-0.16}^{+0.28}$   \\ 
$\sigma_{\mathrm{GP}}$ (ppm) & log U(1, $10^4$)  & $247_{-24}^{+32}$ ppm  \\ 
$\rho_{\mathrm{GP}}$ (days) & log U(0.01, 10.0)  & $0.82_{-0.13}^{+0.17}$   \\ \hline
\end{tabular}
\caption{Parameters estimated from the light curve analysis of TESS. The period P, mid-transit time T$_0$ (BJD $-2400000.5$ days) and GP time-scale $\rho_{\mathrm{GP}}$ are given in days. The priors on P, T$_0$, $a / R_\star$, $b$, and $R_\mathrm{p} / R_\star$ are taken from the constraints by \cite{Kemmer2020}.
}
\label{tab:TESS_params}
\vspace{-4mm}
\end{table}

\section{Aperture radius selection} \label{app:aperture}

An important consideration when performing photometry is to appropriately select an aperture size. If we assume that each pixel is independent and that the noise follows a normal distribution, we can effectively address this issue by instead using optimal extraction \citep{horne_optimal_1986}. However, if these assumptions are not valid, it becomes challenging to determine how to properly weigh each pixel without a realistic model of the system. Therefore, we consider aperture extraction and explore a range of different aperture radii.

\begin{figure}
	\includegraphics[width=0.49\textwidth]{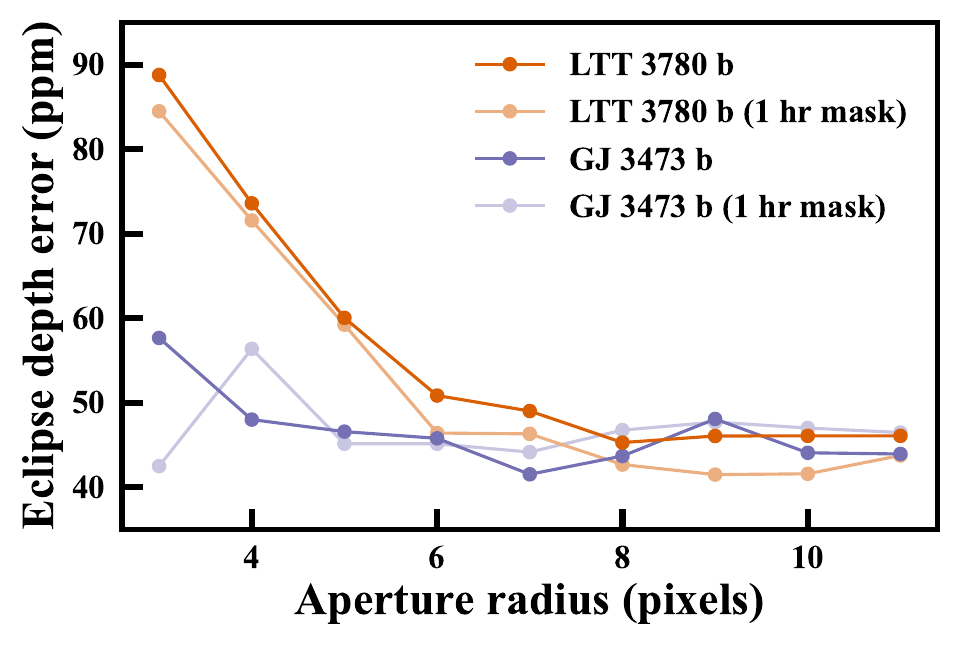}
    \caption{Eclipse depth uncertainty for a range of aperture radii. The orange points correspond to the data of LTT~3780~b, while the purple points correspond to GJ~3473~b. The lighter-coloured points represent uncertainties where we masked the first hour of each observation to minimise detector settling effects.}
    \label{fig:Aperture_Errors} 
\end{figure}

Two competing factors influence the determination of the optimal aperture size. On the one hand, if the aperture is too small we might get inaccurate results due to PSF jitter or other effects that could cause neighbouring pixels to become related, such as wavefront distortions or charge migration. For this reason, we consider an aperture larger than approximately 5-6 pixels, which encompasses the Airy disk, to minimize these systematic errors. Conversely, an aperture that is too large would introduce unnecessary read noise. To identify the optimal size, we investigate the uncertainty in eclipse depth for a variety of aperture sizes, as shown in Figure~\ref{fig:Aperture_Errors}. In this analysis, we not only include the target of this study but also two additional secondary eclipse observations of LTT~3780~b \citep{Allen2025}. LTT~3780~b is the brightest target in our program using the SUB256 subarray and has 22 groups per integration, while GJ~3473~b is the dimmest target in the same subarray, with 39 groups per integration. We observe that the uncertainty in eclipse depth is larger for smaller apertures, particularly for LTT~3780~b, and stabilizes after around 6-8 pixels. This trend remains consistent even when we mask the first hour of the observations, suggesting that the underlying mechanism does not strongly depend on the effects of detector settling. Consequently, we select an aperture size of 6 pixels for our analysis.

\begin{figure}
	\includegraphics[width=0.49\textwidth]{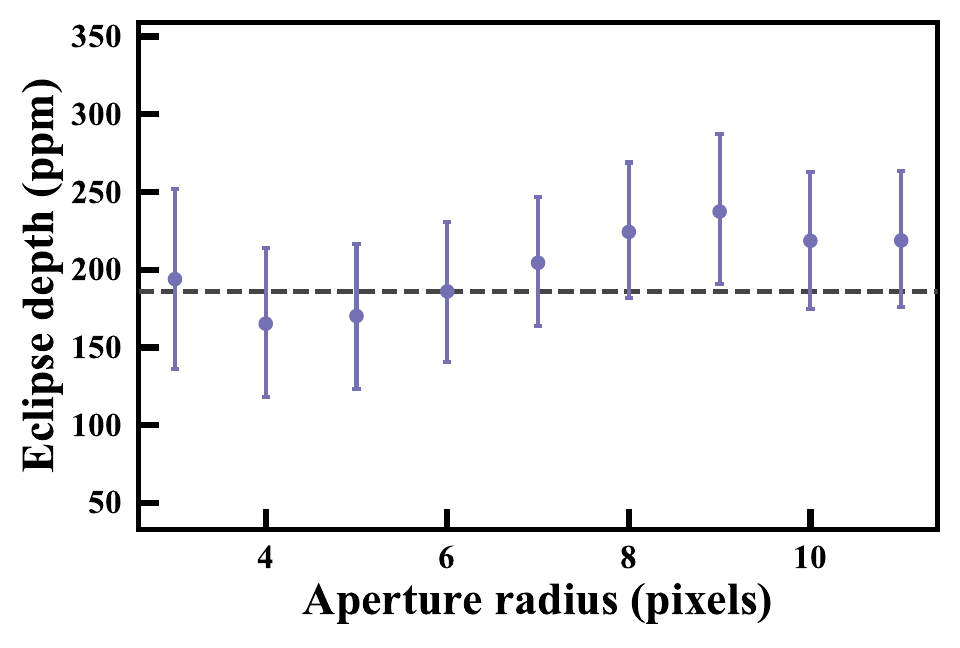}
    \caption{ Eclipse depth of GJ~3473~b as a function of aperture size. The dashed horizontal line indicates the eclipse depth for an aperture radius of 6 pixels, which is the standard size used in our analysis. }
    \label{fig:Aperture_Values} 
    \vspace{-2mm}
\end{figure}

For completeness, we present the eclipse depths as a function of aperture radius in Figure~\ref{fig:Aperture_Values}. We find that the inferred eclipse depths are largely consistent across different apertures, given the measurement uncertainty. The standard deviation among these inferences is 24~ppm, which, when added in quadrature, increases the eclipse depth uncertainty of the canonical case from 45~ppm to 51~ppm. However, this argument only works if we consider these choices equally valid which, as discussed above, is not the case. Additionally, a change of just 6 ppm is likely within the uncertainty of the uncertainty itself, given the range of 34-51 ppm from different model assumptions that we cannot rule out with high confidence, as shown in Table~\ref{tab:eclipse_depth}.

\section{A potential flare} \label{app:flare}

We observe a temporary increase in flux just before the secondary eclipse of the second visit, as seen in Figure~\ref{fig:MIRI_all}. The peak flux of this flare-like feature is around 700~ppm, which exceeds the depth of the secondary eclipse. Therefore, we consider it unlikely that this increase is of planetary origin. Another possibility is that it could be caused by instrument systematics. However, we did not detect any changes in the point spread function (PSF), either in its position or width, that correlate with the temporary increase in flux. Additionally, the background flux remained stable for the entire duration of the observation. We also visually inspected each image for signs of asteroids, which could potentially contaminate the extracted stellar flux, but found no evidence of this. Another potential cause could be a persistence signal resulting from a large cosmic ray hit, as noted by \cite{Fortune2025}. In fact, we do detect a persistence signal in a pixel at the edge of the aperture, occurring toward the end of the second observation, as described in Section~\ref{sec:obs}. However, the timing does not align, and we find that the overall contribution of this effect is negligible. Therefore, we conclude that the signal is likely astrophysical, possibly resulting from a stellar flare \citep[similar events are reported by][]{Gillon2025}.

In this section, we examine various methods for addressing this flare-like feature, either by incorporating a flare model or by masking the affected integrations. For the first approach, we adopt the analytic single-flare model proposed by \cite{TovarMendoza2022}, which we add to the overall eclipse model following equation~\eqref{eq:obs_flux}. This flare model consists of the convolution of a Gaussian function (representing heating) with one or more exponential functions (representing cooling, with different time scales). Here, we consider a single time scale for the cooling rate. The model is given by

\begin{equation} \label{eq:flare_model}
\begin{aligned}
    f_\mathrm{flare}(t) &= \int_{-\infty}^t g(\tau) h(t - \tau) d\tau \\[0.3em]
    &= \frac{A D}{2} e^{D(B - t) + \left(\frac{C D}{2}\right)^2}  \mathrm{erfc} \left( \frac{B - t}{C} + \frac{D C}{2} \right)\,,
\end{aligned}
\end{equation}
with
\begin{equation}
    g(t) = \frac{A}{\sqrt{\pi} \, C} e^{-\frac{(t-B)^2}{C^2}}\,\,, \hspace{5mm} h(t) = D \, e^{-D t}\,.
\end{equation}
Expressed in this form, $A = \int f_\mathrm{flare}\, dt$ represents the integral of the relative flux excess, also called the equivalent duration. Parameters $B$ and $C$  correspond to the midpoint and width of the Gaussian heating impulse, respectively, while $D$ denotes the inverse time scale of the exponential cooling ($C, D > 0$). Note that we identify a minor inconsistency in the previous reporting of this expression, leading to equation~\ref{eq:flare_model} having a somewhat different normalisation from what is presented in \cite{TovarMendoza2022}.

\begin{table}
\renewcommand{\arraystretch}{1.25} % Default value: 1
\centering
\begin{tabular}{lccc}
\hline \hline
Case & $\ln B$ & Eclipse depth \\ \hline
Masked region (canonical case) & - & $186_{-45}^{+45}$ ppm \\ 
Gaussian model & $0$ & $192_{-48}^{+49}$ ppm\\ 
\cite{TovarMendoza2022} model & $-0.3$ & $195_{-45}^{+46}$ ppm\\ 
No flare model & $-2.8$ &  $235_{-64}^{+62}$ ppm\\ \hline
% No flare model, no GP & $-15.5$ &  $217_{-32}^{+33}$ ppm\\ \hline
\end{tabular}
\vspace{2mm}
\caption{The eclipse depth of GJ~3473~b under different treatments of the flare-like feature observed during the second visit (see Figure~\ref{fig:MIRI_all}). The affected integrations are included in all cases except the canonical one, where they are masked. As a result, Bayes factor comparisons between the canonical and other cases are not meaningful.
}
\label{tab:flare}
\vspace{-4mm}
\end{table}

We report the eclipse depth obtained by jointly fitting the light curves and the above flare model in Table~\ref{tab:flare}. For this, we do not mask the affected integrations, unlike in the canonical case. For the priors on $C$ and $D$, we adopt uniform priors between 1.4–84.9~min (corresponding to a standard deviation of 1–60~min) and between 1~min$^{-1}$ and $1/12$~hr$^{-1}$, respectively. Given the apparent symmetric shape of the flare-like feature, we also fit a Gaussian function alone, without including exponential cooling. We find that the resulting difference in evidence between these two approaches is negligible and that the eclipse depths are consistent with the canonical case. We obtain consistent values for the equivalent duration $A$ of $1.17_{-0.31}^{+0.39}$~seconds and $1.16_{-0.31}^{+0.41}$~seconds for the \cite{TovarMendoza2022} model and Gaussian model, respectively. Additionally, we estimate the full-width at half-maximum (FWHM) to be $23_{-7}^{+9}$~minutes and $26_{-6}^{+10}$~minutes for the two scenarios, respectively. 

In contrast, if we opt not to fit a flare model while still including the affected integrations, we obtain lower evidence and a somewhat higher eclipse depth, as detailed in Table~\ref{tab:flare}. In this scenario, the uncertainty in the eclipse depth increases due to larger residuals, which necessitates an increase in the amplitude of the Gaussian Process (GP) kernel. Specifically, the evidence for the inclusion of a GP kernel is $|\ln B\,| = 15.5$ when we do not include a flare model and do not mask the affected integrations. This is significantly higher than what we obtain when masking the affected integrations ($|\ln B\,| = 1.8$), as outlined in Table~\ref{tab:eclipse_depth}. Overall, we find that the eclipse depth is not sensitive to the treatment of the flare-like feature, whether we model the excess flux or mask the affected integrations. For this reason, we adopt the latter approach in this work, which involves fewer free parameters.

Using the absolute-calibrated flux described in Section~\ref{sec:obs}, and assuming that the excess flux is indeed caused by a flare, we determine that it had an energy of ${3.24_{-0.86}^{+1.08}}\times10^{28}$~erg in the F1500W bandpass. We calculated this using \citep[e.g.,][]{Loyd2018}
\begin{equation}
    E_\mathrm{flare, \, F1500W} = 4\pi d^2 \, F_\star \, \Delta \nu\,A\,,
\end{equation}
where $d = 27.315 \pm 0.018$~pc is the distance \citep{Gaia2016, Gaia2023}, $F_\star = 7.976 \pm 0.038$~mJy is the stellar flux density during the second visit, and $\Delta \nu = 3.89\times10^{12}$~Hz is the effective bandwidth of the total MIRI F1500W throughput.

\section{List of parameters} \label{app:parameters}

\begin{table}
\renewcommand{\arraystretch}{1.25} % Default value: 1
\setlength{\tabcolsep}{2.5pt}
\centering
\begin{tabular}{lcccc}
\hline \hline
Parameter & Prior & Value \\ \hline
P (days) & N(1.19800484, $5.3\times10^{-7}$) & $1.19800453_{-4.6\times10^{-7}}^{+5.3\times10^{-7}}$ \\
T$_0$ (BJD) & N(58491.70415, 0.00042) &  $58491.70404_{-0.00040}^{+0.00040}$   \\ 
$a / R_\star$ & N(9.32, 0.19) & $9.23_{-0.18}^{+0.18}$ \\ 
$i$ (deg) & N(88.16, 0.38) &  $88.32_{-0.36}^{+0.35}$ \\ 
$F_p/F_\star$ (ppm) & U(0, 500) & $186_{-45}^{+45}$ \\
$R_\mathrm{p} / R_\star$ & fixed & 0.03220  \\ 
$R_\star$ ($R_\odot$) & fixed & 0.364  \\ 
$e$ & fixed & 0  \\ 
$\omega$ (deg) & fixed & 90  \\
$F_\star^{(1)}$ (DN/s) & U($0.98 \bar F$, $1.02\bar F$) & $51766.0_{-9.0}^{+7.8}$ \\ 
$F_\star^{(2)}$ (DN/s) & U($0.98 \bar F$, $1.02\bar F$) & $51769.0_{-8.7}^{+8.0}$ \\ 
$F_\star^{(3)}$ (DN/s) & U($0.98 \bar F$, $1.02\bar F$) & $51742.1_{-9.2}^{+7.6}$ \\ 
$F_\star^{(4)}$ (DN/s) & U($0.98 \bar F$, $1.02\bar F$) & $51674.9_{-8.9}^{+8.4}$ \\ 
$p_1^{(1)}$ & U($-0.02$, $0.02$) & $0.00195_{-0.00022}^{+0.00022}$ \\ 
$p_1^{(2)}$ & U($-0.02$, $0.02$) & $0.00145_{-0.00024}^{+0.00022}$ \\ 
$p_1^{(3)}$ & U($-0.02$, $0.02$) & $0.00190_{-0.00022}^{+0.00023}$ \\ 
$p_1^{(4)}$ & U($-0.02$, $0.02$) & $-0.00294_{-0.00024}^{+0.00024}$ \\ 
log($p_2$/days) & U(-2.86, -0.60) & $-1.779_{-0.042}^{+0.047}$ \\ 
log($\alpha$) & U(-4, 0.3) & $-1.38_{-0.59}^{+1.03}$ \\
log($\beta$/days) & U(-2.86, 1.0) & $-0.96_{-0.85}^{+1.13}$ \\
log($\gamma^{(1)}$) & U(-4, 0.3) & $-0.221_{-0.040}^{+0.034}$ \\
log($\gamma^{(2)}$) & U(-4, 0.3) & $-0.199_{-0.041}^{+0.037}$ \\
log($\gamma^{(3)}$) & U(-4, 0.3) & $-0.180_{-0.035}^{+0.031}$ \\
log($\gamma^{(4)}$) & U(-4, 0.3) & $-0.226_{-0.040}^{+0.036}$ \\ \hline

\end{tabular}
\vspace{2mm}
\caption{Parameters estimates and priors for the MIRI light curve analysis. These parameters correspond to our canonical case, described in Section~\ref{sec:light_curves}. Here, $\bar F$ represent the median flux (in DN/s) of each visit, used to construct the prior of the stellar flux.
}
\label{tab:MIRI_params}
\end{table}

Here we provide a list of all priors and constraints for the canonical case described in Section~\ref{sec:light_curves}. These correspond to 20 free parameters, which are detailed in Table~\ref{tab:MIRI_params}. The priors for the period $P$, mid-transit time $T_0$, normalized semi-major axis $a / R_\star$, and inclination $i$ are derived from the analysis of the TESS data in Appendix~\ref{app:TESS}. Similarly, we fix the planet-to-star radius ratio $R_\mathrm{p} / R_\star$ to the value obtained from TESS. The stellar radius $R_\star$, which is used to compute the light travel time, is obtained from \cite{Kemmer2020}. The remaining fitted parameters are used to describe the stellar flux, the systematic model (as described in equation \eqref{eq:sys}), and the noise characteristics.

The prior for the settling time scale, $p_2$, follows a log-uniform distribution ranging from 2 minutes (equivalent to 10 integrations) to 6 hours (roughly twice the observation time per visit). For the noise model (i.e., Gaussian process), we use the following parameterisation of the covariance matrix
\begin{equation}
    \Sigma_{i,j} = \sigma_{i}^2\delta_{i,j} + \bar \sigma^2 \left( \alpha^2 \,e^{-|t_i-t_j| / \beta} +  \gamma^2 \delta_{i,j} \right).
\end{equation}
Here, $\sigma_i$ represents the predicted uncertainty, derived from the propagation of photon and read noise; $\bar \sigma$ is the median of $\sigma_i$; $\alpha$ and $\beta$ denote the amplitude and time-scale of the time-correlated noise (exponential kernel), respectively; and $\gamma$ indicates the strength of additional white noise (jitter term). We set the prior of $\beta$ to be log-uniform between 2 minutes and 1 day. Although this upper limit exceeds the observation duration, we still consider such long time scales due to the constraints from TESS, which indicate time-correlated noise with a time scale of about 1 day, as presented in Table~\ref{tab:TESS_params}. 

\bibliography{ms.bib}{}

\begin{thebibliography}{}
\expandafter\ifx\csname natexlab\endcsname\relax\def\natexlab#1{#1}\fi
\providecommand{\url}[1]{\href{#1}{#1}}
\providecommand{\dodoi}[1]{doi:~\href{http://doi.org/#1}{\nolinkurl{#1}}}
\providecommand{\doeprint}[1]{\href{http://ascl.net/#1}{\nolinkurl{http://ascl.net/#1}}}
\providecommand{\doarXiv}[1]{\href{https://arxiv.org/abs/#1}{\nolinkurl{https://arxiv.org/abs/#1}}}

% type= article
\bibitem[{M.~K. {Alam} {et~al.}(2025){Alam}, {Gao}, {Adams Redai}, {Wallack}, {Wogan}, {Aguichine}, {Dattilo}, {Alderson}, {Batalha}, {Batalha}, {Kirk}, {L{\'o}pez-Morales}, {Meech}, {Moran}, {Teske}, {Wakeford}, \& {Wolfgang}}]{Alam2025}
{Alam}, M.~K., {Gao}, P., {Adams Redai}, J., {et~al.} 2025, \bibinfo{title}{{JWST COMPASS: The First Near- to Mid-infrared Transmission Spectrum of the Hot Super-Earth L 168-9 b},} \aj, 169, 15, \dodoi{10.3847/1538-3881/ad8eb5}

% type= article
\bibitem[{L. {Alderson} {et~al.}(2024){Alderson}, {Batalha}, {Wakeford}, {Wallack}, {Aguichine}, {Teske}, {Adams Redai}, {Alam}, {Batalha}, {Gao}, {Kirk}, {L{\'o}pez-Morales}, {Moran}, {Scarsdale}, {Wogan}, \& {Wolfgang}}]{Alderson2024}
{Alderson}, L., {Batalha}, N.~E., {Wakeford}, H.~R., {et~al.} 2024, \bibinfo{title}{{JWST COMPASS: NIRSpec/G395H Transmission Observations of the Super-Earth TOI-836b},} \aj, 167, 216, \dodoi{10.3847/1538-3881/ad32c9}

% type= article
\bibitem[{L. {Alderson} {et~al.}(2025){Alderson}, {Moran}, {Wallack}, {Batalha}, {Wogan}, {Dattilo}, {Wakeford}, {Redai}, {Alam}, {Aguichine}, {Batalha}, {Gagnebin}, {Gao}, {Kirk}, {L{\'o}pez-Morales}, {Meech}, {Teske}, \& {Wolfgang}}]{Alderson2025}
{Alderson}, L., {Moran}, S.~E., {Wallack}, N.~L., {et~al.} 2025, \bibinfo{title}{{JWST COMPASS: NIRSpec/G395H Transmission Observations of the Super-Earth TOI-776 b},} \aj, 169, 142, \dodoi{10.3847/1538-3881/adad64}

% type= inproceedings
\bibitem[{F. {Allard}(2014){Allard}}]{Allard2014}
{Allard}, F. 2014, \bibinfo{title}{{The BT-Settl Model Atmospheres for Stars, Brown Dwarfs and Planets},} in IAU Symposium, Vol. 299, Exploring the Formation and Evolution of Planetary Systems, ed. M.~{Booth}, B.~C. {Matthews}, \& J.~R. {Graham}, 271--272, \dodoi{10.1017/S1743921313008545}

% type= article
\bibitem[{N.~H. {Allen} {et~al.}(2025){Allen}, {Espinoza}, {Diamond-Lowe}, {Mendon{\c{c}}a}, {Demory}, {Gressier}, {Ih}, {Fortune}, {August}, {Holmberg}, {Meier Vald{\'e}s}, {Zgraggen}, {Buchhave}, {Burgasser}, {Fisher}, {Gibson}, {Heng}, {Hoeijmakers}, {Kitzmann}, {Prinoth}, {Rathcke}, \& {Morris}}]{Allen2025}
{Allen}, N.~H., {Espinoza}, N., {Diamond-Lowe}, H., {et~al.} 2025, \bibinfo{title}{{Hot Rocks Survey. IV. Emission from LTT 3780 b Is Consistent with a Bare Rock},} \aj, 170, 240, \dodoi{10.3847/1538-3881/adfc51}

% type= article
\bibitem[{P. {Auclair-Desrotour} \& K. {Heng}(2020){Auclair-Desrotour} \& {Heng}}]{Auclair-Desrotour2020}
{Auclair-Desrotour}, P., \& {Heng}, K. 2020, \bibinfo{title}{{Atmospheric stability and collapse on tidally locked rocky planets},} \aap, 638, A77, \dodoi{10.1051/0004-6361/202037513}

% type= article
\bibitem[{P.~C. {August} {et~al.}(2025){August}, {Buchhave}, {Diamond-Lowe}, {Mendon{\c{c}}a}, {Gressier}, {Rathcke}, {Allen}, {Fortune}, {Jones}, {Meier Vald{\'e}s}, {Demory}, {Espinoza}, {Fisher}, {Gibson}, {Heng}, {Hoeijmakers}, {Hooton}, {Kitzmann}, {Prinoth}, {Eastman}, \& {Barnes}}]{August2025}
{August}, P.~C., {Buchhave}, L.~A., {Diamond-Lowe}, H., {et~al.} 2025, \bibinfo{title}{{Hot Rocks Survey I: A possible shallow eclipse for LHS 1478 b},} \aap, 695, A171, \dodoi{10.1051/0004-6361/202452611}

% type= article
\bibitem[{A. {Banerjee} {et~al.}(2024){Banerjee}, {Barstow}, {Gressier}, {Espinoza}, {Sing}, {Allen}, {Birkmann}, {Challener}, {Crouzet}, {Haswell}, {Lewis}, {Lewis}, \& {Yang}}]{Banerjee2024}
{Banerjee}, A., {Barstow}, J.~K., {Gressier}, A., {et~al.} 2024, \bibinfo{title}{{Atmospheric Retrievals Suggest the Presence of a Secondary Atmosphere and Possible Sulfur Species on L98-59 d from JWST Nirspec G395H Transmission Spectroscopy},} \apjl, 975, L11, \dodoi{10.3847/2041-8213/ad73d0}

% type= article
\bibitem[{Y.~I. {Baranov} {et~al.}(2004){Baranov}, {Lafferty}, \& {Fraser}}]{Baranov2004}
{Baranov}, Y.~I., {Lafferty}, W.~J., \& {Fraser}, G.~T. 2004, \bibinfo{title}{{Infrared spectrum of the continuum and dimer absorption in the vicinity of the O $_{2}$ vibrational fundamental in O $_{2}$/CO $_{2}$ mixtures},} Journal of Molecular Spectroscopy, 228, 432, \dodoi{10.1016/j.jms.2004.04.010}

% type= article
\bibitem[{R.~J. {Barber} {et~al.}(2006){Barber}, {Tennyson}, {Harris}, \& {Tolchenov}}]{Barber2006}
{Barber}, R.~J., {Tennyson}, J., {Harris}, G.~J., \& {Tolchenov}, R.~N. 2006, \bibinfo{title}{{A high-accuracy computed water line list},} Mon. Notices Royal Astron. Soc., 368, 1087, \dodoi{10.1111/j.1365-2966.2006.10184.x}

% type= article
\bibitem[{R. {Barnes}(2017){Barnes}}]{Barnes2017}
{Barnes}, R. 2017, \bibinfo{title}{{Tidal locking of habitable exoplanets},} Celestial Mechanics and Dynamical Astronomy, 129, 509, \dodoi{10.1007/s10569-017-9783-7}

% type= article
\bibitem[{T.~J. {Bell} {et~al.}(2024){Bell}, {Crouzet}, {Cubillos}, {Kreidberg}, {Piette}, {Roman}, {Barstow}, {Blecic}, {Carone}, {Coulombe}, {Ducrot}, {Hammond}, {Mendon{\c{c}}a}, {Moses}, {Parmentier}, {Stevenson}, {Teinturier}, {Zhang}, {Batalha}, {Bean}, {Benneke}, {Charnay}, {Chubb}, {Demory}, {Gao}, {Lee}, {L{\'o}pez-Morales}, {Morello}, {Rauscher}, {Sing}, {Tan}, {Venot}, {Wakeford}, {Aggarwal}, {Ahrer}, {Alam}, {Baeyens}, {Barrado}, {Caceres}, {Carter}, {Casewell}, {Challener}, {Crossfield}, {Decin}, {D{\'e}sert}, {Dobbs-Dixon}, {Dyrek}, {Espinoza}, {Feinstein}, {Gibson}, {Harrington}, {Helling}, {Hu}, {Iro}, {Kempton}, {Kendrew}, {Komacek}, {Krick}, {Lagage}, {Leconte}, {Lendl}, {Lewis}, {Lothringer}, {Malsky}, {Mancini}, {Mansfield}, {Mayne}, {Evans-Soma}, {Molaverdikhani}, {Nikolov}, {Nixon}, {Palle}, {Petit dit de la Roche}, {Piaulet}, {Powell}, {Rackham}, {Schneider}, {Steinrueck}, {Taylor}, {Welbanks}, {Yurchenko}, {Zhang}, \& {Zieba}}]{Bell2024}
{Bell}, T.~J., {Crouzet}, N., {Cubillos}, P.~E., {et~al.} 2024, \bibinfo{title}{{Nightside clouds and disequilibrium chemistry on the hot Jupiter WASP-43b},} Nature Astronomy, 8, 879, \dodoi{10.1038/s41550-024-02230-x}

% type= article
\bibitem[{A. {Bello-Arufe} {et~al.}(2025){Bello-Arufe}, {Damiano}, {Bennett}, {Hu}, {Welbanks}, {MacDonald}, {Seligman}, {Sing}, {Tokadjian}, {Oza}, \& {Yang}}]{BelloArufe2025}
{Bello-Arufe}, A., {Damiano}, M., {Bennett}, K.~A., {et~al.} 2025, \bibinfo{title}{{Evidence for a Volcanic Atmosphere on the Sub-Earth L 98-59 b},} \apjl, 980, L26, \dodoi{10.3847/2041-8213/adaf22}

% type= article
\bibitem[{J. {Bouwman} {et~al.}(2023){Bouwman}, {Kendrew}, {Greene}, {Bell}, {Lagage}, {Schreiber}, {Dicken}, {Sloan}, {Espinoza}, {Scheithauer}, {Coulais}, {Fox}, {Gastaud}, {Glauser}, {Jones}, {Labiano}, {Lahuis}, {Morrison}, {Murray}, {Mueller}, {Nayak}, {Wright}, {Glasse}, \& {Rieke}}]{Bouwman2023}
{Bouwman}, J., {Kendrew}, S., {Greene}, T.~P., {et~al.} 2023, \bibinfo{title}{{Spectroscopic Time Series Performance of the Mid-infrared Instrument on the JWST},} \pasp, 135, 038002, \dodoi{10.1088/1538-3873/acbc49}

% type= inproceedings
\bibitem[{H. {Bushouse}(2020){Bushouse}}]{Bushouse2020}
{Bushouse}, H. 2020, \bibinfo{title}{{The JWST Science Calibration Pipeline},} in Astronomical Society of the Pacific Conference Series, Vol. 527, Astronomical Data Analysis Software and Systems XXIX, ed. R.~{Pizzo}, E.~R. {Deul}, J.~D. {Mol}, J.~{de Plaa}, \& H.~{Verkouter}, 583

% type= article
\bibitem[{W. {Cassidy} \& B. {Hapke}(1975){Cassidy} \& {Hapke}}]{Cassidy1975}
{Cassidy}, W., \& {Hapke}, B. 1975, \bibinfo{title}{{Effects of Darkening Processes on Surfaces of Airless Bodies},} \icarus, 25, 371, \dodoi{10.1016/0019-1035(75)90002-0}

% type= inproceedings
\bibitem[{L.~C. {Cheek} {et~al.}(2009){Cheek}, {Pieters}, {Dyar}, \& {Milam}}]{Cheek2009}
{Cheek}, L.~C., {Pieters}, C.~M., {Dyar}, M.~D., \& {Milam}, K.~A. 2009, \bibinfo{title}{{Revisiting Plagioclase Optical Properties for Lunar Exploration},} in 40th Annual Lunar and Planetary Science Conference, Lunar and Planetary Science Conference, 1928

% type= book
\bibitem[{A.~N. {Cox}(2000){Cox}}]{Cox2000}
{Cox}, A.~N. 2000, {Allen's astrophysical quantities}

% type= article
\bibitem[{B.-O. Demory {et~al.}(2015)Demory, Gillon, Madhusudhan, \& Queloz}]{Demory2015}
Demory, B.-O., Gillon, M., Madhusudhan, N., \& Queloz, D. 2015, \bibinfo{title}{Variability in the super-Earth 55 Cnc e,} Monthly Notices of the Royal Astronomical Society, 455, 2018, \dodoi{10.1093/mnras/stv2239}

% type= article
\bibitem[{D. {Dicken} {et~al.}(2024){Dicken}, {Mar{\'\i}n}, {Shivaei}, {Guillard}, {Libralato}, {Glasse}, {Gordon}, {Cossou}, {Kavanagh}, {Temim}, {Flagey}, {Klaassen}, {Rieke}, {Wright}, {Alberts}, {Azzollini}, {{\'A}lvarez-M{\'a}rquez}, {Bouchet}, {Bright}, {Cracraft}, {Coulais}, {Detre}, {Engesser}, {Fox}, {Gaspar}, {Gastaud}, {Glauser}, {Hines}, {Kendrew}, {Labiano}, {Lagage}, {Lee}, {Law}, {Morrison}, {Noriega-Crespo}, {Jones}, {Patapis}, {Scheithauer}, {Sloan}, \& {Tamas}}]{Dicken2024}
{Dicken}, D., {Mar{\'\i}n}, M.~G., {Shivaei}, I., {et~al.} 2024, \bibinfo{title}{{JWST MIRI flight performance: Imaging},} \aap, 689, A5, \dodoi{10.1051/0004-6361/202449451}

% type= article
\bibitem[{E. {Ducrot} {et~al.}(2025){Ducrot}, {Lagage}, {Min}, {Gillon}, {Bell}, {Tremblin}, {Greene}, {Dyrek}, {Bouwman}, {Waters}, {G{\"u}del}, {Henning}, {Vandenbussche}, {Absil}, {Barrado}, {Boccaletti}, {Coulais}, {Decin}, {Edwards}, {Gastaud}, {Glasse}, {Kendrew}, {Olofsson}, {Patapis}, {Pye}, {Rouan}, {Whiteford}, {Argyriou}, {Cossou}, {Glauser}, {Krause}, {Lahuis}, {Royer}, {Scheithauer}, {Colina}, {van Dishoeck}, {Ostlin}, {Ray}, \& {Wright}}]{Ducrot2025}
{Ducrot}, E., {Lagage}, P.-O., {Min}, M., {et~al.} 2025, \bibinfo{title}{{Combined analysis of the 12.8 and 15 {\ensuremath{\mu}}m JWST/MIRI eclipse observations of TRAPPIST-1 b},} Nature Astronomy, 9, 358, \dodoi{10.1038/s41550-024-02428-z}

% type= article
\bibitem[{N. {Espinoza} {et~al.}(2019){Espinoza}, {Kossakowski}, \& {Brahm}}]{Espinoza_juliet2019}
{Espinoza}, N., {Kossakowski}, D., \& {Brahm}, R. 2019, \bibinfo{title}{{juliet: a versatile modelling tool for transiting and non-transiting exoplanetary systems},} \mnras, 490, 2262, \dodoi{10.1093/mnras/stz2688}

% type= article
\bibitem[{F. Feroz {et~al.}(2009)Feroz, Hobson, \& Bridges}]{Feroz2009}
Feroz, F., Hobson, M.~P., \& Bridges, M. 2009, \bibinfo{title}{{MultiNest: an efficient and robust Bayesian inference tool for cosmology and particle physics},} Monthly Notices of the Royal Astronomical Society, 398, 1601, \dodoi{10.1111/j.1365-2966.2009.14548.x}

% type= article
\bibitem[{D. {Foreman-Mackey} {et~al.}(2017){Foreman-Mackey}, {Agol}, {Ambikasaran}, \& {Angus}}]{ForemanMackey2017}
{Foreman-Mackey}, D., {Agol}, E., {Ambikasaran}, S., \& {Angus}, R. 2017, \bibinfo{title}{{Fast and Scalable Gaussian Process Modeling with Applications to Astronomical Time Series},} \aj, 154, 220, \dodoi{10.3847/1538-3881/aa9332}

% type= article
\bibitem[{M. {Fortune} {et~al.}(2025){Fortune}, {Gibson}, {Diamond-Lowe}, {Mendon{\c{c}}a}, {Gressier}, {Kitzmann}, {Allen}, {August}, {Ih}, {Meier Vald{\'e}s}, {Zgraggen}, {Buchhave}, {Demory}, {Espinoza}, {Heng}, {Jones}, \& {Rathcke}}]{Fortune2025}
{Fortune}, M., {Gibson}, N.~P., {Diamond-Lowe}, H., {et~al.} 2025, \bibinfo{title}{{Hot Rocks Survey III: A deep eclipse for LHS 1140c and a new Gaussian process method to account for correlated noise in individual pixels},} arXiv e-prints, arXiv:2505.22186, \dodoi{10.48550/arXiv.2505.22186}

% type= article
\bibitem[{ {Gaia Collaboration} {et~al.}(2016){Gaia Collaboration}, {Prusti}, {de Bruijne}, {Brown}, {Vallenari}, {Babusiaux}, {Bailer-Jones}, {Bastian}, {Biermann}, {Evans}, {Eyer}, {Jansen}, {Jordi}, {Klioner}, {Lammers}, {Lindegren}, {Luri}, {Mignard}, {Milligan}, {Panem}, {Poinsignon}, {Pourbaix}, {Randich}, {Sarri}, {Sartoretti}, {Siddiqui}, {Soubiran}, {Valette}, {van Leeuwen}, {Walton}, {Aerts}, {Arenou}, {Cropper}, {Drimmel}, {H{\o}g}, {Katz}, {Lattanzi}, {O'Mullane}, {Grebel}, {Holland}, {Huc}, {Passot}, {Bramante}, {Cacciari}, {Casta{\~n}eda}, {Chaoul}, {Cheek}, {De Angeli}, {Fabricius}, {Guerra}, {Hern{\'a}ndez}, {Jean-Antoine-Piccolo}, {Masana}, {Messineo}, {Mowlavi}, {Nienartowicz}, {Ord{\'o}{\~n}ez-Blanco}, {Panuzzo}, {Portell}, {Richards}, {Riello}, {Seabroke}, {Tanga}, {Th{\'e}venin}, {Torra}, {Els}, {Gracia-Abril}, {Comoretto}, {Garcia-Reinaldos}, {Lock}, {Mercier}, {Altmann}, {Andrae}, {Astraatmadja}, {Bellas-Velidis}, {Benson}, {Berthier}, {Blomme}, {Busso}, {Carry}, {Cellino},
  {Clementini}, {Cowell}, {Creevey}, {Cuypers}, {Davidson}, {De Ridder}, {de Torres}, {Delchambre}, {Dell'Oro}, {Ducourant}, {Fr{\'e}mat}, {Garc{\'\i}a-Torres}, {Gosset}, {Halbwachs}, {Hambly}, {Harrison}, {Hauser}, {Hestroffer}, {Hodgkin}, {Huckle}, {Hutton}, {Jasniewicz}, {Jordan}, {Kontizas}, {Korn}, {Lanzafame}, {Manteiga}, {Moitinho}, {Muinonen}, {Osinde}, {Pancino}, {Pauwels}, {Petit}, {Recio-Blanco}, {Robin}, {Sarro}, {Siopis}, {Smith}, {Smith}, {Sozzetti}, {Thuillot}, {van Reeven}, {Viala}, {Abbas}, {Abreu Aramburu}, {Accart}, {Aguado}, {Allan}, {Allasia}, {Altavilla}, {{\'A}lvarez}, {Alves}, {Anderson}, {Andrei}, {Anglada Varela}, {Antiche}, {Antoja}, {Ant{\'o}n}, {Arcay}, {Atzei}, {Ayache}, {Bach}, {Baker}, {Balaguer-N{\'u}{\~n}ez}, {Barache}, {Barata}, {Barbier}, {Barblan}, {Baroni}, {Barrado y Navascu{\'e}s}, {Barros}, {Barstow}, {Becciani}, {Bellazzini}, {Bellei}, {Bello Garc{\'\i}a}, {Belokurov}, {Bendjoya}, {Berihuete}, {Bianchi}, {Bienaym{\'e}}, {Billebaud}, {Blagorodnova}, {Blanco-Cuaresma},
  {Boch}, {Bombrun}, {Borrachero}, {Bouquillon}, {Bourda}, {Bouy}, {Bragaglia}, {Breddels}, {Brouillet}, {Br{\"u}semeister}, {Bucciarelli}, {Budnik}, {Burgess}, {Burgon}, {Burlacu}, {Busonero}, {Buzzi}, {Caffau}, {Cambras}, {Campbell}, {Cancelliere}, {Cantat-Gaudin}, {Carlucci}, {Carrasco}, {Castellani}, {Charlot}, {Charnas}, {Charvet}, {Chassat}, {Chiavassa}, {Clotet}, {Cocozza}, {Collins}, {Collins}, \& {Costigan}}]{Gaia2016}
{Gaia Collaboration}, {Prusti}, T., {de Bruijne}, J.~H.~J., {et~al.} 2016, \bibinfo{title}{{The Gaia mission},} \aap, 595, A1, \dodoi{10.1051/0004-6361/201629272}

% type= article
\bibitem[{ {Gaia Collaboration} {et~al.}(2023){Gaia Collaboration}, {Vallenari}, {Brown}, {Prusti}, {de Bruijne}, {Arenou}, {Babusiaux}, {Biermann}, {Creevey}, {Ducourant}, {Evans}, {Eyer}, {Guerra}, {Hutton}, {Jordi}, {Klioner}, {Lammers}, {Lindegren}, {Luri}, {Mignard}, {Panem}, {Pourbaix}, {Randich}, {Sartoretti}, {Soubiran}, {Tanga}, {Walton}, {Bailer-Jones}, {Bastian}, {Drimmel}, {Jansen}, {Katz}, {Lattanzi}, {van Leeuwen}, {Bakker}, {Cacciari}, {Casta{\~n}eda}, {De Angeli}, {Fabricius}, {Fouesneau}, {Fr{\'e}mat}, {Galluccio}, {Guerrier}, {Heiter}, {Masana}, {Messineo}, {Mowlavi}, {Nicolas}, {Nienartowicz}, {Pailler}, {Panuzzo}, {Riclet}, {Roux}, {Seabroke}, {Sordo}, {Th{\'e}venin}, {Gracia-Abril}, {Portell}, {Teyssier}, {Altmann}, {Andrae}, {Audard}, {Bellas-Velidis}, {Benson}, {Berthier}, {Blomme}, {Burgess}, {Busonero}, {Busso}, {C{\'a}novas}, {Carry}, {Cellino}, {Cheek}, {Clementini}, {Damerdji}, {Davidson}, {de Teodoro}, {Nu{\~n}ez Campos}, {Delchambre}, {Dell'Oro}, {Esquej},
  {Fern{\'a}ndez-Hern{\'a}ndez}, {Fraile}, {Garabato}, {Garc{\'\i}a-Lario}, {Gosset}, {Haigron}, {Halbwachs}, {Hambly}, {Harrison}, {Hern{\'a}ndez}, {Hestroffer}, {Hodgkin}, {Holl}, {Jan{\ss}en}, {Jevardat de Fombelle}, {Jordan}, {Krone-Martins}, {Lanzafame}, {L{\"o}ffler}, {Marchal}, {Marrese}, {Moitinho}, {Muinonen}, {Osborne}, {Pancino}, {Pauwels}, {Recio-Blanco}, {Reyl{\'e}}, {Riello}, {Rimoldini}, {Roegiers}, {Rybizki}, {Sarro}, {Siopis}, {Smith}, {Sozzetti}, {Utrilla}, {van Leeuwen}, {Abbas}, {{\'A}brah{\'a}m}, {Abreu Aramburu}, {Aerts}, {Aguado}, {Ajaj}, {Aldea-Montero}, {Altavilla}, {{\'A}lvarez}, {Alves}, {Anders}, {Anderson}, {Anglada Varela}, {Antoja}, {Baines}, {Baker}, {Balaguer-N{\'u}{\~n}ez}, {Balbinot}, {Balog}, {Barache}, {Barbato}, {Barros}, {Barstow}, {Bartolom{\'e}}, {Bassilana}, {Bauchet}, {Becciani}, {Bellazzini}, {Berihuete}, {Bernet}, {Bertone}, {Bianchi}, {Binnenfeld}, {Blanco-Cuaresma}, {Blazere}, {Boch}, {Bombrun}, {Bossini}, {Bouquillon}, {Bragaglia}, {Bramante}, {Breedt},
  {Bressan}, {Brouillet}, {Brugaletta}, {Bucciarelli}, {Burlacu}, {Butkevich}, {Buzzi}, {Caffau}, {Cancelliere}, {Cantat-Gaudin}, {Carballo}, {Carlucci}, {Carnerero}, {Carrasco}, {Casamiquela}, {Castellani}, {Castro-Ginard}, {Chaoul}, {Charlot}, {Chemin}, {Chiaramida}, {Chiavassa}, {Chornay}, {Comoretto}, {Contursi}, {Cooper}, {Cornez}, {Cowell}, {Crifo}, {Cropper}, {Crosta}, {Crowley}, {Dafonte}, {Dapergolas}, {David}, {David}, {de Laverny}, {De Luise}, \& {De March}}]{Gaia2023}
{Gaia Collaboration}, {Vallenari}, A., {Brown}, A.~G.~A., {et~al.} 2023, \bibinfo{title}{{Gaia Data Release 3. Summary of the content and survey properties},} \aap, 674, A1, \dodoi{10.1051/0004-6361/202243940}

% type= article
\bibitem[{P. {Gao} {et~al.}(2023){Gao}, {Piette}, {Steinrueck}, {Nixon}, {Zhang}, {Kempton}, {Bean}, {Rauscher}, {Parmentier}, {Batalha}, {Savel}, {Arnold}, {Roman}, {Malsky}, \& {Taylor}}]{Gao2023}
{Gao}, P., {Piette}, A. A.~A., {Steinrueck}, M.~E., {et~al.} 2023, \bibinfo{title}{{The Hazy and Metal-rich Atmosphere of GJ 1214 b Constrained by Near- and Mid-infrared Transmission Spectroscopy},} \apj, 951, 96, \dodoi{10.3847/1538-4357/acd16f}

% type= article
\bibitem[{J.~P. {Gardner} {et~al.}(2006){Gardner}, {Mather}, {Clampin}, {Doyon}, {Greenhouse}, {Hammel}, {Hutchings}, {Jakobsen}, {Lilly}, {Long}, {Lunine}, {McCaughrean}, {Mountain}, {Nella}, {Rieke}, {Rieke}, {Rix}, {Smith}, {Sonneborn}, {Stiavelli}, {Stockman}, {Windhorst}, \& {Wright}}]{Gardner2006}
{Gardner}, J.~P., {Mather}, J.~C., {Clampin}, M., {et~al.} 2006, \bibinfo{title}{{The James Webb Space Telescope},} \ssr, 123, 485, \dodoi{10.1007/s11214-006-8315-7}

% type= article
\bibitem[{M. {Gillon} {et~al.}(2025){Gillon}, {Ducrot}, {Bell}, {Huang}, {Lincowski}, {Lyu}, {Maurel}, {Revol}, {Agol}, {Bolmont}, {Dong}, {Fauchez}, {Koll}, {Leconte}, {Meadows}, {Selsis}, {Turbet}, {Charnay}, {Delre}, {Demory}, {Householder}, {Zieba}, {Berardo}, {Dyrek}, {Edwards}, {de Wit}, {Greene}, {Hu}, {Iro}, {Kreidberg}, {Lagage}, {Lustig-Yaeger}, \& {Iyer}}]{Gillon2025}
{Gillon}, M., {Ducrot}, E., {Bell}, T.~J., {et~al.} 2025, \bibinfo{title}{{First JWST thermal phase curves of temperate terrestrial exoplanets reveal no thick atmosphere around TRAPPIST-1 b and c},} arXiv e-prints, arXiv:2509.02128, \dodoi{10.48550/arXiv.2509.02128}

% type= article
\bibitem[{S. {Ginzburg} {et~al.}(2018){Ginzburg}, {Schlichting}, \& {Sari}}]{Ginzburg2018}
{Ginzburg}, S., {Schlichting}, H.~E., \& {Sari}, R. 2018, \bibinfo{title}{{Core-powered mass-loss and the radius distribution of small exoplanets},} \mnras, 476, 759, \dodoi{10.1093/mnras/sty290}

% type= article
\bibitem[{K.~D. Gordon {et~al.}(2024)Gordon, Sloan, Garcia~Marin, Libralato, Rieke, Aguilar, Bohlin, Cracraft, Decleir, Gaspar, Kendrew, Law, Noriega-Crespo, \& Regan}]{Gordon2025}
Gordon, K.~D., Sloan, G.~C., Garcia~Marin, M., {et~al.} 2024, \bibinfo{title}{The James Webb Space Telescope Absolute Flux Calibration. II. Mid-infrared Instrument Imaging and Coronagraphy,} The Astronomical Journal, 169, 6, \dodoi{10.3847/1538-3881/ad8cd4}

% type= article
\bibitem[{T.~P. {Greene} {et~al.}(2023){Greene}, {Bell}, {Ducrot}, {Dyrek}, {Lagage}, \& {Fortney}}]{Greene2023}
{Greene}, T.~P., {Bell}, T.~J., {Ducrot}, E., {et~al.} 2023, \bibinfo{title}{{Thermal emission from the Earth-sized exoplanet TRAPPIST-1 b using JWST},} \nat, 618, 39, \dodoi{10.1038/s41586-023-05951-7}

% type= article
\bibitem[{A. {Gressier} {et~al.}(2024){Gressier}, {Espinoza}, {Allen}, {Sing}, {Banerjee}, {Barstow}, {Valenti}, {Lewis}, {Birkmann}, {Challener}, {Manjavacas}, {Alves de Oliveira}, {Crouzet}, \& {Beck}}]{Gressier2024}
{Gressier}, A., {Espinoza}, N., {Allen}, N.~H., {et~al.} 2024, \bibinfo{title}{{Hints of a Sulfur-rich Atmosphere around the 1.6 R $_{{\ensuremath{\oplus}}}$ Super-Earth L98-59 d from JWST NIRspec G395H Transmission Spectroscopy},} \apjl, 975, L10, \dodoi{10.3847/2041-8213/ad73d1}

% type= article
\bibitem[{S.~L. {Grimm} \& K. {Heng}(2015){Grimm} \& {Heng}}]{Grimm2015}
{Grimm}, S.~L., \& {Heng}, K. 2015, \bibinfo{title}{{HELIOS-K: An Ultrafast, Open-source Opacity Calculator for Radiative Transfer},} \apj, 808, 182, \dodoi{10.1088/0004-637X/808/2/182}

% type= article
\bibitem[{S.~L. {Grimm} {et~al.}(2021){Grimm}, {Malik}, {Kitzmann}, {Guzm{\'a}n-Mesa}, {Hoeijmakers}, {Fisher}, {Mendon{\c{c}}a}, {Yurchenko}, {Tennyson}, {Alesina}, {Buchschacher}, {Burnier}, {Segransan}, {Kurucz}, \& {Heng}}]{Grimm2021}
{Grimm}, S.~L., {Malik}, M., {Kitzmann}, D., {et~al.} 2021, \bibinfo{title}{{HELIOS-K 2.0 Opacity Calculator and Open-source Opacity Database for Exoplanetary Atmospheres},} \apjs, 253, 30, \dodoi{10.3847/1538-4365/abd773}

% type= article
\bibitem[{M. {Gruszka} \& A. {Borysow}(1997){Gruszka} \& {Borysow}}]{Gruszka1997}
{Gruszka}, M., \& {Borysow}, A. 1997, \bibinfo{title}{{Roto-Translational Collision-Induced Absorption of CO $_{2}$for the Atmosphere of Venus at Frequencies from 0 to 250 cm $^{-1}$, at Temperatures from 200 to 800 K},} \icarus, 129, 172, \dodoi{10.1006/icar.1997.5773}

% type= article
\bibitem[{M. {Hammond} {et~al.}(2025){Hammond}, {Guimond}, {Lichtenberg}, {Nicholls}, {Fisher}, {Luque}, {Meier}, {Taylor}, {Changeat}, {Dang}, {Hay}, {Herbort}, \& {Teske}}]{Hammond2025}
{Hammond}, M., {Guimond}, C.~M., {Lichtenberg}, T., {et~al.} 2025, \bibinfo{title}{{Reliable Detections of Atmospheres on Rocky Exoplanets with Photometric JWST Phase Curves},} \apjl, 978, L40, \dodoi{10.3847/2041-8213/ada0bc}

% type= article
\bibitem[{B. {Hapke}(2001){Hapke}}]{Hapke2001}
{Hapke}, B. 2001, \bibinfo{title}{{Space weathering from Mercury to the asteroid belt},} \jgr, 106, 10039, \dodoi{10.1029/2000JE001338}

% type= article
\bibitem[{K. {Heng}(2023){Heng}}]{Heng2023}
{Heng}, K. 2023, \bibinfo{title}{{The Transient Outgassed Atmosphere of 55 Cancri e},} \apjl, 956, L20, \dodoi{10.3847/2041-8213/acfe05}

% type= article
\bibitem[{M. {Holmberg} \& N. {Madhusudhan}(2023){Holmberg} \& {Madhusudhan}}]{holmberg2023}
{Holmberg}, M., \& {Madhusudhan}, N. 2023, \bibinfo{title}{{Exoplanet spectroscopy with JWST NIRISS: diagnostics and case studies},} \mnras, 524, 377, \dodoi{10.1093/mnras/stad1580}

% type= article
\bibitem[{K. Horne(1986)Horne}]{horne_optimal_1986}
Horne, K. 1986, \bibinfo{title}{An optimal extraction algorithm for {CCD} spectroscopy,} Publications of the Astronomical Society of the Pacific, 98, 609, \dodoi{10.1086/131801}

% type= article
\bibitem[{R. {Hu} {et~al.}(2012){Hu}, {Ehlmann}, \& {Seager}}]{Hu2012}
{Hu}, R., {Ehlmann}, B.~L., \& {Seager}, S. 2012, \bibinfo{title}{{Theoretical Spectra of Terrestrial Exoplanet Surfaces},} \apj, 752, 7, \dodoi{10.1088/0004-637X/752/1/7}

% type= article
\bibitem[{R. {Hu} {et~al.}(2024){Hu}, {Bello-Arufe}, {Zhang}, {Paragas}, {Zilinskas}, {van Buchem}, {Bess}, {Patel}, {Ito}, {Damiano}, {Scheucher}, {Oza}, {Knutson}, {Miguel}, {Dragomir}, {Brandeker}, \& {Demory}}]{Hu2024}
{Hu}, R., {Bello-Arufe}, A., {Zhang}, M., {et~al.} 2024, \bibinfo{title}{{A secondary atmosphere on the rocky exoplanet 55 Cancri e},} \nat, 630, 609, \dodoi{10.1038/s41586-024-07432-x}

% type= article
\bibitem[{J. {Ih} {et~al.}(2023){Ih}, {Kempton}, {Whittaker}, \& {Lessard}}]{Ih2023}
{Ih}, J., {Kempton}, E. M.~R., {Whittaker}, E.~A., \& {Lessard}, M. 2023, \bibinfo{title}{{Constraining the Thickness of TRAPPIST-1 b's Atmosphere from Its JWST Secondary Eclipse Observation at 15 {\ensuremath{\mu}}m},} \apjl, 952, L4, \dodoi{10.3847/2041-8213/ace03b}

% type= article
\bibitem[{A.~R. {Iyer} {et~al.}(2023){Iyer}, {Line}, {Muirhead}, {Fortney}, \& {Gharib-Nezhad}}]{Iyer2023}
{Iyer}, A.~R., {Line}, M.~R., {Muirhead}, P.~S., {Fortney}, J.~J., \& {Gharib-Nezhad}, E. 2023, \bibinfo{title}{{The SPHINX M-dwarf Spectral Grid. I. Benchmarking New Model Atmospheres to Derive Fundamental M-dwarf Properties},} \apj, 944, 41, \dodoi{10.3847/1538-4357/acabc2}

% type= article
\bibitem[{X. {Ji} {et~al.}(2025){Ji}, {Chatterjee}, {Park Coy}, \& {Kite}}]{Ji2025}
{Ji}, X., {Chatterjee}, R.~D., {Park Coy}, B., \& {Kite}, E.~S. 2025, \bibinfo{title}{{The Cosmic Shoreline Revisited: A Metric for Atmospheric Retention Informed by Hydrodynamic Escape},} arXiv e-prints, arXiv:2504.19872, \dodoi{10.48550/arXiv.2504.19872}

% type= article
\bibitem[{J.~F. {Kasting} {et~al.}(1993){Kasting}, {Whitmire}, \& {Reynolds}}]{Kasting1993}
{Kasting}, J.~F., {Whitmire}, D.~P., \& {Reynolds}, R.~T. 1993, \bibinfo{title}{{Habitable Zones around Main Sequence Stars},} Icarus, 101, 108, \dodoi{10.1006/icar.1993.1010}

% type= article
\bibitem[{J. {Kemmer} {et~al.}(2020){Kemmer}, {Stock}, {Kossakowski}, {Kaminski}, {Molaverdikhani}, {Schlecker}, {Caballero}, {Amado}, {Astudillo-Defru}, {Bonfils}, {Ciardi}, {Collins}, {Espinoza}, {Fukui}, {Hirano}, {Jenkins}, {Latham}, {Matthews}, {Narita}, {Pall{\'e}}, {Parviainen}, {Quirrenbach}, {Reiners}, {Ribas}, {Ricker}, {Schlieder}, {Seager}, {Vanderspek}, {Winn}, {Almenara}, {B{\'e}jar}, {Bluhm}, {Bouchy}, {Boyd}, {Christiansen}, {Cifuentes}, {Cloutier}, {Collins}, {Cort{\'e}s-Contreras}, {Crossfield}, {Crouzet}, {de Leon}, {Della-Rose}, {Delfosse}, {Dreizler}, {Esparza-Borges}, {Essack}, {Forveille}, {Figueira}, {Galad{\'\i}-Enr{\'\i}quez}, {Gan}, {Glidden}, {Gonzales}, {Guerra}, {Harakawa}, {Hatzes}, {Henning}, {Herrero}, {Hodapp}, {Hori}, {Howell}, {Ikoma}, {Isogai}, {Jeffers}, {K{\"u}rster}, {Kawauchi}, {Kimura}, {Klagyivik}, {Kotani}, {Kurokawa}, {Kusakabe}, {Kuzuhara}, {Lafarga}, {Livingston}, {Luque}, {Matson}, {Morales}, {Mori}, {Muirhead}, {Murgas}, {Nishikawa}, {Nishiumi}, {Omiya},
  {Reffert}, {Rodr{\'\i}guez L{\'o}pez}, {Santos}, {Sch{\"o}fer}, {Schwarz}, {Shiao}, {Tamura}, {Terada}, {Twicken}, {Ueda}, {Vievard}, {Watanabe}, \& {Zechmeister}}]{Kemmer2020}
{Kemmer}, J., {Stock}, S., {Kossakowski}, D., {et~al.} 2020, \bibinfo{title}{{Discovery of a hot, transiting, Earth-sized planet and a second temperate, non-transiting planet around the M4 dwarf GJ 3473 (TOI-488)},} \aap, 642, A236, \dodoi{10.1051/0004-6361/202038967}

% type= article
\bibitem[{D.~D.~B. {Koll}(2022){Koll}}]{Koll2022}
{Koll}, D. D.~B. 2022, \bibinfo{title}{{A Scaling for Atmospheric Heat Redistribution on Tidally Locked Rocky Planets},} \apj, 924, 134, \dodoi{10.3847/1538-4357/ac3b48}

% type= article
\bibitem[{D.~D.~B. {Koll} {et~al.}(2019){Koll}, {Malik}, {Mansfield}, {Kempton}, {Kite}, {Abbot}, \& {Bean}}]{Koll2019b}
{Koll}, D. D.~B., {Malik}, M., {Mansfield}, M., {et~al.} 2019, \bibinfo{title}{{Identifying Candidate Atmospheres on Rocky M Dwarf Planets via Eclipse Photometry},} \apj, 886, 140, \dodoi{10.3847/1538-4357/ab4c91}

% type= article
\bibitem[{L. Kreidberg(2015)Kreidberg}]{kreidberg_batman_2015}
Kreidberg, L. 2015, \bibinfo{title}{batman : {BAsic} {Transit} {Model} {cAlculatioN} in {Python},} Publications of the Astronomical Society of the Pacific, 127, 1161, \dodoi{10.1086/683602}

% type= article
\bibitem[{J.~S.~V. {Lagerros}(1998){Lagerros}}]{Lagerros1998}
{Lagerros}, J. S.~V. 1998, \bibinfo{title}{{Thermal physics of asteroids. IV. Thermal infrared beaming},} \aap, 332, 1123

% type= article
\bibitem[{J. {Leconte} {et~al.}(2015){Leconte}, {Wu}, {Menou}, \& {Murray}}]{Leconte2015}
{Leconte}, J., {Wu}, H., {Menou}, K., \& {Murray}, N. 2015, \bibinfo{title}{{Asynchronous rotation of Earth-mass planets in the habitable zone of lower-mass stars},} Science, 347, 632, \dodoi{10.1126/science.1258686}

% type= article
\bibitem[{O. {Lim} {et~al.}(2023){Lim}, {Benneke}, {Doyon}, {MacDonald}, {Piaulet}, {Artigau}, {Coulombe}, {Radica}, {L'Heureux}, {Albert}, {Rackham}, {de Wit}, {Salhi}, {Roy}, {Flagg}, {Fournier-Tondreau}, {Taylor}, {Cook}, {Lafreni{\`e}re}, {Cowan}, {Kaltenegger}, {Rowe}, {Espinoza}, {Dang}, \& {Darveau-Bernier}}]{Lim2023}
{Lim}, O., {Benneke}, B., {Doyon}, R., {et~al.} 2023, \bibinfo{title}{{Atmospheric Reconnaissance of TRAPPIST-1 b with JWST/NIRISS: Evidence for Strong Stellar Contamination in the Transmission Spectra},} \apjl, 955, L22, \dodoi{10.3847/2041-8213/acf7c4}

% type= article
\bibitem[{A.~P. {Lincowski} {et~al.}(2023){Lincowski}, {Meadows}, {Zieba}, {Kreidberg}, {Morley}, {Gillon}, {Selsis}, {Agol}, {Bolmont}, {Ducrot}, {Hu}, {Koll}, {Lyu}, {Mandell}, {Suissa}, \& {Tamburo}}]{Lincowski2023}
{Lincowski}, A.~P., {Meadows}, V.~S., {Zieba}, S., {et~al.} 2023, \bibinfo{title}{{Potential Atmospheric Compositions of TRAPPIST-1 c Constrained by JWST/MIRI Observations at 15 {\ensuremath{\mu}}m},} \apjl, 955, L7, \dodoi{10.3847/2041-8213/acee02}

% type= article
\bibitem[{E.~D. {Lopez} \& J.~J. {Fortney}(2013){Lopez} \& {Fortney}}]{Lopez2013}
{Lopez}, E.~D., \& {Fortney}, J.~J. 2013, \bibinfo{title}{{The Role of Core Mass in Controlling Evaporation: The Kepler Radius Distribution and the Kepler-36 Density Dichotomy},} \apj, 776, 2, \dodoi{10.1088/0004-637X/776/1/2}

% type= article
\bibitem[{R.~O.~P. Loyd {et~al.}(2018)Loyd, France, Youngblood, Schneider, Brown, Hu, Segura, Linsky, Redfield, Tian, Rugheimer, Miguel, \& Froning}]{Loyd2018}
Loyd, R. O.~P., France, K., Youngblood, A., {et~al.} 2018, \bibinfo{title}{The MUSCLES Treasury Survey. V. FUV Flares on Active and Inactive M Dwarfs,} 867, 71, \dodoi{10.3847/1538-4357/aae2bd}

% type= article
\bibitem[{R. {Luger} {et~al.}(2015){Luger}, {Barnes}, {Lopez}, {Fortney}, {Jackson}, \& {Meadows}}]{luger2015}
{Luger}, R., {Barnes}, R., {Lopez}, E., {et~al.} 2015, \bibinfo{title}{{Habitable Evaporated Cores: Transforming Mini-Neptunes into Super-Earths in the Habitable Zones of M Dwarfs},} Astrobiology, 15, 57, \dodoi{10.1089/ast.2014.1215}

% type= article
\bibitem[{J. {Lustig-Yaeger} {et~al.}(2023){Lustig-Yaeger}, {Fu}, {May}, {Ceballos}, {Moran}, {Peacock}, {Stevenson}, {Kirk}, {L{\'o}pez-Morales}, {MacDonald}, {Mayorga}, {Sing}, {Sotzen}, {Valenti}, {Redai}, {Alam}, {Batalha}, {Bennett}, {Gonzalez-Quiles}, {Kruse}, {Lothringer}, {Rustamkulov}, \& {Wakeford}}]{LustigYaeger2023}
{Lustig-Yaeger}, J., {Fu}, G., {May}, E.~M., {et~al.} 2023, \bibinfo{title}{{A JWST transmission spectrum of the nearby Earth-sized exoplanet LHS 475 b},} Nature Astronomy, 7, 1317, \dodoi{10.1038/s41550-023-02064-z}

% type= article
\bibitem[{X. {Lyu} {et~al.}(2024){Lyu}, {Koll}, {Cowan}, {Hu}, {Kreidberg}, \& {Rose}}]{Lyu2024}
{Lyu}, X., {Koll}, D. D.~B., {Cowan}, N.~B., {et~al.} 2024, \bibinfo{title}{{Super-Earth LHS3844b is Tidally Locked},} \apj, 964, 152, \dodoi{10.3847/1538-4357/ad2077}

% type= article
\bibitem[{N. {Madhusudhan} {et~al.}(2025){Madhusudhan}, {Constantinou}, {Holmberg}, {Sarkar}, {Piette}, \& {Moses}}]{Madhusudhan2025}
{Madhusudhan}, N., {Constantinou}, S., {Holmberg}, M., {et~al.} 2025, \bibinfo{title}{{New Constraints on DMS and DMDS in the Atmosphere of K2-18 b from JWST MIRI},} \apjl, 983, L40, \dodoi{10.3847/2041-8213/adc1c8}

% type= article
\bibitem[{M. {Malik} {et~al.}(2019{\natexlab{a}}){Malik}, {Kempton}, {Koll}, {Mansfield}, {Bean}, \& {Kite}}]{Malik2019b}
{Malik}, M., {Kempton}, E. M.~R., {Koll}, D. D.~B., {et~al.} 2019{\natexlab{a}}, \bibinfo{title}{{Analyzing Atmospheric Temperature Profiles and Spectra of M Dwarf Rocky Planets},} \apj, 886, 142, \dodoi{10.3847/1538-4357/ab4a05}

% type= article
\bibitem[{M. {Malik} {et~al.}(2019{\natexlab{b}}){Malik}, {Kitzmann}, {Mendon{\c{c}}a}, {Grimm}, {Marleau}, {Linder}, {Tsai}, \& {Heng}}]{Malik2019a}
{Malik}, M., {Kitzmann}, D., {Mendon{\c{c}}a}, J.~M., {et~al.} 2019{\natexlab{b}}, \bibinfo{title}{{Self-luminous and Irradiated Exoplanetary Atmospheres Explored with HELIOS},} \aj, 157, 170, \dodoi{10.3847/1538-3881/ab1084}

% type= article
\bibitem[{M. {Malik} {et~al.}(2017){Malik}, {Grosheintz}, {Mendon{\c{c}}a}, {Grimm}, {Lavie}, {Kitzmann}, {Tsai}, {Burrows}, {Kreidberg}, {Bedell}, {Bean}, {Stevenson}, \& {Heng}}]{Malik2017}
{Malik}, M., {Grosheintz}, L., {Mendon{\c{c}}a}, J.~M., {et~al.} 2017, \bibinfo{title}{{HELIOS: An Open-source, GPU-accelerated Radiative Transfer Code for Self-consistent Exoplanetary Atmospheres},} \aj, 153, 56, \dodoi{10.3847/1538-3881/153/2/56}

% type= article
\bibitem[{M. {Mansfield} {et~al.}(2019){Mansfield}, {Kite}, {Hu}, {Koll}, {Malik}, {Bean}, \& {Kempton}}]{Mansfield2019}
{Mansfield}, M., {Kite}, E.~S., {Hu}, R., {et~al.} 2019, \bibinfo{title}{{Identifying Atmospheres on Rocky Exoplanets through Inferred High Albedo},} \apj, 886, 141, \dodoi{10.3847/1538-4357/ab4c90}

% type= article
\bibitem[{E.~M. {May} {et~al.}(2023){May}, {MacDonald}, {Bennett}, {Moran}, {Wakeford}, {Peacock}, {Lustig-Yaeger}, {Highland}, {Stevenson}, {Sing}, {Mayorga}, {Batalha}, {Kirk}, {L{\'o}pez-Morales}, {Valenti}, {Alam}, {Alderson}, {Fu}, {Gonzalez-Quiles}, {Lothringer}, {Rustamkulov}, \& {Sotzen}}]{May2023}
{May}, E.~M., {MacDonald}, R.~J., {Bennett}, K.~A., {et~al.} 2023, \bibinfo{title}{{Double Trouble: Two Transits of the Super-Earth GJ 1132 b Observed with JWST NIRSpec G395H},} \apjl, 959, L9, \dodoi{10.3847/2041-8213/ad054f}

% type= article
\bibitem[{E.~A. {Meier Vald{\'e}s} {et~al.}(2025){Meier Vald{\'e}s}, {Demory}, {Diamond-Lowe}, {Mendon{\c{c}}a}, {August}, {Fortune}, {Allen}, {Kitzmann}, {Gressier}, {Hooton}, {Jones}, {Buchhave}, {Espinoza}, {Fisher}, {Gibson}, {Heng}, {Hoeijmakers}, {Prinoth}, {Rathcke}, \& {Eastman}}]{MeierValdes2025}
{Meier Vald{\'e}s}, E.~A., {Demory}, B.~O., {Diamond-Lowe}, H., {et~al.} 2025, \bibinfo{title}{{Hot Rocks Survey: II. The thermal emission of TOI-1468 b reveals a bare hot rock},} \aap, 698, A68, \dodoi{10.1051/0004-6361/202453449}

% type= article
\bibitem[{E.~J. {Mlawer} {et~al.}(2023){Mlawer}, {Cady-Pereira}, {Mascio}, \& {Gordon}}]{Mlawer2023}
{Mlawer}, E.~J., {Cady-Pereira}, K.~E., {Mascio}, J., \& {Gordon}, I.~E. 2023, \bibinfo{title}{{The inclusion of the MT\_CKD water vapor continuum model in the HITRAN molecular spectroscopic database},} \jqsrt, 306, 108645, \dodoi{10.1016/j.jqsrt.2023.108645}

% type= article
\bibitem[{S.~E. {Moran} {et~al.}(2023){Moran}, {Stevenson}, {Sing}, {MacDonald}, {Kirk}, {Lustig-Yaeger}, {Peacock}, {Mayorga}, {Bennett}, {L{\'o}pez-Morales}, {May}, {Rustamkulov}, {Valenti}, {Adams Redai}, {Alam}, {Batalha}, {Fu}, {Gonzalez-Quiles}, {Highland}, {Kruse}, {Lothringer}, {Ortiz Ceballos}, {Sotzen}, \& {Wakeford}}]{moran_high_2023}
{Moran}, S.~E., {Stevenson}, K.~B., {Sing}, D.~K., {et~al.} 2023, \bibinfo{title}{{High Tide or Riptide on the Cosmic Shoreline? A Water-rich Atmosphere or Stellar Contamination for the Warm Super-Earth GJ 486b from JWST Observations},} \apjl, 948, L11, \dodoi{10.3847/2041-8213/accb9c}

% type= article
\bibitem[{J.~E. {Owen} \& Y. {Wu}(2017){Owen} \& {Wu}}]{Owen2017}
{Owen}, J.~E., \& {Wu}, Y. 2017, \bibinfo{title}{{The Evaporation Valley in the Kepler Planets},} \apj, 847, 29, \dodoi{10.3847/1538-4357/aa890a}

% type= article
\bibitem[{K. {Paragas} {et~al.}(2025){Paragas}, {Knutson}, {Hu}, {Ehlmann}, {Alemanno}, {Helbert}, {Maturilli}, {Zhang}, {Iyer}, \& {Rossman}}]{Paragas2025}
{Paragas}, K., {Knutson}, H.~A., {Hu}, R., {et~al.} 2025, \bibinfo{title}{{A New Spectral Library for Modeling the Surfaces of Hot, Rocky Exoplanets},} \apj, 981, 130, \dodoi{10.3847/1538-4357/ada9eb}

% type= article
\bibitem[{E.~K. {Pass} {et~al.}(2025){Pass}, {Charbonneau}, \& {Vanderburg}}]{Pass2025}
{Pass}, E.~K., {Charbonneau}, D., \& {Vanderburg}, A. 2025, \bibinfo{title}{{The Receding Cosmic Shoreline of Mid-to-late M Dwarfs: Measurements of Active Lifetimes Worsen Challenges for Atmosphere Retention by Rocky Exoplanets},} \apjl, 986, L3, \dodoi{10.3847/2041-8213/adda39}

% type= article
\bibitem[{J.~A. {Patel} {et~al.}(2024){Patel}, {Brandeker}, {Kitzmann}, {Petit dit de la Roche}, {Bello-Arufe}, {Heng}, {Meier Vald{\'e}s}, {Persson}, {Zhang}, {Demory}, {Bourrier}, {Deline}, {Ehrenreich}, {Fridlund}, {Hu}, {Lendl}, {Oza}, {Alibert}, \& {Hooton}}]{Patel2024}
{Patel}, J.~A., {Brandeker}, A., {Kitzmann}, D., {et~al.} 2024, \bibinfo{title}{{JWST reveals the rapid and strong day-side variability of 55 Cancri e},} \aap, 690, A159, \dodoi{10.1051/0004-6361/202450748}

% type= article
\bibitem[{C.~M. {Pieters} \& S.~K. {Noble}(2016){Pieters} \& {Noble}}]{Pieters2016}
{Pieters}, C.~M., \& {Noble}, S.~K. 2016, \bibinfo{title}{{Space weathering on airless bodies},} Journal of Geophysical Research (Planets), 121, 1865, \dodoi{10.1002/2016JE005128}

% type= article
\bibitem[{C.~M. Pieters {et~al.}(2000)Pieters, Taylor, Noble, Keller, Hapke, Morris, Allen, McKay, \& Wentworth}]{Pieters2000}
Pieters, C.~M., Taylor, L.~A., Noble, S.~K., {et~al.} 2000, \bibinfo{title}{Space weathering on airless bodies: Resolving a mystery with lunar samples,} Meteoritics \& Planetary Science, 35, 1101, \dodoi{https://doi.org/10.1111/j.1945-5100.2000.tb01496.x}

% type= article
\bibitem[{A.~A.~A. {Piette} {et~al.}(2022){Piette}, {Madhusudhan}, \& {Mandell}}]{Piette2022}
{Piette}, A. A.~A., {Madhusudhan}, N., \& {Mandell}, A.~M. 2022, \bibinfo{title}{{HyDRo: atmospheric retrieval of rocky exoplanets in thermal emission},} \mnras, 511, 2565, \dodoi{10.1093/mnras/stab3612}

% type= article
\bibitem[{M.~N. Polyanskiy(2024)Polyanskiy}]{Polyanskiy2024}
Polyanskiy, M.~N. 2024, \bibinfo{title}{Refractiveindex.info database of optical constants,} Scientific Data, 11, 94, \dodoi{10.1038/s41597-023-02898-2}

% type= inproceedings
\bibitem[{K.~M. {Pontoppidan} {et~al.}(2016){Pontoppidan}, {Pickering}, {Laidler}, {Gilbert}, {Sontag}, {Slocum}, {Sienkiewicz}, {Hanley}, {Earl}, {Pueyo}, {Ravindranath}, {Karakla}, {Robberto}, {Noriega-Crespo}, \& {Barker}}]{Pontoppidan2016}
{Pontoppidan}, K.~M., {Pickering}, T.~E., {Laidler}, V.~G., {et~al.} 2016, \bibinfo{title}{{Pandeia: a multi-mission exposure time calculator for JWST and WFIRST},} in Society of Photo-Optical Instrumentation Engineers (SPIE) Conference Series, Vol. 9910, Observatory Operations: Strategies, Processes, and Systems VI, ed. A.~B. {Peck}, R.~L. {Seaman}, \& C.~R. {Benn}, 991016, \dodoi{10.1117/12.2231768}

% type= article
\bibitem[{L.~S. {Rothman} {et~al.}(2010){Rothman}, {Gordon}, {Barber}, {Dothe}, {Gamache}, {Goldman}, {Perevalov}, {Tashkun}, \& {Tennyson}}]{Rothman2010}
{Rothman}, L.~S., {Gordon}, I.~E., {Barber}, R.~J., {et~al.} 2010, \bibinfo{title}{{HITEMP, the high-temperature molecular spectroscopic database},} \jqsrt, 111, 2139, \dodoi{10.1016/j.jqsrt.2010.05.001}

% type= article
\bibitem[{N. {Scarsdale} {et~al.}(2024){Scarsdale}, {Wogan}, {Wakeford}, {Wallack}, {Batalha}, {Alderson}, {Aguichine}, {Wolfgang}, {Teske}, {Moran}, {L{\'o}pez-Morales}, {Kirk}, {Gordon}, {Gao}, {Batalha}, {Alam}, \& {Adams Redai}}]{Scarsdale2024}
{Scarsdale}, N., {Wogan}, N., {Wakeford}, H.~R., {et~al.} 2024, \bibinfo{title}{{JWST COMPASS: The 3{\textendash}5 {\ensuremath{\mu}}m Transmission Spectrum of the Super-Earth L 98-59 c},} \aj, 168, 276, \dodoi{10.3847/1538-3881/ad73cf}

% type= inproceedings
\bibitem[{J. {Skilling}(2004){Skilling}}]{Skilling2004}
{Skilling}, J. 2004, \bibinfo{title}{{Nested Sampling},} in American Institute of Physics Conference Series, Vol. 735, Bayesian Inference and Maximum Entropy Methods in Science and Engineering: 24th International Workshop on Bayesian Inference and Maximum Entropy Methods in Science and Engineering, ed. R.~{Fischer}, R.~{Preuss}, \& U.~V. {Toussaint} (AIP), 395--405, \dodoi{10.1063/1.1835238}

% type= article
\bibitem[{M. {Sneep} \& W. {Ubachs}(2005){Sneep} \& {Ubachs}}]{Sneep2005}
{Sneep}, M., \& {Ubachs}, W. 2005, \bibinfo{title}{{Direct measurement of the Rayleigh scattering cross section in various gases},} \jqsrt, 92, 293, \dodoi{10.1016/j.jqsrt.2004.07.025}

% type= article
\bibitem[{J.~R. {Spencer}(1990){Spencer}}]{Spencer1990}
{Spencer}, J.~R. 1990, \bibinfo{title}{{A rough-surface thermophysical model for airless planets},} \icarus, 83, 27, \dodoi{10.1016/0019-1035(90)90004-S}

% type= article
\bibitem[{M.~B. {Syal} {et~al.}(2015){Syal}, {Schultz}, \& {Riner}}]{Syal2015}
{Syal}, M.~B., {Schultz}, P.~H., \& {Riner}, M.~A. 2015, \bibinfo{title}{{Darkening of Mercury's surface by cometary carbon},} Nature Geoscience, 8, 352, \dodoi{10.1038/ngeo2397}

% type= article
\bibitem[{P. {Tamburo} {et~al.}(2018){Tamburo}, {Mandell}, {Deming}, \& {Garhart}}]{Tamburo2018}
{Tamburo}, P., {Mandell}, A., {Deming}, D., \& {Garhart}, E. 2018, \bibinfo{title}{{Confirming Variability in the Secondary Eclipse Depth of the Super-Earth 55 Cancri e},} \aj, 155, 221, \dodoi{10.3847/1538-3881/aabd84}

% type= article
\bibitem[{R. {Thalman} {et~al.}(2014){Thalman}, {Zarzana}, {Tolbert}, \& {Volkamer}}]{Thalman2014}
{Thalman}, R., {Zarzana}, K.~J., {Tolbert}, M.~A., \& {Volkamer}, R. 2014, \bibinfo{title}{{Rayleigh scattering cross-section measurements of nitrogen, argon, oxygen and air},} \jqsrt, 147, 171, \dodoi{10.1016/j.jqsrt.2014.05.030}

% type= article
\bibitem[{G. {Tovar Mendoza} {et~al.}(2022){Tovar Mendoza}, {Davenport}, {Agol}, {Jackman}, \& {Hawley}}]{TovarMendoza2022}
{Tovar Mendoza}, G., {Davenport}, J. R.~A., {Agol}, E., {Jackman}, J. A.~G., \& {Hawley}, S.~L. 2022, \bibinfo{title}{{Llamaradas Estelares: Modeling the Morphology of White-light Flares},} \aj, 164, 17, \dodoi{10.3847/1538-3881/ac6fe6}

% type= article
\bibitem[{R. {Trotta}(2008){Trotta}}]{Trotta2008}
{Trotta}, R. 2008, \bibinfo{title}{{Bayes in the sky: Bayesian inference and model selection in cosmology},} Contemporary Physics, 49, 71, \dodoi{10.1080/00107510802066753}

% type= article
\bibitem[{M. {Turbet} {et~al.}(2018){Turbet}, {Bolmont}, {Leconte}, {Forget}, {Selsis}, {Tobie}, {Caldas}, {Naar}, \& {Gillon}}]{Turbet2018}
{Turbet}, M., {Bolmont}, E., {Leconte}, J., {et~al.} 2018, \bibinfo{title}{{Modeling climate diversity, tidal dynamics and the fate of volatiles on TRAPPIST-1 planets},} \aap, 612, A86, \dodoi{10.1051/0004-6361/201731620}

% type= article
\bibitem[{N. {Tusay} {et~al.}(2025){Tusay}, {Wright}, {Beatty}, {Desch}, {Col{\'o}n}, {Mittal}, {Osborn}, {Campos Estrada}, {Owen}, {Libby-Roberts}, {Gupta}, {Foley}, {Meier Vald{\'e}s}, {Stevens}, \& {Herbst}}]{Tusay2025}
{Tusay}, N., {Wright}, J.~T., {Beatty}, T.~G., {et~al.} 2025, \bibinfo{title}{{A Disintegrating Rocky World Shrouded in Dust and Gas: Mid-IR Observations of K2-22b using JWST},} arXiv e-prints, arXiv:2501.08301, \dodoi{10.48550/arXiv.2501.08301}

% type= article
\bibitem[{G. {Van Looveren} {et~al.}(2025){Van Looveren}, {Boro Saikia}, {Herbort}, {Schleich}, {G{\"u}del}, {Johnstone}, \& {Kislyakova}}]{VanLooveren2025}
{Van Looveren}, G., {Boro Saikia}, S., {Herbort}, O., {et~al.} 2025, \bibinfo{title}{{Habitable Zone and Atmosphere Retention Distance (HaZARD): Stellar-evolution-dependent loss models of secondary atmospheres},} \aap, 694, A310, \dodoi{10.1051/0004-6361/202452998}

% type= article
\bibitem[{P. {Wachiraphan} {et~al.}(2025){Wachiraphan}, {Berta-Thompson}, {Diamond-Lowe}, {Winters}, {Murray}, {Zhang}, {Xue}, {Morley}, {Rosario-Franco}, \& {Duvvuri}}]{Wachiraphan2025}
{Wachiraphan}, P., {Berta-Thompson}, Z.~K., {Diamond-Lowe}, H., {et~al.} 2025, \bibinfo{title}{{The Thermal Emission Spectrum of the Nearby Rocky Exoplanet LTT 1445A b from JWST MIRI/LRS},} \aj, 169, 311, \dodoi{10.3847/1538-3881/adc990}

% type= article
\bibitem[{M. {Weiner Mansfield} {et~al.}(2024){Weiner Mansfield}, {Xue}, {Zhang}, {Mahajan}, {Ih}, {Koll}, {Bean}, {Coy}, {Eastman}, {Kempton}, \& {Kite}}]{WeinerMansfield2024}
{Weiner Mansfield}, M., {Xue}, Q., {Zhang}, M., {et~al.} 2024, \bibinfo{title}{{No Thick Atmosphere on the Terrestrial Exoplanet Gl 486b},} \apjl, 975, L22, \dodoi{10.3847/2041-8213/ad8161}

% type= incollection
\bibitem[{J.~N. {Winn}(2010){Winn}}]{Winn2010}
{Winn}, J.~N. 2010, \bibinfo{title}{{Exoplanet Transits and Occultations},} in Exoplanets, ed. S.~{Seager}, 55--77, \dodoi{10.48550/arXiv.1001.2010}

% type= article
\bibitem[{R. {Wordsworth}(2015){Wordsworth}}]{Wordsworth2015}
{Wordsworth}, R. 2015, \bibinfo{title}{{Atmospheric Heat Redistribution and Collapse on Tidally Locked Rocky Planets},} \apj, 806, 180, \dodoi{10.1088/0004-637X/806/2/180}

% type= article
\bibitem[{G.~S. {Wright} {et~al.}(2023){Wright}, {Rieke}, {Glasse}, {Ressler}, {Garc{\'\i}a Mar{\'\i}n}, {Aguilar}, {Alberts}, {{\'A}lvarez-M{\'a}rquez}, {Argyriou}, {Banks}, {Baudoz}, {Boccaletti}, {Bouchet}, {Bouwman}, {Brandl}, {Breda}, {Bright}, {Cale}, {Colina}, {Cossou}, {Coulais}, {Cracraft}, {De Meester}, {Dicken}, {Engesser}, {Etxaluze}, {Fox}, {Friedman}, {Fu}, {Gasman}, {G{\'a}sp{\'a}r}, {Gastaud}, {Geers}, {Glauser}, {Gordon}, {Greene}, {Greve}, {Grundy}, {G{\"u}del}, {Guillard}, {Haderlein}, {Hashimoto}, {Henning}, {Hines}, {Holler}, {Detre}, {Jahromi}, {James}, {Jones}, {Justtanont}, {Kavanagh}, {Kendrew}, {Klaassen}, {Krause}, {Labiano}, {Lagage}, {Lambros}, {Larson}, {Law}, {Lee}, {Libralato}, {Lorenzo Alverez}, {Meixner}, {Morrison}, {Mueller}, {Murray}, {Mycroft}, {Myers}, {Nayak}, {Naylor}, {Nickson}, {Noriega-Crespo}, {{\"O}stlin}, {O'Sullivan}, {Ottens}, {Patapis}, {Penanen}, {Pietraszkiewicz}, {Ray}, {Regan}, {Roteliuk}, {Royer}, {Samara-Ratna}, {Samuelson}, {Sargent}, {Scheithauer},
  {Schneider}, {Schreiber}, {Shaughnessy}, {Sheehan}, {Shivaei}, {Sloan}, {Tamas}, {Teague}, {Temim}, {Tikkanen}, {Tustain}, {van Dishoeck}, {Vandenbussche}, {Weilert}, {Whitehouse}, \& {Wolff}}]{Wright2023}
{Wright}, G.~S., {Rieke}, G.~H., {Glasse}, A., {et~al.} 2023, \bibinfo{title}{{The Mid-infrared Instrument for JWST and Its In-flight Performance},} \pasp, 135, 048003, \dodoi{10.1088/1538-3873/acbe66}

% type= article
\bibitem[{Q. {Xue} {et~al.}(2024){Xue}, {Bean}, {Zhang}, {Mahajan}, {Ih}, {Eastman}, {Lunine}, {Mansfield}, {Coy}, {Kempton}, {Koll}, \& {Kite}}]{Xue2024}
{Xue}, Q., {Bean}, J.~L., {Zhang}, M., {et~al.} 2024, \bibinfo{title}{{JWST Thermal Emission of the Terrestrial Exoplanet GJ 1132b},} \apjl, 973, L8, \dodoi{10.3847/2041-8213/ad72e9}

% type= article
\bibitem[{K.~J. {Zahnle} \& D.~C. {Catling}(2017){Zahnle} \& {Catling}}]{Zahnle2017}
{Zahnle}, K.~J., \& {Catling}, D.~C. 2017, \bibinfo{title}{{The Cosmic Shoreline: The Evidence that Escape Determines which Planets Have Atmospheres, and what this May Mean for Proxima Centauri B},} \apj, 843, 122, \dodoi{10.3847/1538-4357/aa7846}

% type= article
\bibitem[{M. {Zhang} {et~al.}(2024){Zhang}, {Hu}, {Inglis}, {Dai}, {Bean}, {Knutson}, {Lam}, {Goffo}, \& {Gandolfi}}]{Zhang2024}
{Zhang}, M., {Hu}, R., {Inglis}, J., {et~al.} 2024, \bibinfo{title}{{GJ 367b Is a Dark, Hot, Airless Sub-Earth},} \apjl, 961, L44, \dodoi{10.3847/2041-8213/ad1a07}

% type= article
\bibitem[{S. {Zieba} {et~al.}(2023){Zieba}, {Kreidberg}, {Ducrot}, {Gillon}, {Morley}, {Schaefer}, {Tamburo}, {Koll}, {Lyu}, {Acu{\~n}a}, {Agol}, {Iyer}, {Hu}, {Lincowski}, {Meadows}, {Selsis}, {Bolmont}, {Mandell}, \& {Suissa}}]{Zieba2023}
{Zieba}, S., {Kreidberg}, L., {Ducrot}, E., {et~al.} 2023, \bibinfo{title}{{No thick carbon dioxide atmosphere on the rocky exoplanet TRAPPIST-1 c},} \nat, 620, 746, \dodoi{10.1038/s41586-023-06232-z}

\end{thebibliography}
\bibliographystyle{aasjournalv7}



\end{document}